\documentclass[a4paper,10pt,twoside,bibnote]{cpc-hepnp}

\usepackage{graphicx}
\usepackage{amssymb}
\usepackage{amsmath}
\usepackage{color}
\usepackage{float}
\usepackage{epstopdf}
\usepackage{lastpage}
\usepackage{CJK}

\allowdisplaybreaks[4]

\newcommand{\lsim}{ \mathop{}_ {\textstyle \sim}^{\textstyle <} }

\newcommand{\bear}{\begin{array}}
\newcommand {\eear}{\end{array}}

\newcommand{\beq}{\begin {equation}}
\newcommand{\eeq}{\end   {equation}}
\newcommand{\bea}{\begin {eqnarray}}
\newcommand{\eea}{\end   {eqnarray}}
\newcommand{\baa}{\begin {array}   }
\newcommand{\eaa}{\end   {array}   }
\newcommand{\bit}{\begin {itemize} }
\newcommand{\eit}{\end   {itemize} }
\newcommand{\be }{\begin {equation}}
\newcommand{\ee }{\end   {equation}}
\newcommand{\nn }{\nonumber        }

\newcommand{\bef}{\begin{figure}}
\newcommand {\eef}{\end{figure}}
\newcommand{\bec}{\begin{center}}
\newcommand {\eec}{\end{center}}

\newcommand{\slu}{s_L^t}
\newcommand{\sld}{s_L^b}

\newcommand{\clu}{c_L^t}
\newcommand{\cld}{c_L^b}

\newcommand{\sru}{s_R^t}
\newcommand{\srd}{s_R^b}

\newcommand{\Zm}{Z_{\mu}}
\newcommand{\ts}{$(T)$}
\newcommand{\bs}{$(B)$}
\newcommand{\xt}{$(X,T)$}
\newcommand{\tb}{$(T,B)$}
\newcommand{\by}{$(B,Y)$}
\newcommand{\xtb}{$(X,T,B)$}
\newcommand{\tby}{$(T,B,Y)$}
\newcommand{\sqt}{\sqrt{2}}

\begin{document}

\title{A General Analysis of $Wtb$ anomalous Couplings}

\author{
Qing-Hong Cao$^{1,2,3;1)}$\email{qinghongcao@pku.edu.cn}%
\quad Bin Yan$^{1;2)}$\email{binyan@pku.edu.cn}%
\quad Jiang-Hao Yu$^{4;3}$\email{jhyu@physics.umass.edu}%
\quad Chen Zhang$^{1;4}$\email{larry@pku.edu.cn}
}

\maketitle

\address{%
$^1$ Department of Physics and State Key Laboratory of Nuclear Physics and Technology, Peking University, Beijing 100871, China\\
$^2$ Collaborative Innovation Center of Quantum Matter, Beijing 100871, China\\
$^3$ Center for High Energy Physics, Peking University, Beijing 100871, China\\
$^4$ Amherst Center for Fundamental Interactions, Department of Physics,
University of Massachusetts-Amherst, Amherst, MA 01003, U.S.A.
}

\begin{abstract}
We investigate new physics effects on the $Wtb$ effective couplings in the model-independent framework. The new physics effects can be parametrized by four independent couplings $f_1^L$, $f_1^R$, $f_2^L$ and $f_2^R$. We further introduce a set of parameters $x_0$, $x_m$, $x_p$ and $x_5$ which exhibit the linear relation to the single top production cross sections.  Using the recent data of $t$-channel single top production cross section $\sigma_t$, $tW$ associated production cross section $\sigma_{tW}$, $s$-channel single top production cross section $\sigma_s$ and $W$-helicity fractions $F_0$, $F_L$ and $F_R$ collected at the 8 TeV LHC and Tevatron, we perform a global fit to impose constraints on the top quark effective couplings. Our global fitting results show that the top quark effective couplings are strongly correlated. We show that (i) improving the measurements of $\sigma_t$ and $\sigma_{tW}$ is important in constraining the correlation of $(f_1^R,f_2^R)$ and $(f_2^L,f_2^R)$; (ii) $f_1^L$ and $f_2^R$ are anti-correlated, which are  sensitive to all the four experiments; (iii) $f_1^R$ and $f_2^L$ are also anti-correlated, which are sensitive to the $F_0$ and $F_L$ measurements; (iv) the correlation between $f_2^L$ and $f_2^R$ is sensitive to the precision of $\sigma_t$, $\sigma_{tW}$ and $F_0$ measurements. The effective $Wtb$ couplings are studied in three kinds of new physics models: the $G(221)=SU(2)_1\otimes SU(2)_2\otimes U(1)_X$ models, the vector-like quark models and the Littlest Higgs model with and without $T$-parity. We show that the $Wtb$ couplings in the left-right model and the un-unified model are sensitive to the ratio of gauge couplings  when the new heavy gauge boson's mass ($M_{W^\prime}$) is less than several hundred GeV, but the constraint is loose if  $M_{W^\prime}>1$ TeV.
On the other hand, the $Wtb$ couplings in vector-like quark models and the Littlest Higgs models are sensitive to the mixing angles of new heavy particles and SM particles. 
\end{abstract}



\section{INTRODUCTION \label{sec:1}}

The discovery of the Higgs boson at the Large Hadron Collider (LHC)~\cite{Aad:2012tfa,Chatrchyan:2012ufa} completes the Standard Model (SM) of particle physics.  So far there is no new physics (NP) evidence observed at the LHC. It is possible that the scale of new physics $\Lambda_{\rm NP}$ is much higher than the electroweak symmetry breaking (EWSB) scale ($v=246$ GeV).
If so, we expect to look for indirect effects of NP with the SM particles which we have known.
The top quark, discovered at the Tevatron~\cite{Abe:1995hr,Abachi:1995iq}, is the heaviest particle of the SM. With its mass around the EWSB scale, the top quark is believed to play an important role to connect SM and NP. 
The $Wtb$ coupling plays a pivotal role in top quark physics. The top quark decay and single top quark production processes are sensitive to the $Wtb$ coupling. In addition, it offers a very promising way to probe the NP at the LHC.
For instance, the production rate of single top quark and the polarization of the top quark can be modified by NP beyond the SM, like new gauge boson $W^{\prime}$s, vector-like fermions, etc.
Those heavy particles are predicted in many NP models, such as $SU(2)_1\otimes SU(2)_2\otimes U(1)_X$, often denoted as $G(221)$ models~\cite{Hsieh:2010zr,Cao:2012ng}, vector-like quark models~\cite{delAguila:2000rc,AguilarSaavedra:2009es,Cacciapaglia:2010vn,Aguilar-Saavedra:2013qpa}, little Higgs models~\cite{Belyaev:2006jh,Han:2003wu,Penunuri:2008pb}, warped/composite simplified models~\cite{Contino:2006nn} and many others.
It is convenient to classify the underlying theories according to  different approaches that modify the $Wtb$ coupling. A simple case is that the new heavy particles mixing with SM particles at tree-level, like a new gauge boson $W^{\prime}$ or a new fermion $T (B)$. Another way to generate the anomalous $Wtb$ coupling is through the loop-level. A typical example is supersymmetric models~\cite{Dabelstein:1995jt,Cao:2003yk} and two Higgs doublet models~\cite{Grzadkowski:1991nj}.

The top quark decays before hadronization as its lifetime  is much smaller than the typical hadronization time scale.
In the SM, the dominant decay mode of top quark is $t\rightarrow W^+b$, so the $Wtb$ coupling governs top quark decay process. One way to probe structure of the $Wtb$ coupling is to study the helicity fractions of the $W$ boson in top quark decay.
The $W$-boson helicity fractions are defined as the partial rate for a given helicity state divided by the total decay rate: $F_{L,R,0}\equiv \Gamma_{L,R,0}/\Gamma$, where $F_L$, $F_R$, $F_0$ are the left-handed, right-handed and longitudinal helicity fractions, respectively.
The $W$ helicity fractions calculated with an accuracy of
the next-to-next-to-leading order (NNLO) QCD corrections in the SM are $F_0=0.687\pm 0.005$, $F_L=0.311\pm0.005$, $F_R=0.0017\pm 0.0001$ for a top quark mass of $m_t=172.8\pm1.3$ GeV~\cite{Czarnecki:2010gb,CMS:2013pfa}.
Recently, ATLAS and CMS collaborations have measured the $W$ helicity fractions in top quark decay at $\sqrt{s}=7$ TeV while CMS $8$ TeV results are also available.
These results are consistent with the SM predictions~\cite{ATLAS:2013tla,CMS:2013pfa}.

At hadron colliders, the top quark can be produced singly in three channels:  $t$-channel process ($ub\rightarrow td$), $tW$ associate production process ($bg\rightarrow tW^-$) and $s$-channel process ($u\bar{d}\rightarrow W^+\rightarrow t\bar{b}$). The $t$-channel cross section is predicted at next-to-next-to-leading order plus the contribution due to the resummation of soft-gluon bremsstrahlung (NNLO+NNLL) to be $\sigma_t=87.8^{+3.4}_{-1.9}$ pb~\cite{Kidonakis:2011wy,Kidonakis:2012db} at the LHC with $\sqrt{s}=8$ TeV.
The theoretical prediction for $tW$-channel and $s$-channel at $\sqrt{s}=8$ TeV at NLO$+$NNLL precision in QCD are $\sigma_{Wt}=22.4\pm 1.5$ pb~\cite{Kidonakis:2010ux}, $\sigma_s=5.6\pm 0.2$ pb~\cite{Kidonakis:2010tc}, respectively.
While, the cross section at $\sqrt{s}=13$ TeV are $\sigma_t=221^{+6}_{-2}\pm3$ pb, $\sigma_{Wt}=72.6\pm 1.3\pm1.3$ pb and $\sigma_{s}=11.29\pm0.18\pm0.26$ pb~\cite{Kidonakis:2016smr}. Recently, both ATLAS and CMS collaborations have measured the single top quark production cross sections at $\sqrt{s}=7$ TeV,  $\sqrt{s}=8$ TeV and $\sqrt{s}=13$ TeV, which are summarized in Sec.~{\ref{sec:fitting}}.

In this work, we utilize the effective field theory (EFT) to calculate the single top quark production cross sections, $W$ helicity fractions, and use the recent experimental data ( denotes as $Wtb$ measurements) to determine the general $Wtb$ couplings.
Model independent analyses of the $Wtb$ couplings have been performed using the EFT approach, see for example in Refs.~\cite{Kane:1991bg,Malkawi:1994tg,Carlson:1994bg,Whisnant:1997qu,Yang:1997iv,Cao:1998at,Hikasa:1998wx,Larios:1999au,Lin:2001yq,Espriu:2001vj,Chen:2005vr,Batra:2006iq,
Cao:2007ea,AguilarSaavedra:2008gt,Berger:2009hi,Zhang:2010dr,AguilarSaavedra:2010nx,Rindani:2011pk,Rindani:2011gt,Bach:2012fb,Fabbrichesi:2014wva,Bernardo:2014vha,Sarmiento-Alvarado:2014eha,Bach:2014zca}.
In our work, we compute the deviation from the SM by including the dimension-6 operators for top quark decay and single top quark production processes.
We incorporate the quadratic terms of dimension-6 operators  to obtain the correlations among different operators in the single top quark cross section and $W$ helicity fraction calculations.
The constraints on $Wtb$ couplings based on some of the recent experimental data were studied in Refs.~\cite{Fabbrichesi:2014wva,Bernardo:2014vha}.
We perform a global fit of the general $Wtb$ couplings by analysing the correlations among different couplings and discuss the implication of the top quark effective theory on serveral NP models.

The paper is organized as follows. In Sec.~{\ref{sec:efttop}} we calculate the single top quark production cross sections via an effective field theory approach.
In Sec.~{\ref{sec:fitting}} we present the allowed parameter space of the general $Wtb$ couplings after incorporating the most recent ATLAS and CMS results. In Sec.~{\ref{sec:couplings}} we discuss the constraints on various new physics models from the $Wtb$ couplings. Finally, we conclude in Sec.~{\ref{sec:conclusion}}.

\section{Top quark effective field theory}
\label{sec:efttop}

\subsection{Dimension-6 Operators and Effective $Wtb$ Couplings}

Using the EFT approach to explore the possible NP effects has been discussed widely, see for example in Refs.~\cite{Buchmuller:1985jz,Peccei:1989kr,Malkawi:1994tg,Georgi:1994qn,Larios:1996ib,Whisnant:1997qu,Yang:1997iv,
Hikasa:1998wx,Tait:2000sh,AguilarSaavedra:2008zc,Berger:2009hi,Drobnak:2010ej,Degrande:2014tta}.
A model independent way to parametrize the low energy effects of NP theories is the linearly realized effective Lagrangian, which incorporates the particle content and symmetries of the SM.
The nonlinearly realized mechanism of the electroweak symmetry is studied in Refs.~\cite{Peccei:1989kr,Malkawi:1994tg,Larios:1996ib,Tait:2000sh}. In this paper,  we assume the new scalar particle which observed at the LHC is the SM Higgs boson and use the linear realization to parametrize the NP effects~\cite{Buchmuller:1985jz,AguilarSaavedra:2008zc}.
The effective Lagrangian before the electroweak symmetry breaking is
\bea
\mathcal{L}_{\rm eff} =\mathcal{L}_{\rm SM}+ \sum_i \frac{ C_i}{\Lambda^2} O_i+
\mathcal{O}(\dfrac{1}{\Lambda^3}),
\eea
where $\mathcal{L}_{\rm SM}$ is the SM Lagrangian, $\Lambda$ is the characteristic scale of new physics, $O_i$ denotes $SU(3)_c\otimes SU(2)_L\otimes U(1)_Y$ gauge invariant dimension-6 operators, and $C_i$ is corresponding to Wilson coefficient which represents the strength of the effective operator $O_i$. The dimension-5 operator violate the lepton number and is not considered in this work. The great agreements between the experimental measurements  and the SM predictions indicate that the NP effects should be small. Hence, we restrict ourselves to the dimension-6 operators in this work. The complete set of dimension-6 effective operators generating the anomalous $Wtb$ couplings is~\cite{Buchmuller:1985jz,AguilarSaavedra:2008zc}
\begin{align}\label{D6Op}
 O_{\phi q}^{(3)} & =i(\phi^\dag\tau^ID_\mu\phi)(\bar{q}_{L}\gamma^\mu\tau^Iq_{L}), &
 O_{\phi \phi} & =i(\tilde{\phi}^\dag D_\mu\phi)(\bar{t}_{R}\gamma^\mu b_{R}),\nn\\
 O_{Dt} & =(\bar{q}_{L}D_\mu t_{R})D^{\mu}\tilde{\phi}, &
 O_{\bar{D}t} & =(D_\mu\bar{q}_{L}t_{R})D^{\mu}\tilde{\phi},\nn\\
 O_{Db} & =(\bar{q}_{L}D_\mu b_{R})D^{\mu}\phi, &
  O_{\bar{D}b} & =(D_\mu\bar{q}_{L}b_{R})D^{\mu}\phi, \nn\\
 O_{tW} & =(\bar{q}_{L}\sigma^{\mu\nu}\tau^It_{R})\tilde{\phi} W_{\mu\nu}^I, &
 O_{bW} & =(\bar{q}_{L}\sigma^{\mu\nu}\tau^Ib_{R})\phi W_{\mu\nu}^I,  \nn\\
 O_{qW} & =\bar{q}_{L}\gamma^{\mu}\tau^ID^{\nu}q_{L}W_{\mu\nu}^I.
\end{align}
where\,$q_{L}^T=(t,b)_L$\, denotes the $SU(2)_L$ weak doublet of the third generation left-handed quark fields, $t_{R}$ and $b_{R}$ are $SU(2)_L$ weak singlet of right-handed top and bottom-quark fields, $\phi$ is $SU(2)_L$ weak doublet of Higgs field, defined $\phi^T=\dfrac{1}{\sqrt{2}}(0,v+h)$ with $v=246$ GeV in the unitarity gauge  with $\tilde{\phi}=i\tau^2\phi^*$, and $D_{\mu}=\partial_{\mu}-ig(\tau^I/2)W_{\mu}^I-ig^{\prime}B_{\mu}Y$ is the covariant derivative, where $g$ and $g^\prime$ are gauge couplings of $SU(2)_L$ and $U(1)_Y$, respectively, and $Y$ is the hypercharge of the field to which $D_{\mu}$ is applied. $W_{\mu\nu}^I=\partial_{\mu}W_{\nu}^I-\partial_{\nu}W_{\mu}^I+g\varepsilon_{IJK}W_{\mu}^JW_{\nu}^K$ are the strength tensors of $SU(2)_L$ gauge fields and $\varepsilon_{IJK}$ denote the structure constants, and $\tau^I$ is the usual Pauli matrix.

Three types of the dimension-6 operators contribute to the $Wtb$ couplings: the first type is operators involving scalar field carrying one covariant derivative, the second is the operators involving fermion and scalar fields both carrying one covariant derivative, and the third one is the operators involving field strength tensor.

\begin{figure}
\centering
\includegraphics[scale=0.3]{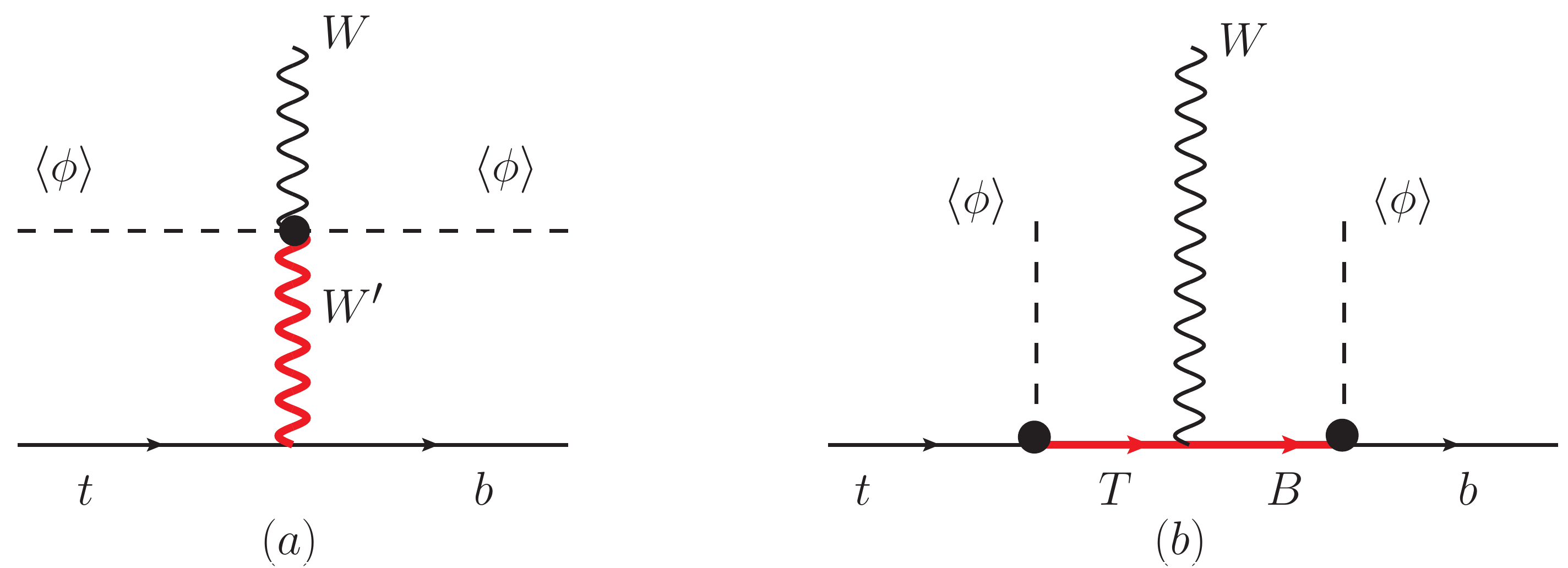}\\
\includegraphics[scale=0.3]{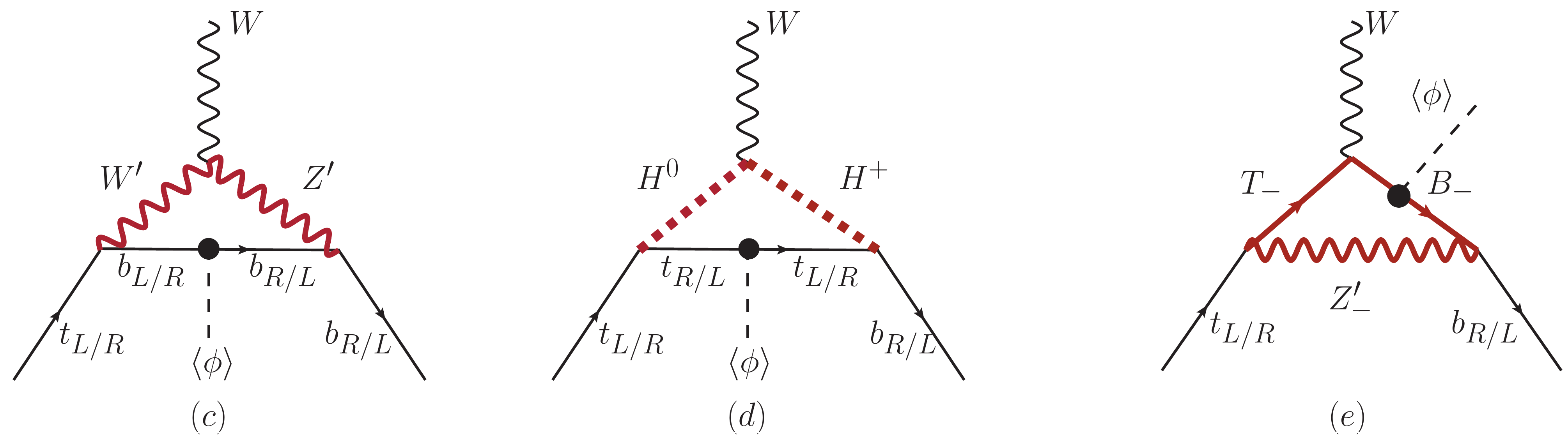}
\caption{Pictorial illustration of relation between effective operators and NP models. (a) and (b) denote the tree level mixing effect between the new heavy particles and SM particles, (c)-(e) denote the possible loop-induced dimension-6 operators diagram of some extension NP models.}
\label{Fig:operator}
\end{figure}

The operators $O_{\phi q}^{(3)}$ and $O_{\phi \phi}$ belong to the first type. It can be generated at the tree-level after integrating out the new heavy particles, such as a heavy charged vector boson ($W^{\prime\pm}$) that mixes with SM gauge boson $W^{\pm}$~\cite{Hsieh:2010zr,Cao:2012ng} or a heavy quark that mixes with top quark or bottom quark~\cite{Cacciapaglia:2010vn,Aguilar-Saavedra:2013qpa}.
A pictorial illustration of the relation between the tree-level effective operators and the possible NP models are shown in Fig.~\ref{Fig:operator}(a) and Fig.~\ref{Fig:operator}(b). We use the bold-red line to denote the NP particles.
The anomalous $Wtb$ couplings arise at tree-level after the spontaneous symmetry breaking with $\langle\phi\rangle=v/\sqrt{2}$.

The operators $O_{Dt}$, $O_{\bar{D}t}$, $O_{Db}$, $O_{\bar{D}b}$ fall into the second category. Those operators give a contribution of order $p^2/\Lambda^2$, where $p$ is the typical momentum scale in the process. Such operators, corresponding to the vertices involving three external lines can be induced only at the loop-level after integrating out the heavy particles~\cite{Arzt:1994gp}.

The operators $O_{tW}$, $O_{bW}$, $O_{qW}$ belong to the third type operators. Usually, the effective operators which involve the field strength tensor are generated only at loop level if the complete theory is a gauge theory~\cite{Arzt:1994gp}. Typical examples include supersymmetric models~\cite{Li:1992ga,Dabelstein:1995jt,Cao:2003yk}, two Higgs doublet models~\cite{Grzadkowski:1991nj}, etc.

A pictorial illustration of the relation between loop-induced effective operators and the possible underlying theories is shown in the Fig.~\ref{Fig:operator}(c-e).
The $Wtb$ coupling can be induced in the NP models with extended gauge structure, which yields extra gauge bosons (see Fig.~\ref{Fig:operator}(c)), or new scalar particles which consist of a new  charge scalar and neutral scalar (see  Fig.~\ref{Fig:operator}(d)), or new fermions which carrying one discrete quantum number to avoid the mixing with top quark and bottom quark at tree-level (see  Fig.~\ref{Fig:operator}(e)). We will comment on the impact of $Wtb$ measurements on several NP models in Sec.~{\ref{sec:couplings}}.

In usual, the effective Lagrangians consist of redundant terms, which could be removed by the classical equations of motion~\cite{Arzt:1993gz}. It is based on the equivalence theorem of the S matrix.
After we use the equations of motion to remove redundant operators, the relevant operators reduce to
\bea
\label{D6inOp}
 O_{\phi q}^{(3)} & =&i(\phi^\dag\tau^ID_\mu\phi)(\bar{q}_{L}\gamma^\mu\tau^Iq_{L}) \nn\\
 O_{\phi \phi} & =& i(\tilde{\phi}^\dag D_\mu\phi)(\bar{t}_{R}\gamma^\mu b_{R}), \nn \\
 O_{tW} & =& (\bar{q}_{L}\sigma^{\mu\nu}\tau^It_{R})\tilde{\phi} W_{\mu\nu}^I, \nn\\
 O_{bW} & =& (\bar{q}_{L}\sigma^{\mu\nu}\tau^Ib_{R})\phi W_{\mu\nu}^I.
\eea
Assuming new anomalous couplings arise from the above dimension-6 operators, we can parametrize the general effective $Wtb$ couplings as~\cite{Chen:2005vr}
\bea
\label{eff_lag}
   \mathcal{L}_{Wtb}&=&\frac{g}{\sqrt{2}}W_{\mu }^-\bar{b}\gamma ^{\mu }\left( (1 + f_1^L) P_L+f_1^RP_R\right)t \nn\\
   &-&\frac{g}{\sqrt{2}m_W}\partial_{\nu }W_{\mu }^-\bar{b}\sigma ^{\mu \nu }\left(f_2^LP_L+f_2^RP_R\right)t+h.c.,
\eea
where $P_{L,R}=(1\mp\gamma_{5})/2$ are the usual chirality projectors, $m_W$ is the $W$-boson's mass and the Cabibbo-Kobayaski-Maskawa matrix element $V_{tb}$ is taken to be 1 in our analysis.
In the SM, the values of the coefficients $f_i$ vanish at the tree-level.
Those couplings could be generated if NP exists.
Although those couplings can in general be complex quantities, we assume the four coefficients are real in our calculation~\footnote{The complex anomalous couplings has studied in Ref.~\cite{Bernardo:2014vha} and ATLAS collaboration in Ref.~\cite{ATLAS:2013ula} also give a constraint on the imaginary part of $f_2^R$.}.
The coefficients of the effective $Wtb$ couplings are related to the Wilson coefficients of the dimensional-6 operators as follows~\cite{Whisnant:1997qu}:
\begin{align}\label{wilson}
   f_1^L & = \frac{C_{\phi q}^{(3)\ast}v^2}{\Lambda^2},   & f_1^R&=\frac{1}{2}C_{\phi\phi}^{\ast}\frac{v^2}{\Lambda^2},\nn\\
   f_2^L & =\sqrt{2}C_{bW}^{\ast}\frac{v^2}{\Lambda^2},  &
   f_2^R&=\sqrt{2}C_{tW}\frac{v^2}{\Lambda^2}.
\end{align}

\subsection{Single Top Production}

In this section we discuss the contribution of general effective $Wtb$ couplings to the total cross sections of the single top production and the $W$ helicity fractions in the top quark decay. In this work, we focus on the four independent operators, see Eq.~\ref{D6inOp}.
The top-quark can be produced singly through the electroweak interaction. Depending on the kinematic of the $W$-boson involved, the single-top production are usually subcategorized into three channels: the $s$-channel production ($q_W^2 >0$), the $t$-channel production ($q_W^2<0$) and the $tW$ associate production ($q_W^2=m_W^2$) where $q_W$ denotes the four momenta of the $W$-boson.

We separate the total cross section of the single-top production into the SM contribution plus the contributions from anomalous $Wtb$ couplings
\beq
\sigma_i = \sigma_i^{\rm SM} + K \Delta \sigma_i,
\eeq
where $\sigma_{i}^{\rm SM}$ denotes the cross section of the $i$-channel ($i=s,t,tW$) single-top production in the SM with the approximate NNLO
 QCD correction and $\Delta \sigma_i$ denotes the variation from the SM prediction induced by the anomalous couplings at the tree level.  The $K$-factor, defined as $K\equiv \sigma_{\rm NNLO}^{\rm SM}/\sigma_{\rm LO}^{\rm SM}$, describes the approximate NNLO QCD corrections in the SM. We assume the anomalous couplings receives exactly the same corrections as the SM processes. Following Ref.~\cite{Chen:2005vr} we write the contributions of anomalous couplings to the single top productions as
\bea
\Delta\sigma_{t}    &=&a_0 x_0 + a_m x_m + a_p x_p +a_5 x_5,\nn\\
\Delta\sigma_{s}   &=&b_0 x_0 + b_m x_m + b_p x_p +b_5 x_5,\nn\\
\Delta\sigma_{tW}  &=&c_0 x_0 + c_m x_m + c_p x_p +c_5 x_5,
\label{eqvtw}
\eea
where we reparametrize the four coefficients $f_{1,2}^{L,R}$ as $x_0,x_m, x_p, x_5$~\cite{Chen:2005vr}
\begin{eqnarray}
\label{xnotation}
  x_0 & =& \left(1+f_1^L+\frac{f_2^R}{a_t}\right)^2+\left(f_1^R+\frac{f_2^L}{a_t}\right)^2-1,  \nn\\
  x_m & =& \left(1+f_1^L+a_tf_2^R\right)^2-1, \nn\\
  x_p & =& \left(f_1^R+a_tf_2^L\right)^2,  \nn\\
  x_5 & =& a_t^2\left[(f_2^{L})^2+(f_2^{R})^2\right],
\end{eqnarray}
with $a_t \equiv m_t/m_W$. Note that the terms proportional to the bottom quark mass have been ignored in our calculation due to the suppressed factor $(m_b/m_W)^2$~\footnote{We deem the negligence of bottom quark mass in the present work reasonable because the inclusion of bottom mass only induces a tiny asymmetry of about $0.01\sim 0.02$~\cite{Bernardo:2014vha,Fabbrichesi:2014wva}  on the allowed region of $f_{1,2}^{L,R}$ whereas the length of the marginal allowed region of one variable in the global fit is typically of $\mathcal{O}(1)$, see Fig.~\ref{Fig:scan3}. Even if only one parameter such as $f_1^R$ is allowed to vary whereas all the other parameters are turned off, the length of the allowed region reaches about 0.2 (see Fig.~\ref{Fig:scan1}) which is still larger than the amount of asymmetry induced by bottom mass.}.

The coefficients ($a_i$, $b_i$ and $c_i$) depend on the collider type and energy and have to be calculated numerically. Ref.~\cite{Chen:2005vr} calculated the $a_i$'s and $b_i$'s at the Tevatron Run II and 14~TeV LHC. In this work we update both the $a_i$'s and $b_i$'s using the CTEQ6L parton distribution functions (PDFs)~\cite{Pumplin:2002vw} at the 7~TeV, 8~TeV and 13~TeV LHC. The numerical values of the coefficients are shown in Table~~\ref{coeff}.

\begin{table}
\caption{The coefficients $a_i$, $b_i$ and $c_i$ in Eq.~\ref{eqvtw} for the single top quark production. All the coefficients are in the unit of Picobarn.
The SM cross section at the approximate NNLO with $m_t=173$ GeV~( $\sigma_i^{SM}$ where $i=\{t,s,tW\}$ channels) and the $K_i$ factors~\cite{Stelzer:1997ns,Zhu:2002uj,Harris:2002md, Campbell:2004ch, Cao:2004ky, Cao:2004ap, Cao:2005pq,Campbell:2009ss, Heim:2009ku,Kidonakis:2010tc, Schwienhorst:2010je,Wang:2012dc,Kidonakis:2012rm,Kidonakis:2016smr} are also shown.}
\begin{center}
\label{coeff}
\begin{tabular}{l |c| c |c| c| c| c}
         \hline
         $t$-channel  & $a_0$    & $a_m$      & $a_p$      & $a_5$   & \quad$\sigma_{t}^{\rm SM}$ \quad&\quad $K$ \quad\\
         \hline
         LHC (7TeV\,$t$)         & 42.355   & -4.290  & -9.700  & 17.514 & 43.0 & 1.13 \\
         \hline
         LHC (7TeV\,$\bar{t}$)   & 24.251   &-5.315   & -2.514  & 9.748  & 22.9  & 1.21 \\
         \hline
         LHC (8TeV\,$t$)         & 56.060   & -5.990  & -12.727  & 23.582 & 56.4  & 1.13\\
         \hline
         LHC (8TeV\,$\bar{t}$)   & 32.846   & -7.139  & -3.594   & 13.423 & 30.7 & 1.19 \\
         \hline
         LHC (13TeV\,$t$)  & 142.763   & -17.718  & -31.487  & 63.636 & 138.0  & 1.10 \\
         \hline
         LHC (13TeV\,$\bar{t}$)  & 90.369   & -19.136  & -11.450   & 39.062 & 83.0 &1.16  \\
         \hline
         \hline
         $s$-channel   & $b_0$    & $b_m$    & $b_p$    & $b_5$   & $\sigma_{s}^{\rm SM}$ & $K$ \\
         \hline
         Tevatron ($t/\bar{t}$)          & $-0.099$   & 0.419    & 0.419   & 0.281  & 0.523 & 1.68 \\
         \hline
         LHC (8 TeV $t$)          & -0.724     &2.917     & 2.917   & 2.873  &3.79  &1.73  \\
         \hline
         LHC (8 TeV $\bar{t}$)     & -0.384     &1.584     & 1.584   & 1.364   &1.76  &1.47  \\
         \hline
         \hline
         $tW$ channel & $c_0$    & $c_m$     & $c_p$     & $c_5$    & $\sigma_{tW}^{\rm SM}$ & $K$\\
         \hline
         LHC (7TeV\,$t/\bar{t}$)        & 7.592 & -2.777 & -2.777 & 5.386 &7.8  & 1.62\\
         \hline
         LHC (8TeV\,$t/\bar{t}$)        & 11.095 & -4.055 & -4.055 & 7.990 & 11.1  &1.58 \\
         \hline
         LHC (13TeV\,$t/\bar{t}$) & 38.622 & -14.076 & -14.076 & 29.339 & 36.3  &1.48 \\
         \hline
        \end{tabular}
\end{center}
\label{coeffxi}
\end{table}

To obtain the coefficients\,($c_0,c_m,c_p,c_5$) of the $tW$-channel single-top production,
we rewrite the cross section in terms of different combinations of the effective couplings
\begin{equation}
\Delta\sigma_{tW}= \left[ (1+ f_1^L)^2 - 1 + (f_1^R)^2\right] \sigma_{0} +
\left[  (1+f_1^L)f_2^R   + f_1^R f_2^L \right] \sigma_{N1}
+ \left[ (f_2^L)^2  + (f_2^R)^2\right] \sigma_{N2},
\end{equation}
where $\sigma_{0}$ denotes the SM LO prediction while $\sigma_{N1}$ and $\sigma_{N2}$ represents the partial cross section that is proportional to the $f_2^R$  and $(f_2^R)^2$, respectively.
The cross sections $\sigma_{0}$, $\sigma_{N1}$ and $\sigma_{N2}$ are obtained by
integrating out the final state phase space, and then convoluting with the initial state PDFs. The coefficients \,($c_0,c_m,c_p,c_5$) are then determined from $\sigma_0$, $\sigma_{N1}$ and $\sigma_{N2}$ as following:
\begin{align}
  c_0 & =\frac{\sigma_{N1}a_t-2\sigma_{0}a_t^2}{2(1-a_t^2)}, &
  c_m & =\frac{2\sigma_{0}-\sigma_{N1}a_t}{2(1-a_t^2)}, \nn \\
  c_p & =\frac{2\sigma_{0}-\sigma_{N1}a_t}{2(1-a_t^2)},&
  c_5 & =\frac{\sigma_{N2}}{a_t^2}-\frac{\sigma_{N1}-2\sigma_{0}a_t}{2(1-a_t^2)a_t^3}-\frac{2\sigma_{0}-\sigma_{N1}a_t}{2(1-a_t^2)}.
\end{align}
The numerical values of the coefficients are given in Table~\ref{coeff} at the 7~TeV, 8~TeV and 13~TeV LHC.

With the help of the coefficients $a_i$, $b_i $ and $c_i$, the contributions of the $Wtb$ anomalous couplings can be written as:
\bea
\Delta\sigma_{\beta} &=& 2\Big(\beta_0+\beta_m\Big)f_1^L+2\left(\frac{1}{a_t}\beta_0+a_t\beta_m\right)f_2^R\nn\\
&+&\Big(\beta_0+\beta_m\Big)\left(f_1^{L}\right)^2 + \Big(\beta_0+\beta_p\Big)\left(f_1^{R}\right)^2\nn\\
&+&2\left(\frac{1}{a_t}\beta_0+a_t\beta_m\right)f_1^Lf_2^R +2\left(\frac{1}{a_t}\beta_0+a_t\beta_p\right)f_1^Rf_2^L\nn \\
&+&\left(\frac{1}{a_t^2}\beta_0+a_t^2\Big[\beta_p+\beta_5\Big]\right)\left(f_2^{L}\right)^2
                      +\left(\frac{1}{a_t^2}\beta_0+a_t^2\Big[\beta_m+\beta_5\Big]\right)\left(f_2^{R}\right)^2,
\label{eq:xsec}
\eea
where $\beta_i=a_i,b_i,c_i$ denotes the single top quark cross section coefficients of different channels.

\subsection{$W$ Helicity fractions in Top Decay}

For completeness, we also list the fraction of the $W$ helicity in the top quark decay in terms of $x_i$~\cite{Chen:2005vr},
\begin{align}\label{wpol}
  F_0 & =\frac{a_t^2(1+x_0)}{a_t^2(1+x_0)+2(1+x_m+x_p)}, \nn \\
  F_L & = \frac{2(1+x_m)}{a_t^2(1+x_0)+2(1+x_m+x_p)},\nn \\
  F_R & =\frac{2x_p}{a_t^2(1+x_0)+2(1+x_m+x_p)},
\end{align}
where $F_0$, $F_L$ and $F_R$ represent the fractions of $W$-boson with longitudinal polarization ($W_0$), left-handed polarization ($W_L$) and right-handed polarization ($W_R$). It is obvious that $x_0$ represents the contribution from the $W$ boson longitudinal polarization, $x_m$ and $x_p$ denote the contribution from $W$ boson left and right handed polarization, respectively.

Neglecting terms which are proportional to the bottom quark mass, the tree level results of the $W$ helicity fractions in the SM are
\bea\label{wpol2}
  F_0^{SM}  =\frac{a_t^2}{a_t^2+2}=0.70, \quad
  F_L^{SM}  = \frac{2}{a_t^2+2}=0.30, \quad
  F_R^{SM}  =0.
\eea
In the SM the top quark decays predominantly into the longitudinal $W$ boson because the coupling of top-quark to the longitudinal $W$ boson is similar to the Yukawa coupling, which is proportional to the top quark mass. The top quark can not decay into a right-handed $W$ boson owing to the purely left-handed $Wtb$ coupling in the SM. When the bottom quark mass is ignored, the right-handed $W$-boson is forbidden by the angular momentum conservation.

However, the anomalous coupling $f_1^R$ or $f_2^L$ can yield a right-handed $W$ boson in top-quark decay.
Different from the $f_1^R$ coupling, the $f_2^L$ contribution to a $W_R$ involves flipping the chirality of top-quark which gives rise to a factor of $a_t$. Therefore, the $F_R$ is proportional to $x_p = (f_1^R+a_tf_2^L)^2$. Similarly, the $f_2^R$ coupling can also produce a $W_L$ in the top-quark decay by flipping the top-quark's chirality.  As a result, the $F_L$ is proportional to $1+x_m = (1+f_1^L+a_tf_2^R)^2$.
All of the four effective $Wtb$ couplings can generate a $W_0$ in the top-quark decay. The $F_0$ is proportional to $a_t^2(1+x_0)=(a_t+a_t f_1^L+f_2^R)^2+(a_tf_1^R+f_2^L)^2$.

\section{Global Fit of the Effective $Wtb$ Couplings}
\label{sec:fitting}

\subsection{Experimental Data and Statistical Analysis}

The single top production cross sections and the $W$ helicity fractions have been measured
at the Tevatron and the LHC. The best measurement of the cross section of the $s$-channel single-top production is given at the Tevatron at $\sqrt{s}=1.96$ TeV with luminosity $9.7$ fb$^{-1}$~\cite{CDF:2014uma}. We also consider the updated experimental results of the $t$-channel and $tW$-channel cross sections and $W$-helicity measurements at both the CMS and ATLAS  collaborations.  All the experimental data are summarized in Table~\ref{measurements}.

\newcommand{\tabincell}[2]{\begin{tabular}{@{}#1@{}}#2\end{tabular}}
\begin{table}
\centering
\caption{Recent measurements of the cross sections for the single top-quark productions and the $W$ helicity fractions at the Tevatron and LHC.} \label{measurements}
\scalebox{0.75}[0.75]{%
\begin{tabular}{c|c|c|c}
\hline
   & \textbf{CMS} & \textbf{ATLAS} &\textbf{Tevatron}  \tabularnewline
\hline
\tabincell{c}{\textbf{ s-channel}\\ \textbf{(1.96 TeV)}} & $-$  & $-$ & $1.29_{-0.24}^{+0.26}$ pb~\cite{CDF:2014uma} \tabularnewline
\hline
\tabincell{c}{\textbf{ s-channel}\\ \textbf{(8 TeV)}} & $-$  & \tabincell{c}{$4.8\pm 1.1^{+2.2}_{-2.0}$ pb~\cite{ATLAS:2015047}\\(value$\pm$ stat$\pm$ sys)}&-\tabularnewline
\hline
\tabincell{c}{\textbf{ t-channel}\\ \textbf{(8 TeV)}} & \tabincell{c}{$83.6 \pm 2.3\pm 7.4$  pb~\cite{Khachatryan:2014iya}\\(value$\pm$ stat$\pm$ sys)} & \tabincell{c}{$82.6\pm 1.2\pm 11.4\pm 3.1\pm 2.3$  pb~\cite{ATLAS-CONF-2014-007}\\(value$\pm$ stat$\pm$ syst$\pm$ PDF$\pm$ lumi)} & $-$\tabularnewline
\hline
\tabincell{c}{\textbf{ t-channel}\\ \textbf{(13 TeV)}} & \tabincell{c}{$227.9 \pm 9.1\pm 14.0 ^{+28.7}_{-27.7}\pm 3.8$  pb~\cite{CMS:2016ayb}\\(value$\pm$ stat$\pm$ sys$\pm$ exp$\pm$ theo$\pm$ lumi)} & \tabincell{c}{$247\pm 6.4\pm 32.5\pm 3.1\pm 3.6$  pb~\cite{Aaboud:2016ymp}\\(value$\pm$ stat$\pm$ syst$\pm$ PDF$\pm$ lumi)} & $-$\tabularnewline
\hline
\tabincell{c}{\textbf{ tW-channel}\\ \textbf{(8 TeV)}}&  \multicolumn{2}{c|}{$25.0\pm 4.7$  pb~\cite{CMS:2014efa}} & $-$ \tabularnewline
\hline
\tabincell{c}{\textbf{ tW-channel}\\ \textbf{(13 TeV)}}& $-$ & 94$\pm 10 ~(\rm{stat.})^{+28}_{-23}~(\rm{syst.})$ pb~\cite{ATLAS:2016lte} & $-$ \tabularnewline
\hline
\tabincell{c}{\textbf{W-helicity}} \textbf{(7 TeV)} &  \multicolumn{2}{c|}{\tabincell{l}{$F_0=0.626\pm 0.034(stat.)\pm 0.048(syst.)$\\$F_L=0.359\pm 0.021(stat.)\pm 0.028(syst.)$\cite{ATLAS:2013tla}\\$F_R=0.015\pm 0.034$}} & $-$\tabularnewline
\hline
\tabincell{c}{\textbf{W-helicity}} \textbf{(8 TeV)}&  \tabincell{l}{$F_0=0.659\pm0.015(stat.)\pm 0.023(syst.)$\\$F_L=0.350\pm 0.010(stat.)\pm 0.024(syst.)$\cite{CMS:2013pfa}\\$F_R=-0.009\pm 0.006(stat.)\pm 0.020(syst.)$} & $-$ & $-$ \tabularnewline
\hline
\end{tabular}}
\end{table}

We perform a global $\chi^2$ test to obtain the present constraints on the effective $Wtb$ couplings. In the statistical analysis, the $\chi^2$ is defined as
\bea
	\chi^2 = \sum_i \frac{ \left( {\mathcal O}^{\rm exp}_i - {\mathcal O}^{\rm th}_i \right)^2 }{\delta\sigma_i^2},
\eea
where ${\mathcal O}^{\rm exp}_i$ and ${\mathcal O}^{\rm th}_i$ are the experimental values and the theoretical predictions for the experimental observable $i$, respectively.
$\delta\sigma_i$ represents the total error of the experimental measurement, which is defined as  $\delta\sigma_i \equiv \sqrt{(\delta\sigma_i^{\rm stat.})^2 + (\delta\sigma_i^{\rm syst.})^2}$.
The CERN library MINUIT~\cite{James:1975dr} is used in our analysis to obtain the best-fit values of the effective $Wtb$ couplings and the contours at different confidence levels (C.L.).

In this work we consider both direct and indirect constraints on the effective $Wtb$ couplings. The direct constrains arise from the experimental measurements of top-quark productions and decays while the indirect constraints arise from the precision measurement of flavor physics. For example, the anomalous $Wtb$ couplings can contribute to the flavor changing neutral current (FCNC) processes through the quantum effects involving top-quark inside the loop. In particular, the inclusive decay $\bar{B}\rightarrow X_s \gamma$ provides very stringent bounds on the anomalous  $Wtb$ couplings~\cite{Grzadkowski:2008mf}.

\subsection{Constraints on the Effective Couplings and Operators}

\begin{figure}
\centering
\includegraphics[width=0.32\textwidth]{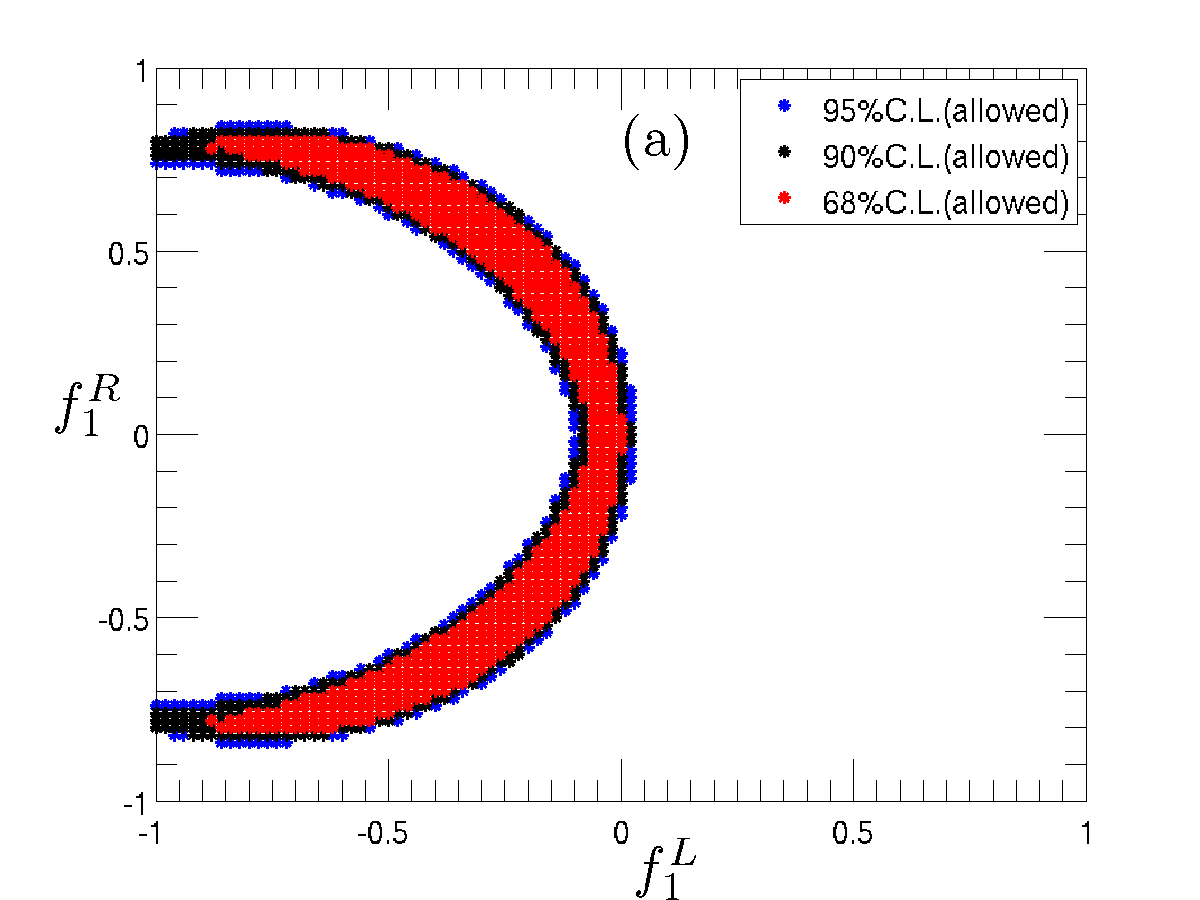}
\includegraphics[width=0.32\textwidth]{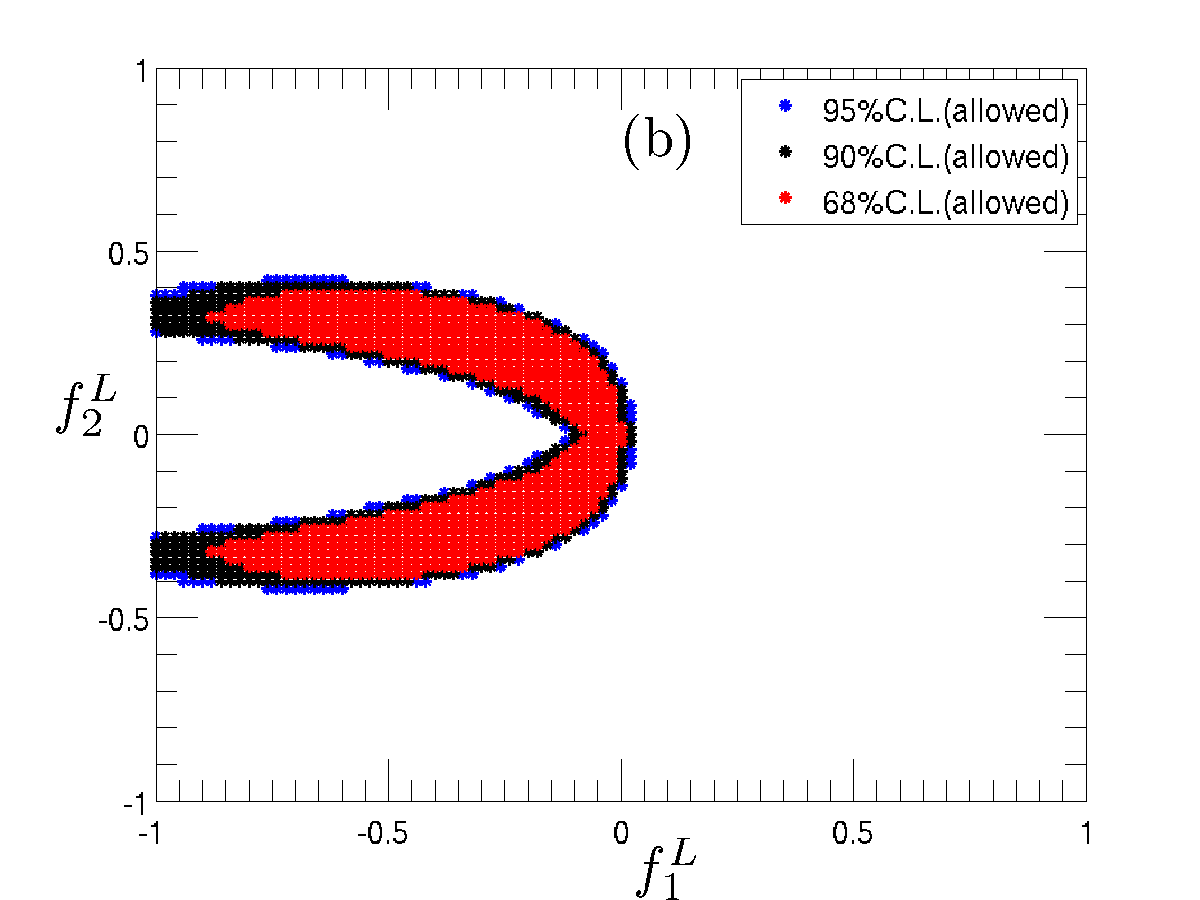}
\includegraphics[width=0.32\textwidth]{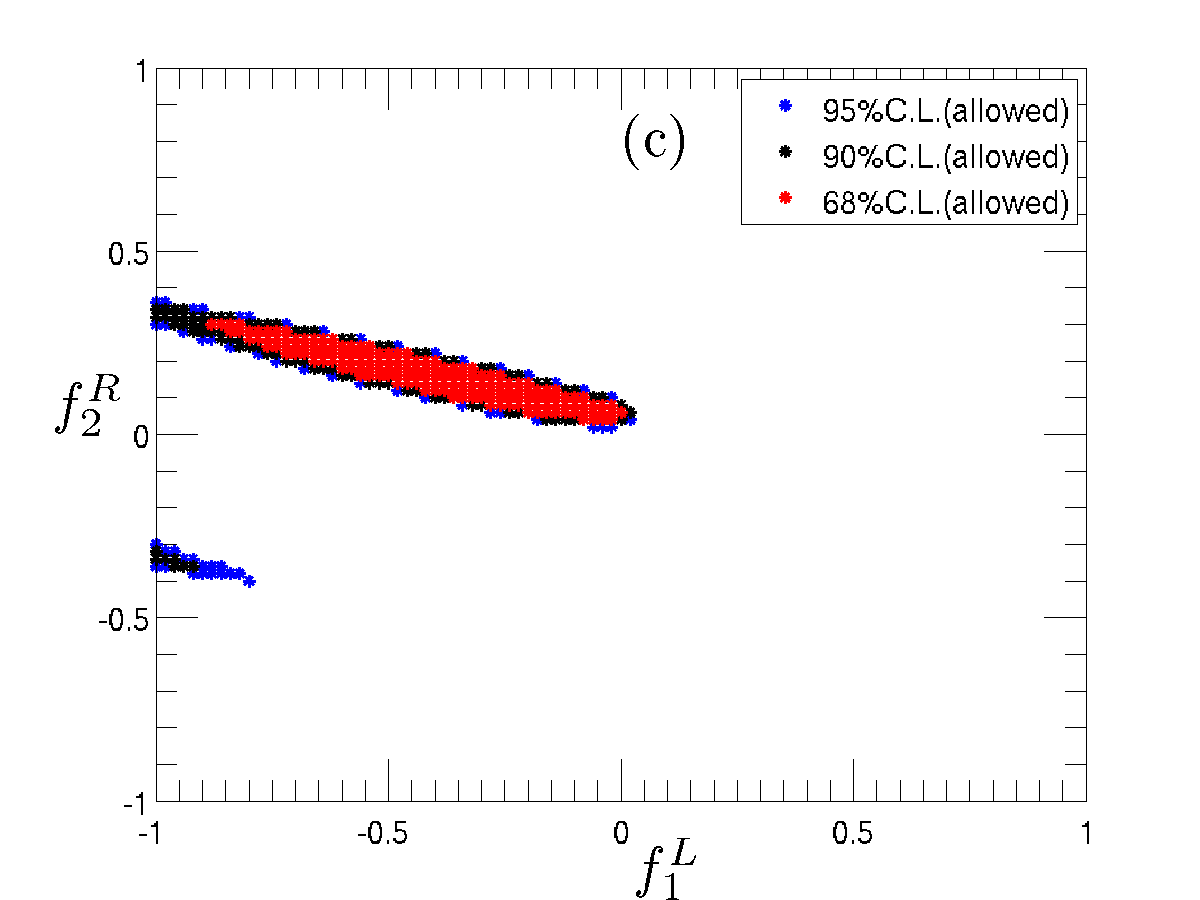}\\
\includegraphics[width=0.32\textwidth]{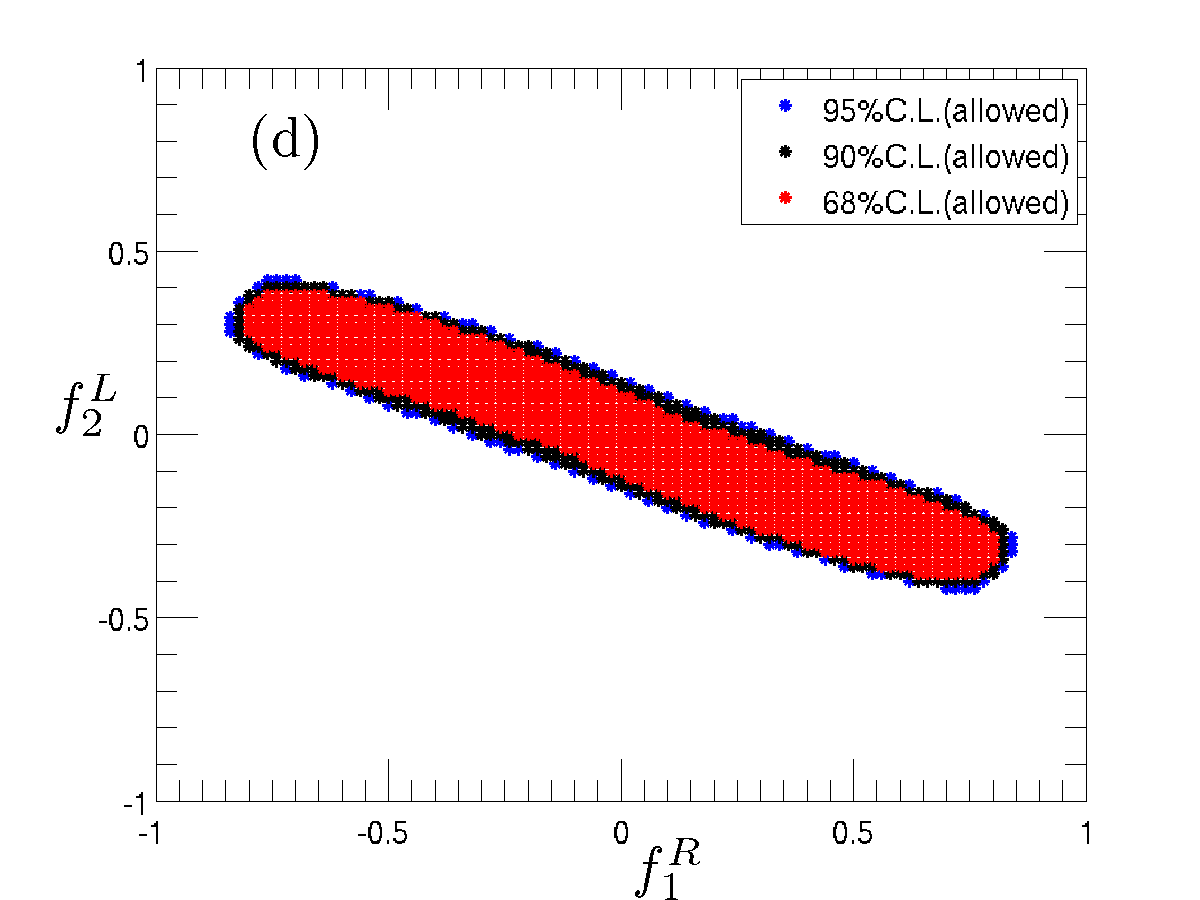}
\includegraphics[width=0.32\textwidth]{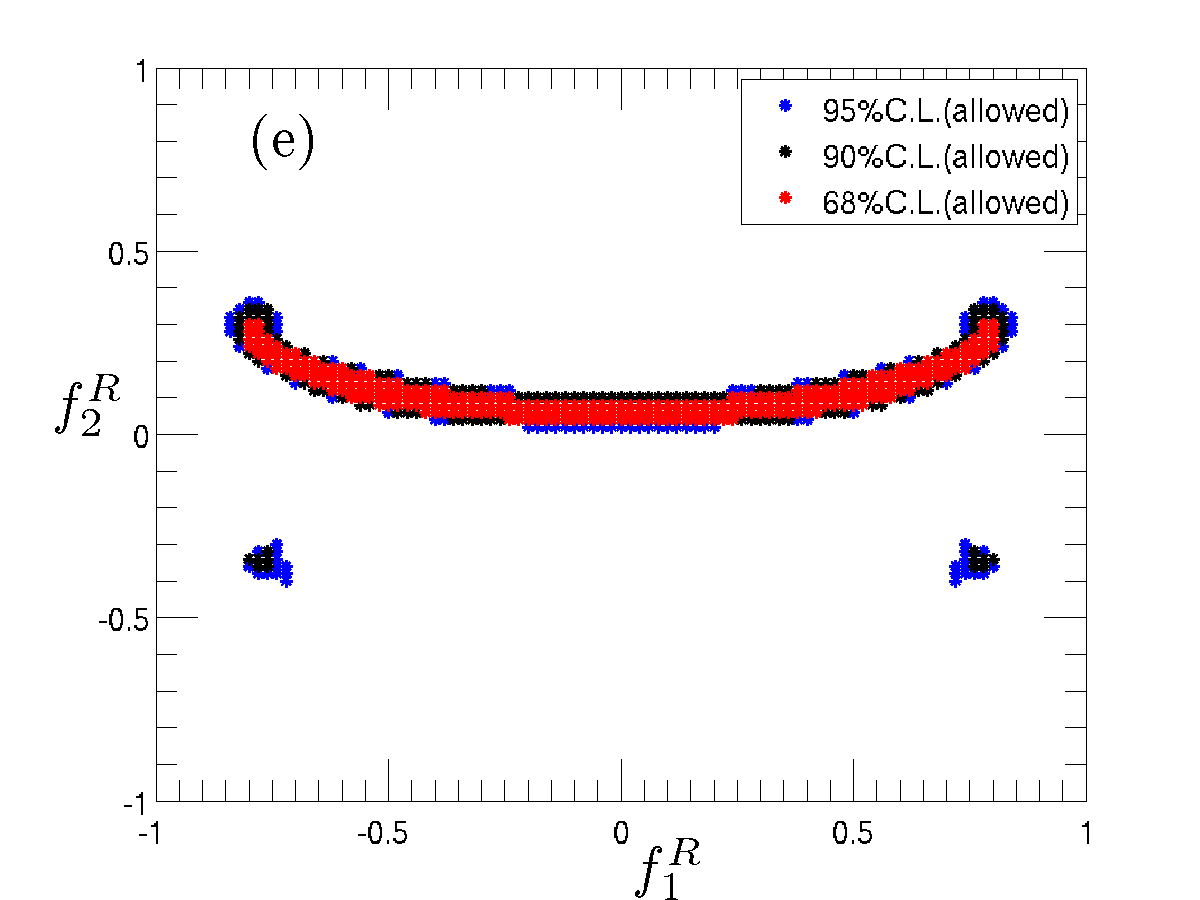}
\includegraphics[width=0.32\textwidth]{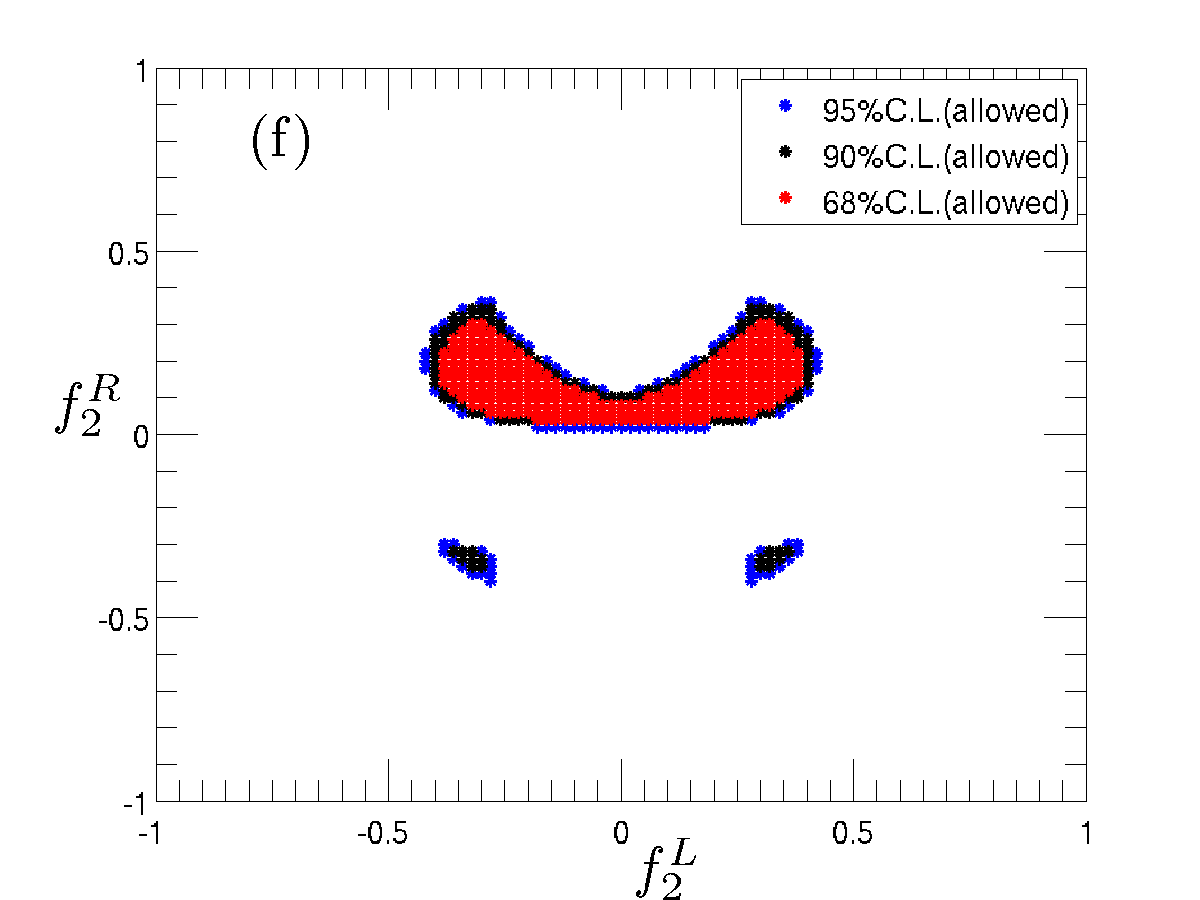}
\caption{ Allowed parameter space on the plane of the effective $Wtb$ couplings at the confidence levels of 68\% (red region), 90\% (black region) and 95\% (blue region). $|f_1^L|\leqslant 1$ is required in our analysis.}
\label{Fig:scan3}
\end{figure}

Below we present the allowed regions of the effective $Wtb$ couplings based on the global-fit analysis of all the four effective $Wtb$ couplings $f_{1,2}^{L,R}$.
We plot in Fig.~\ref{Fig:scan3} the allowed parameter space on the 68\%, 90\% and 95\% C.L., respectively for the effective $Wtb$ couplings and investigate the correlations among the effective $Wtb$ couplings.  We note that the allowed parameter space on the plane of $(f_1^L, f_1^R)$ and $(f_1^L, f_2^L)$ contours are mainly driven by the $t$-channel and $tW$-channel cross sections.
Figure~\ref{Fig:scan3}(a) shows the allowed parameter space of $(f_1^L, f_1^R)$.
The cross sections of the single top quark production processes are proportional to
$(1+f_1^L)^2$ and $\left(f_1^{R}\right)^2$; see  Eq.~\ref{eq:xsec}. It yields the circular contour region on the plane of $f_1^L$ and $ f_1^R$ and favors a negative $f_1^L$.
Figure~\ref{Fig:scan3}(b) displays the contour on the plane of $(f_1^L,f_2^L)$ which exhibits an ellipse shape. Even though the cross sections is proportional to $(1+f_1^L)^2$ and $(f_2^L)^2$,  the $(f_2^L)^2$ term contributes less to the cross section than the  $(1+f_1^L)^2$ term. That generates the ellipse shape.
We also note that the differences among the parameter space at the three confidence levels are very small. It is owing to the fact that the contours of the effective couplings on the plane of two effective $Wtb$ couplings are a projection from the four dimension parameter space down to a two dimension subspace. That projection leads to the small differences.

The $(f_1^L, f_2^R)$ contour shown in Fig.~\ref{Fig:scan3}(c) clearly indicates a strong anti-correlation between $f_1^L$ and $f_2^R$. Such a behavior can be understood from the single-top production cross sections which are approximately proportional to $(1 + f_1^L + a_tf_2^R)^2$.
As a result, the two islands of the allowed parameter space is symmetric around the point ($f_1^L=-1, f_2^R=0$). Note that the region of  $f_1^L<-1$ is also possible.
The correlation in the $(f_1^R, f_2^L)$ contour originates from the relation $(f_1^R + a_t f_2^L)^2$ in the cross sections, yielding the anti-correlation band in Fig.~\ref{Fig:scan3}(d) which is centered around the point ($f_1^R=0, f_2^L=0$).
The tightest constraints on both the $(f_1^L, f_2^R)$ and $(f_1^R, f_2^L)$ contours come from the $tW$-channel cross section and the $W$ helicity fractions.

Figure~\ref{Fig:scan3}(c) shows a positive $f_2^R$ is preferred when demanding $|f_1^L|\leq 1$.
Therefore, both the $(f_1^R, f_2^R)$ and $(f_2^L, f_2^R)$ contours only allow positive  $f_2^R$.
Due to the interference effect between the anomalous $Wtb$ couplings and SM, the linear term which is proportional to $f_2^R$ is valid in the cross sections. Therefore the relation of $f_2^R$ with $f_1^R$ or $f_2^L$
is like $f_2^R\sim a_1(f_1^R)^2$ or $b_1(f_2^L)^2$ as shown in Figs.~\ref{Fig:scan3}(e) and (f).
Still the $tW$ channel cross section determines the shapes of the $(f_1^R, f_2^R)$ and $(f_2^L, f_2^R)$ contours.

The current experimental data of $\sigma_s$ and $\sigma_{tW}$ exhibit $\sim 20\%$ uncertainties. The accuracy of $\sigma_{tW}$ is expected to be improved  at the forthcoming LHC RUN2 , but the precise measurement of $\sigma_s$ is still challenging. We thus consider the $\sigma_t$, $\sigma_{tW}$ and the $W$-helicity fractions in our global analysis to constrain the effective $Wtb$ couplings.
Using the $x_i$'s while neglecting $O(x_i^2)$ and higher order contributions, we can extract the correlation between different channels:
\bea
	\left(
	\begin{array}{cccc}
	a_0^{\rm tot}/\sigma_{t}^{\rm SM} & a_m^{\rm tot}/\sigma_{t}^{\rm SM} & a_p^{\rm tot}/\sigma_{t}^{\rm SM} & a_5^{\rm tot}/\sigma_{t}^{\rm SM} \\
	c_0^{\rm tot}/\sigma_{tW}^{\rm SM} & c_m^{\rm tot}/\sigma_{tW}^{\rm SM} & c_p^{\rm tot}/\sigma_{tW}^{\rm SM} & c_5^{\rm tot}/\sigma_{tW}^{\rm SM}  \\
	F_L^{\rm SM} & -F_L^{\rm SM} & -F_L^{\rm SM} & 0 \\
	-F_0^{\rm SM} & F_0^{\rm SM} & -F_L^{\rm SM} & 0 \\
	\end{array}
	\right)
	\left(
	 \begin {array} {c}
	         x_0 \\ x_m \\  x_p \\ x_5 \\
	    \end {array}
	   \right)
	    = \left(
	 \begin {array} {c}
	         \Delta\sigma_t^{\rm NP}/\sigma_{t}^{\rm SM}  \\ \Delta\sigma_{tW}^{\rm NP}/\sigma_{tW}^{\rm SM}  \\ \Delta F_0^{\rm NP}/F_0^{\rm SM} \\ \Delta F_L^{\rm NP}/F_L^{\rm SM} \\
	    \end {array}
	   \right),
	   \label{eq:operator_xsec}
\eea
where $\Delta\sigma_t^{\rm NP}\equiv (\sigma_t^{\rm exp}-\sigma_t^{\rm SM})$ and $\Delta\sigma_{tW}^{\rm NP}\equiv (\sigma_{tW}^{\rm exp}-\sigma_{tW}^{\rm SM})$ are the variations from the SM prediction. $\Delta F_0^{\rm NP}\equiv (F_0^{\rm exp}-F_0^{\rm SM})$ and $\Delta F_L^{\rm NP}\equiv (F_L^{\rm exp}-F_L^{\rm SM})$ denote the variation of $W_0$ and $W_L$ helicity fractions from the SM theory prediction.
The factor $a_i^{\rm tot}$ and $c_i^{\rm tot}$ are given by:
\begin{align}
  a_i^{\rm tot} &= K(t)~a_i(t)+K(\bar{t})~a_i(\bar{t}), \\
  c_i^{\rm tot} &=2 K(tW)~ c_i(tW),
\end{align}
where  the  coefficients $a_i$ and $c_i$ are given in Table~\ref{coeff}. $K(t)$ denotes the $K$-factor of the $t$-channel single top-quark production, $K(\bar{t})$ denotes the $K$-factor of the $t$-channel single antitop-quark production and the $K(tW)$ is the $K$-factor of the $tW$ associate production.
Solving the systems of linear equations shown in Eq.~\ref{eq:operator_xsec} gives rise to the following relations
\bea
 \label{error}
	\left(
	 \begin {array} {c}
	         x_0 \\ x_m \\  x_p \\ x_5 \\
	    \end {array}
	   \right)
	=\left(
	\begin{array}{cccc}
     ~~1.756 & -0.755 & ~~0.267   & -0.068 \\
     ~~1.756 & -0.755 &    -0.733   &  ~~0.932 \\
         0            &        0       &   -2.315    & -1.000 \\
     -1.547    & ~~1.545  &    -1.918    & ~~0.060
	\end{array}
	\right)
      \left(
	 \begin {array} {c}
	         \sigma_t^{\rm exp(0)}/\sigma_t^{\rm SM}-1 \\
             \sigma_{tW}^{\rm exp(0)}/\sigma_{tW}^{\rm SM}-1  \\	
             F_0^{\rm exp(0)}/F_0^{\rm SM}-1  \\	
	         F_L^{\rm exp(0)}/F_L^{\rm SM}-1  \\	
	    \end {array}
	   \right),
\eea
where $\sigma_t^{\rm exp(0)}$ and $\sigma_{tW}^{\rm exp(0)}$ denote central values of the experimental data of $t$-channel and $tW$-channel cross sections, respectively. Similarly, $F_0^{\rm exp(0)}$ and $F_L^{\rm exp(0)}$ are experimental central values of the longitudinal and left-handed helicity fraction. Note that the central values of the experimental data determine $x_i$'s and the correlations of effective couplings, whereas the experimental errors are translated into the errors of $x_i$'s ($\delta x_i$) which yield the allowed parameter spaces of those effective couplings. Below we employ the error propagation equation of the weighted sums functions to study the dependence of $\delta x_i$'s on experimental errors which are labelled as $\delta\sigma_t^{\rm exp}$,  $\delta\sigma_{tW}^{\rm exp}$, $\delta F_0^{\rm exp}$ and $\delta F_L^{\rm exp}$.

The variance of $x_0$ is
\bea
\label{x0error}
(\delta x_0)^2=3.084\left(\frac{\delta \sigma_t^{\rm exp}}{\sigma_t^{\rm SM}}\right)^2
+0.570\left(\frac{\delta \sigma_{tW}^{\rm exp}}{\sigma_{tW}^{\rm SM}}\right)^2
+0.071\left(\frac{\delta F_0^{\rm exp}}{F_0^{\rm SM}}\right)^2
+0.005\left(\frac{\delta F_L^{\rm exp}}{F_L^{\rm SM}}\right)^2,
\eea
in which the $\sigma_t$ and $\sigma_{tW}$ measurements dominate over the $W$-helicity measurements. Improving the measurements of $\sigma_t$ and $\sigma_{tW}$ is important to test the correlations of $(f_1^R, f_2^R)$ and $(f_2^L, f_2^R)$.

The variance of $x_m$ is
\bea
\label{xmerror}
(\delta x_m)^2=3.084\left(\frac{\delta \sigma_t^{\rm exp}}{\sigma_t^{\rm SM}}\right)^2
+0.570\left(\frac{\delta \sigma_{tW}^{\rm exp}}{\sigma_{tW}^{\rm SM}}\right)^2
+0.537\left(\frac{\delta F_0^{\rm exp}}{F_0^{\rm SM}}\right)^2
+0.869\left(\frac{\delta F_L^{\rm exp}}{F_L^{\rm SM}}\right)^2.
\eea
All coefficients are comparable such that one has to consider all the four experiments to determine  $\delta x_m$. As  $f_1^L$ and $f_2^R$ are anti-correlated in $x_m = (1+f_1^L+a_tf_2^R)^2-1$, improving $\delta x_m$ would further constrain the correlation between $f_1^L$ and $f_2^R$, e.g. the band in Fig.~\ref{Fig:scan3}(c) tends to be narrower.

The $x_p = (f_1^R+a_tf_2^L)^2$ is directly linked to the right-handed $W$-helicity fraction $F_R$, which is inferred from $F_L$ and $F_0$ measurements.  As a result, $\delta x_p$ depends only on the $W$-helicity measurement as following:
\beq
(\delta x_p)^2 = 5.359\left(\frac{\delta F_0^{\rm exp}}{F_0^{\rm SM}}\right)^2
+\left(\frac{\delta F_L^{\rm exp}}{F_L^{\rm SM}}\right)^2.
\eeq
As a result, a strong anti-correlation between $f_1^R$ and $f_2^L$ can be obtained from the $F_0$ and $F_L$ measurements.

The variance of $x_5$ is given by
\bea
(\delta x_5)^2=2.393\left(\frac{\delta \sigma_t^{\rm exp}}{\sigma_t^{\rm SM}}\right)^2
+2.387\left(\frac{\delta \sigma_{tW}^{\rm exp}}{\sigma_{tW}^{\rm SM}}\right)^2
+3.679\left(\frac{\delta F_0^{\rm exp}}{F_0^{\rm SM}}\right)^2
+0.004\left(\frac{\delta F_L^{\rm exp}}{F_L^{\rm SM}}\right)^2.
\eea
It is sensitive to the precision of $\sigma_t$, $\sigma_{tW}$ and $F_0$ measurements.

\begin{figure}
\centering
\includegraphics[width=0.32\textwidth]{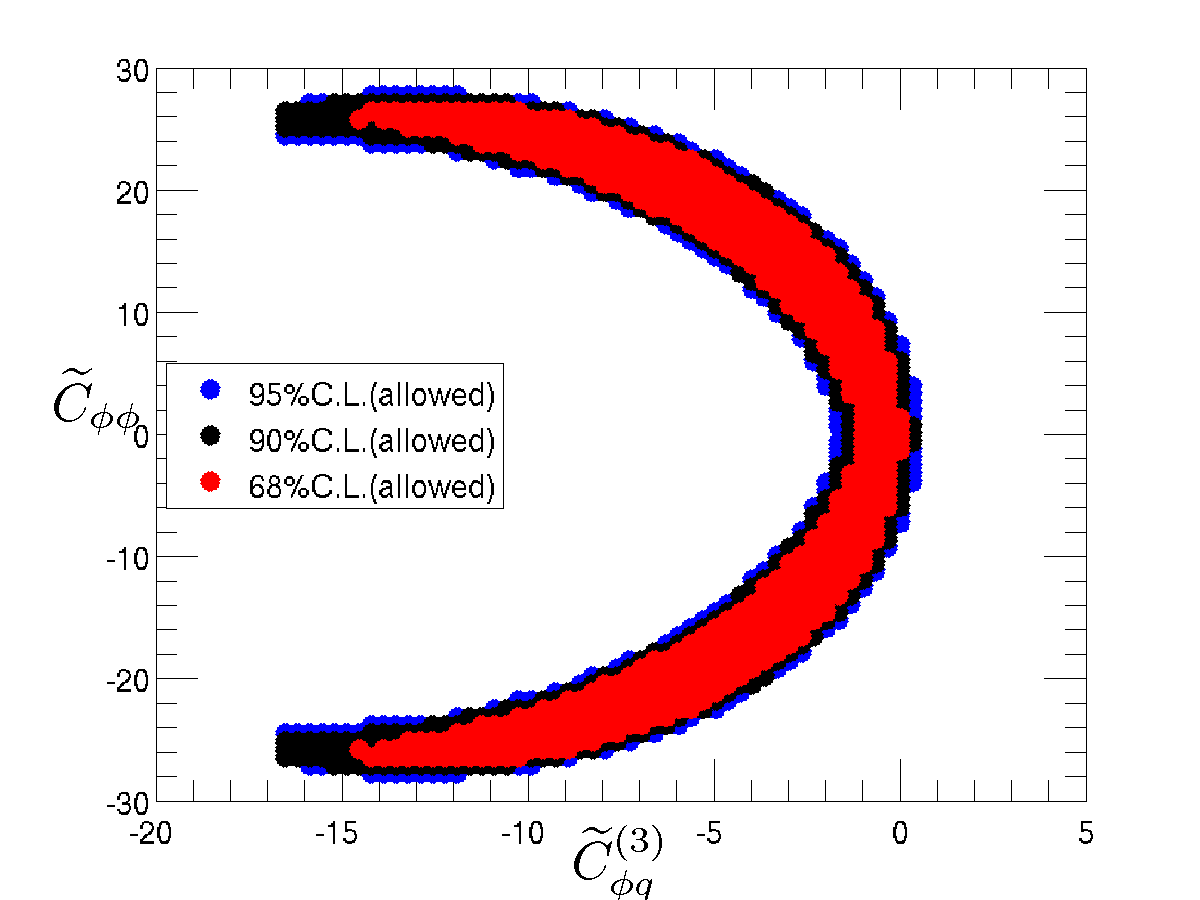}
\includegraphics[width=0.32\textwidth]{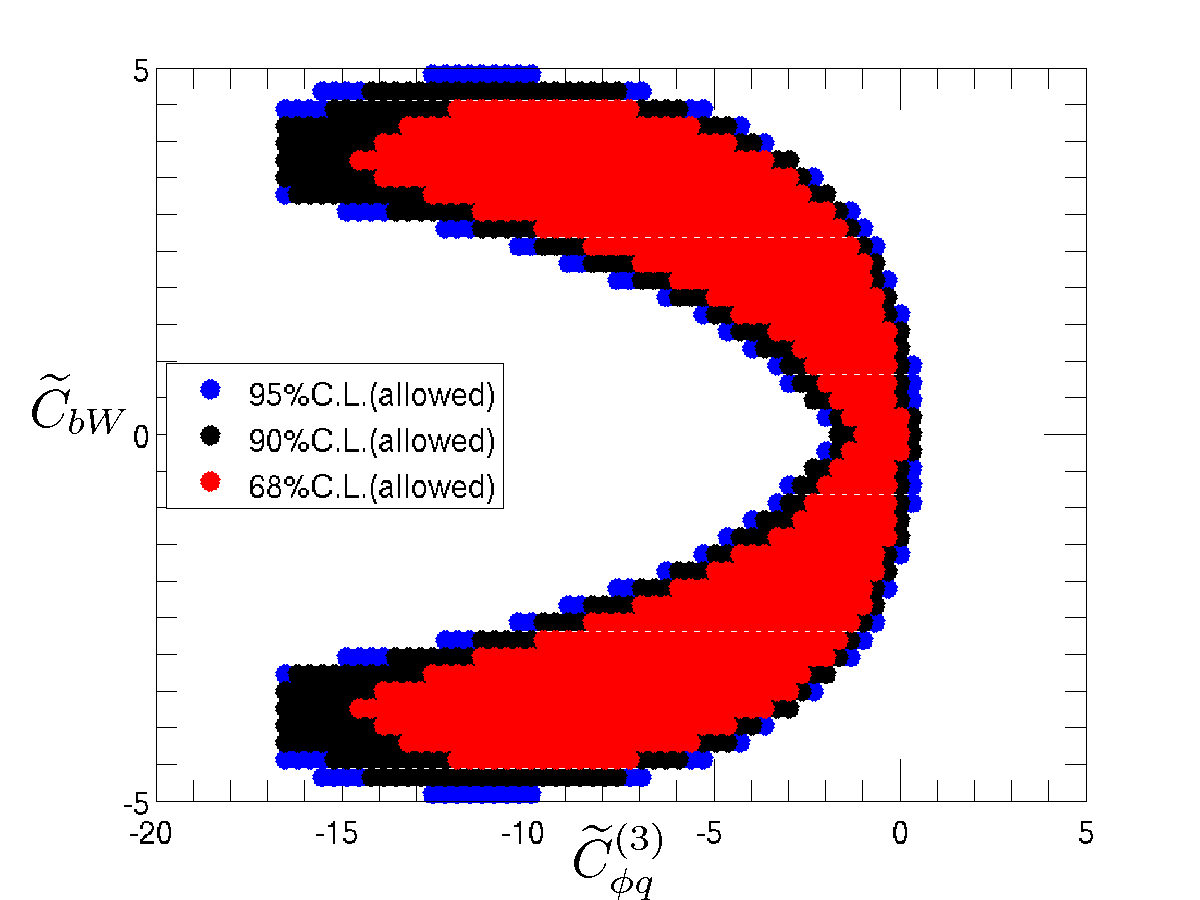}
\includegraphics[width=0.32\textwidth]{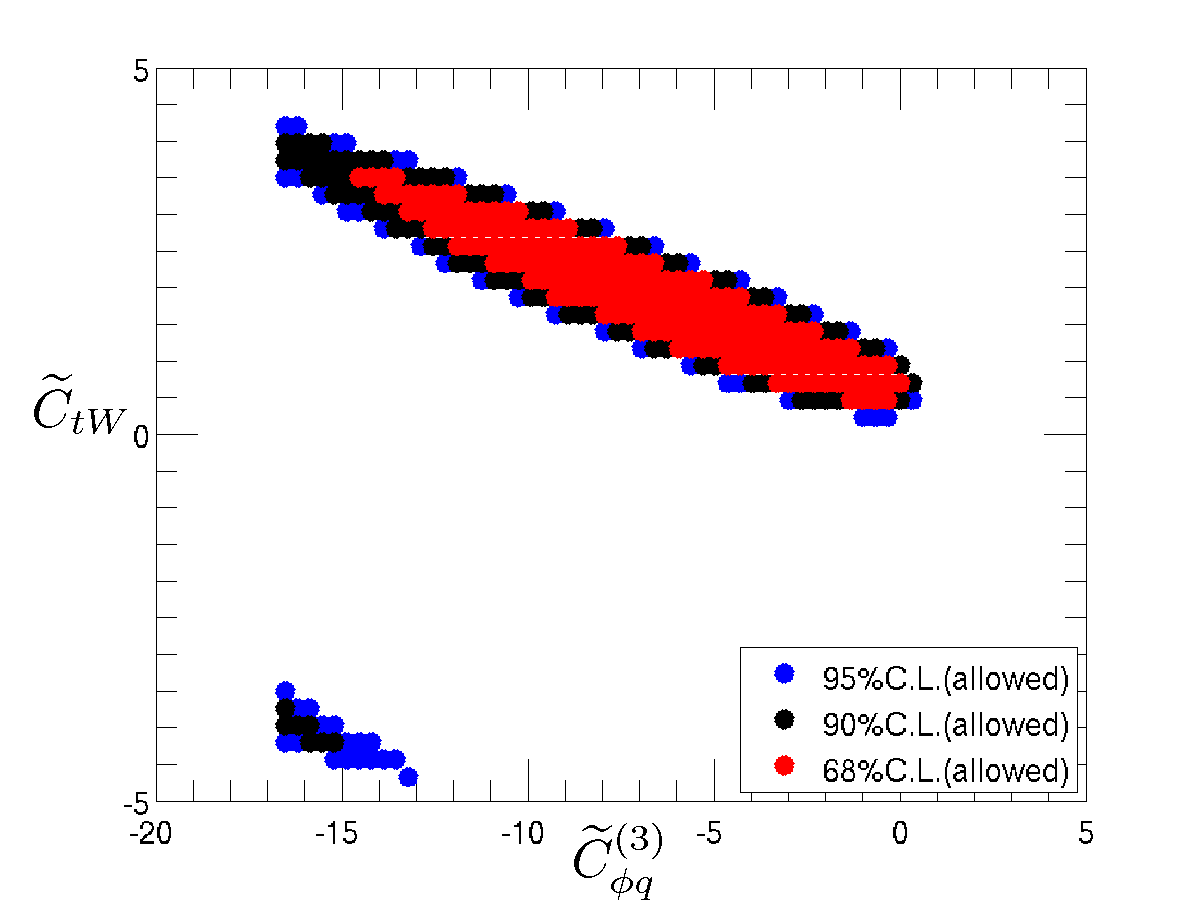}\\
\includegraphics[width=0.32\textwidth]{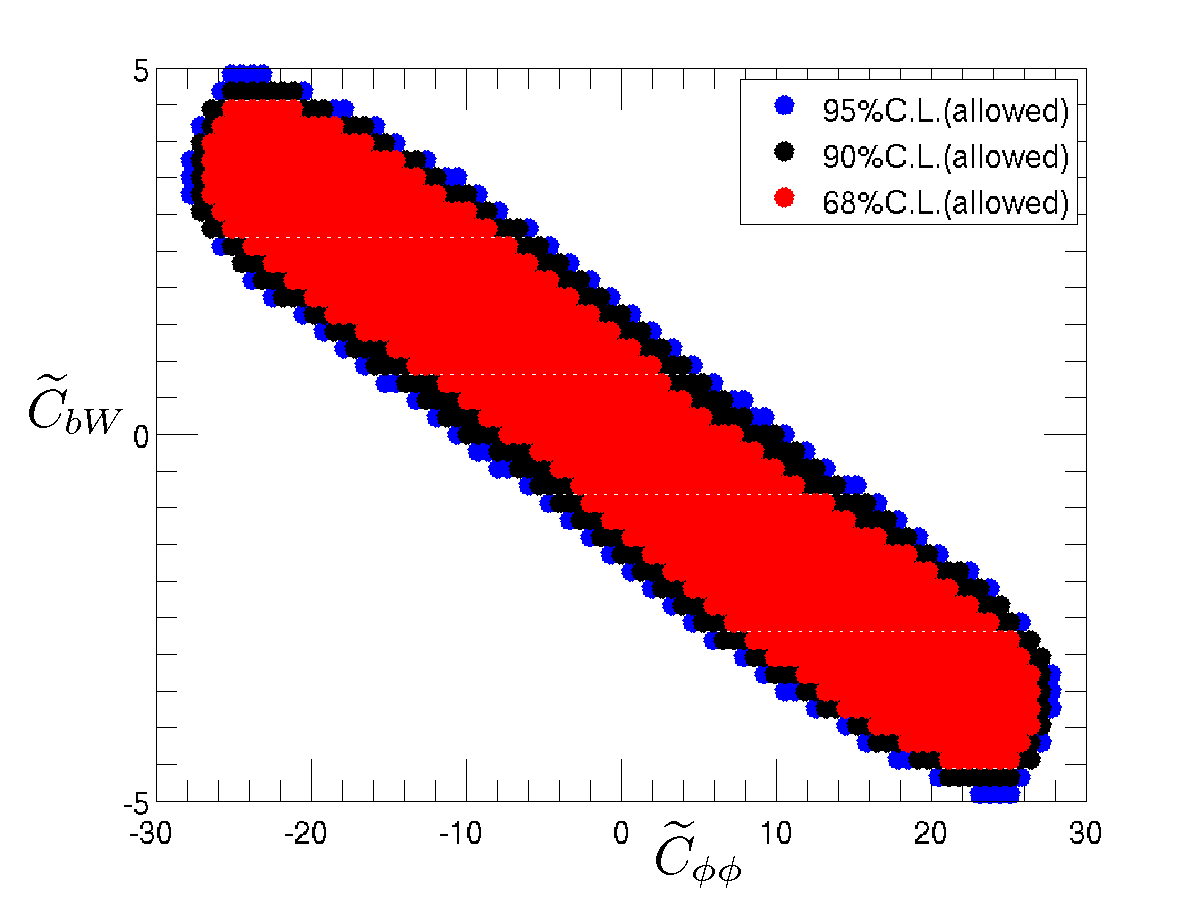}
\includegraphics[width=0.32\textwidth]{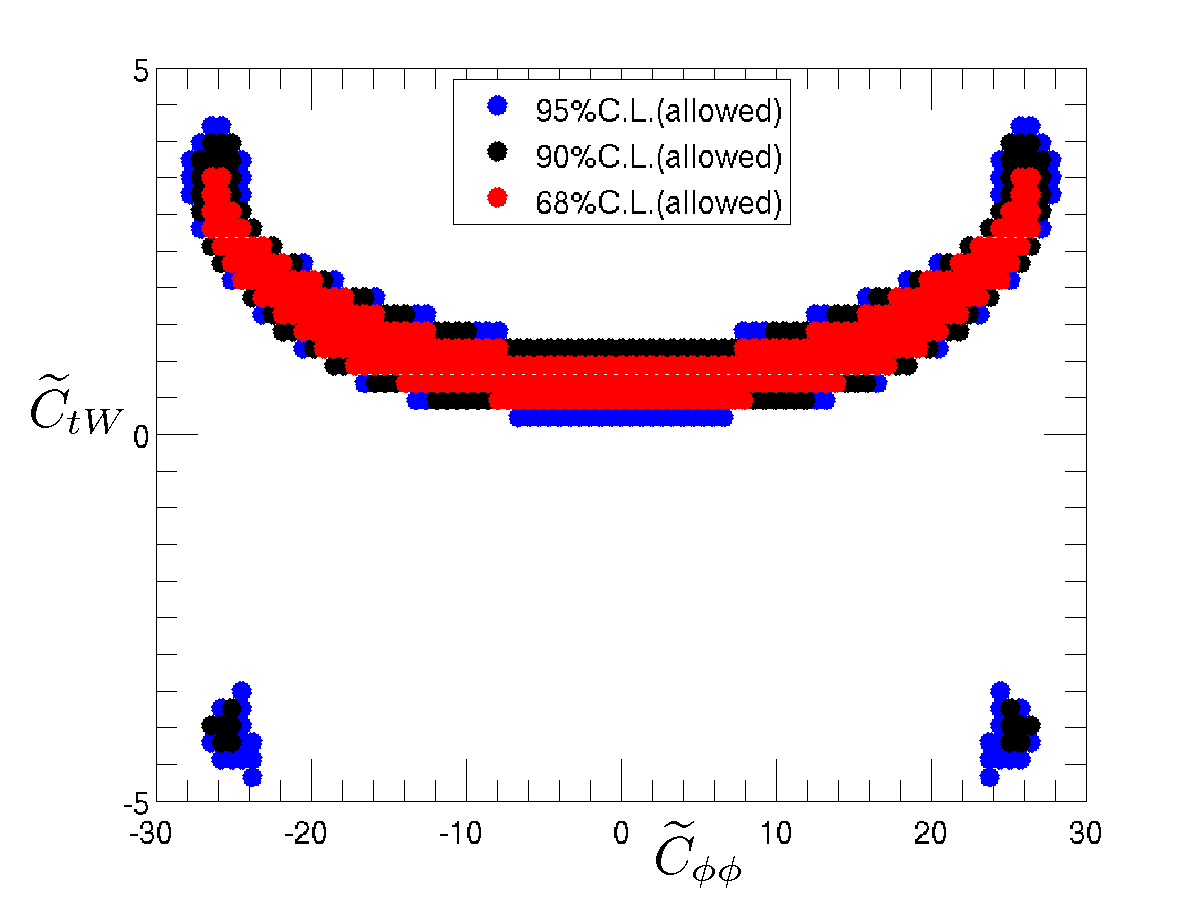}
\includegraphics[width=0.32\textwidth]{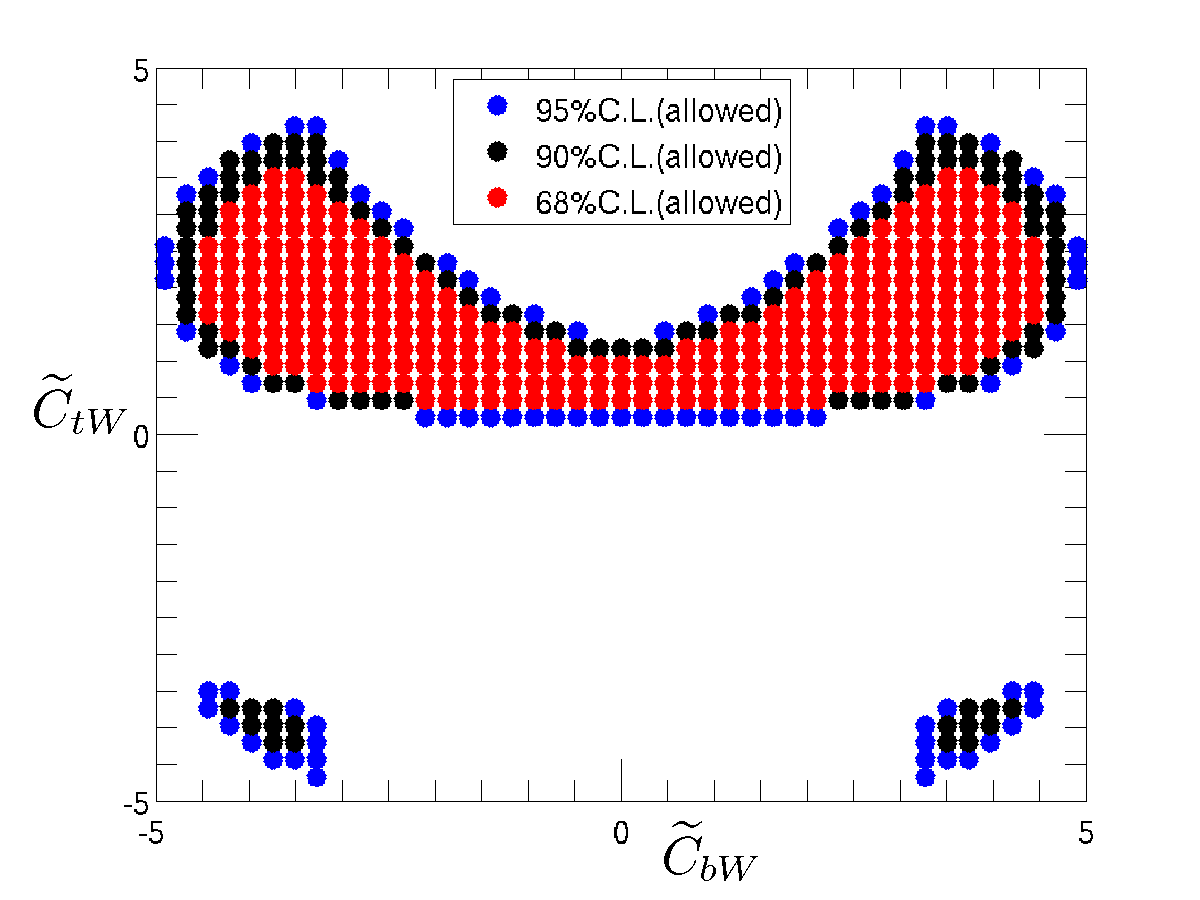}
\caption{Allowed parameter space on 68\%, 90\% and 95\% confidence levels  for parameter $\widetilde{C}_i$, where $\widetilde{C}_i\equiv C_i(1\rm{TeV}/\Lambda)^2$.
}
\label{Fig:scan4}
\end{figure}

From the precision measurement of the $Wtb$ couplings, one can also derive conservative bounds on the NP scales when no deviation is seen compared to SM predictions. Though we expect $C_i=O(1)$, their precise values are unknown. Measurements such as the ones described above can be used to obtain the ratios of these coefficients, but the values of $\Lambda_i$ cannot be obtained separately. Therefore, we define dimensionless parameters $\widetilde{C}_i$ according to the Wilson coefficient $C_i$ and normalize to 1 TeV,
\bea
\widetilde{C}_i\equiv C_i \left(\frac{1\rm{TeV}}{\Lambda}\right)^2.
\eea
The allowed parameter contours for the parameters $\widetilde{C}_{\phi q}^{(3)}$, $\widetilde{C}_{\phi\phi}$, $\widetilde{C}_{bW}$, $\widetilde{C}_{tW}$ are shown in Fig.~\ref{Fig:scan4}.
Although the values of effective couplings $f_1^L$, $f_1^R$, $f_2^L$ and $f_2^R$ are of the same order of magnitude, each individual parameter $\widetilde{C}_i$  is different. For example, the $\widetilde{C}_{\phi\phi}$ equal to 27 also allowed at 95\% C.L., while
 the maximal value for $\widetilde{C}_{bW}$ and $\widetilde{C}_{tW}$ is 5 at the same C.L.. The difference comes from the relation between the
 effective couplings ($f_{1,2}^{L,R}$) and the Wilson coefficients $C_i$'s in Eq.\ref{wilson}. In principle, after we know the range of the Wilson coefficients at low energy,
 we can obtain the values at NP scale $\Lambda$ by renormalization group equations, and further determine the NP parameter space. However,
 in this paper, we focus on the model independent approach to search the NP effects, and will not calculate the mixing of the different operators.

\begin{figure}
\centering
\includegraphics[width=0.4\textwidth]{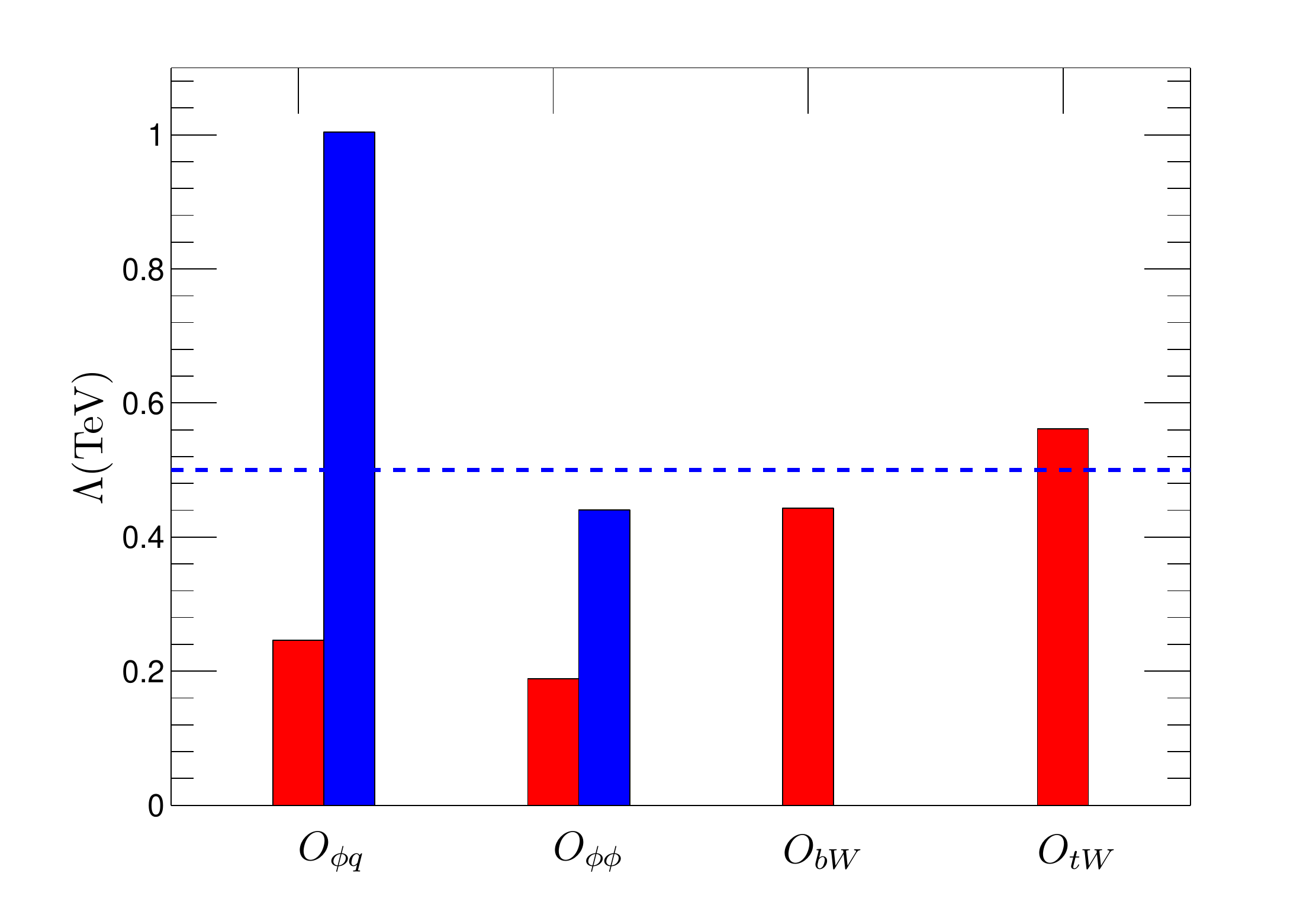}
\caption{ Limits on the cut-off scale $\Lambda_i$ of each individual operator $O_i$ with the Wilson coefficient $|C_i|=1$. The red lines represent those limits when all the four operators contribute simultaneously while the blue lines display those limits obtained when we consider one parameter at a time.  The cut-off scale $\Lambda=500$ GeV is also plotted for reference; see the horizontal blue-dashed line.
}
\label{Fig:cutoff}
\end{figure}

Figure~\ref{Fig:cutoff} displays the lower bound on the NP scale $\Lambda_i$ obtained from the $Wtb$ measurements at the 95\% C.L. with $|C_i |=1$. The red lines represent those limits when all the four operators contribute simultaneously while the blue lines display those limits obtained when we consider one-parameter-at-a-time. The current bound implies $\Lambda_{\phi q}>246$ GeV, $\Lambda_{\phi\phi}>188.6$ GeV, $\Lambda_{bW}>443.2$ GeV, $\Lambda_{tW}>561.7$ GeV when all the four operators contribute simultaneously while $\Lambda_{\phi q}>1000$ GeV, $\Lambda_{\phi\phi}>440.6$ GeV when we consider one parameter at a time.

\subsection{Constraints on the effective $Wtb$ couplings and operators in 2-dimensional subspaces}

\begin{figure}
\centering
\quad\includegraphics[width=0.3\textwidth]{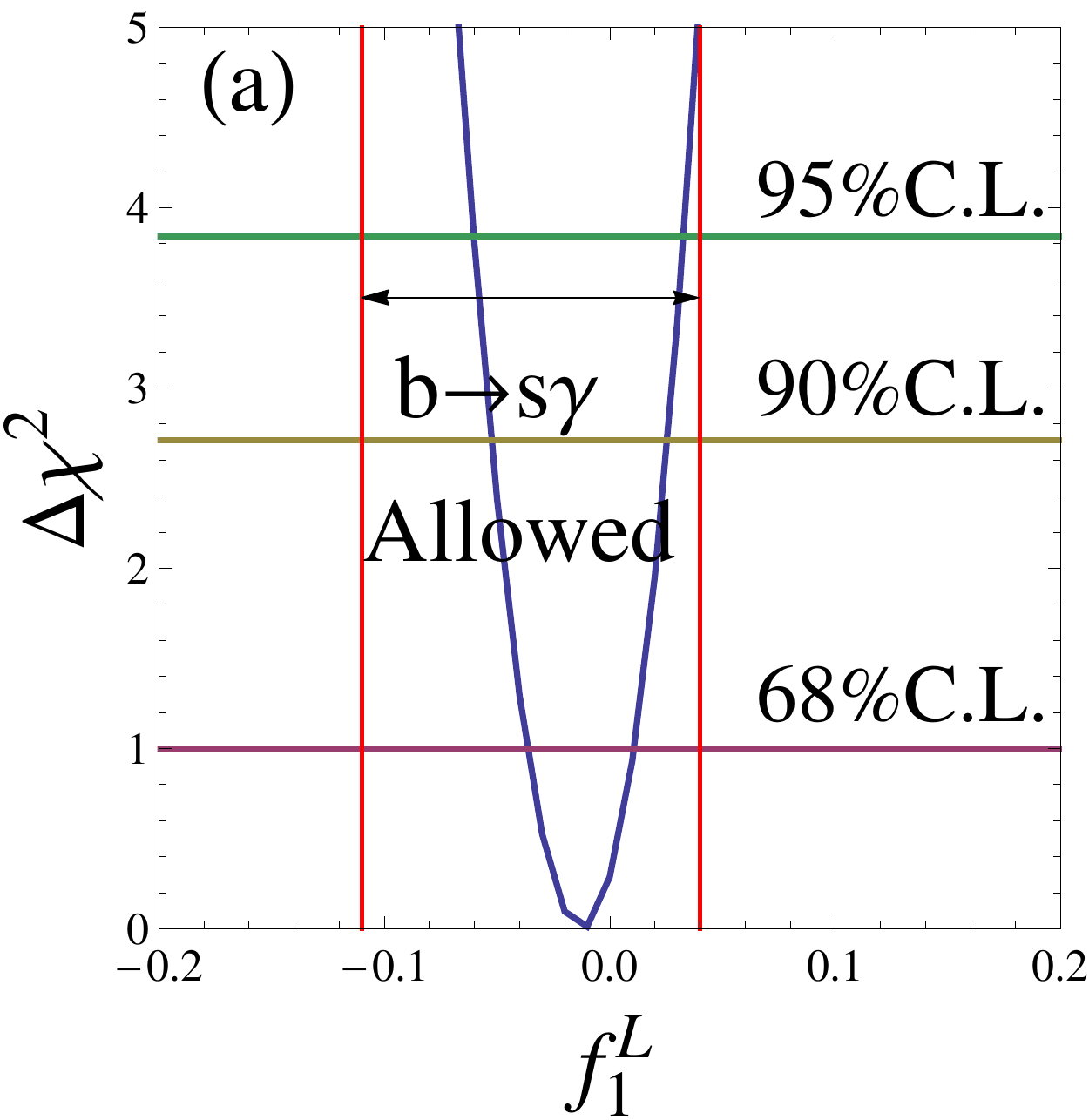}
\quad\includegraphics[width=0.3\textwidth]{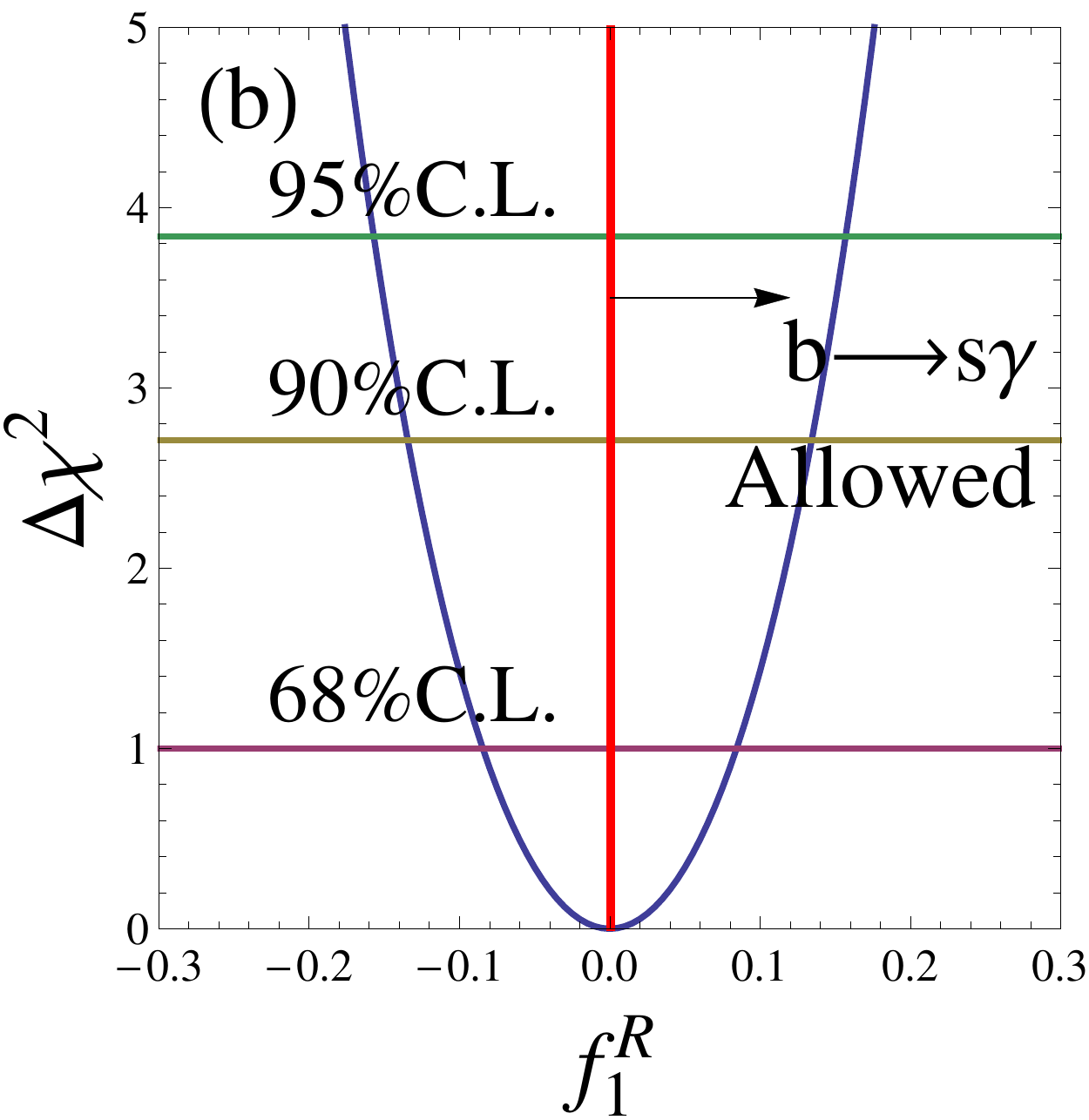}\\
\includegraphics[width=0.32\textwidth]{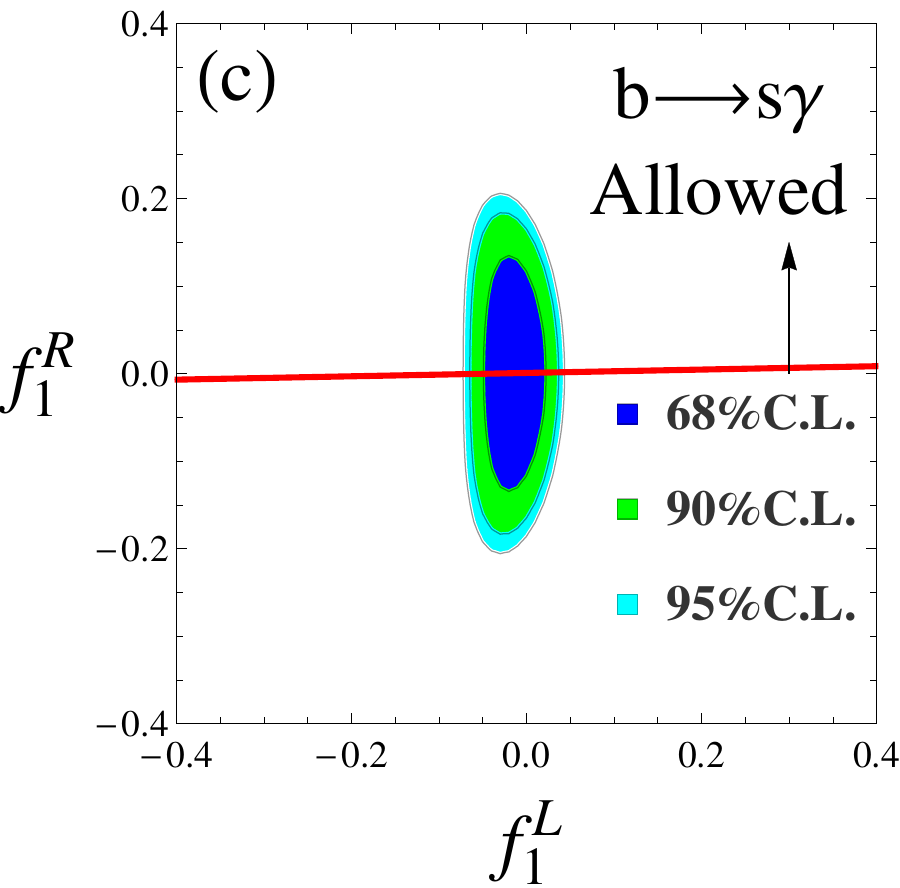}
\includegraphics[width=0.32\textwidth]{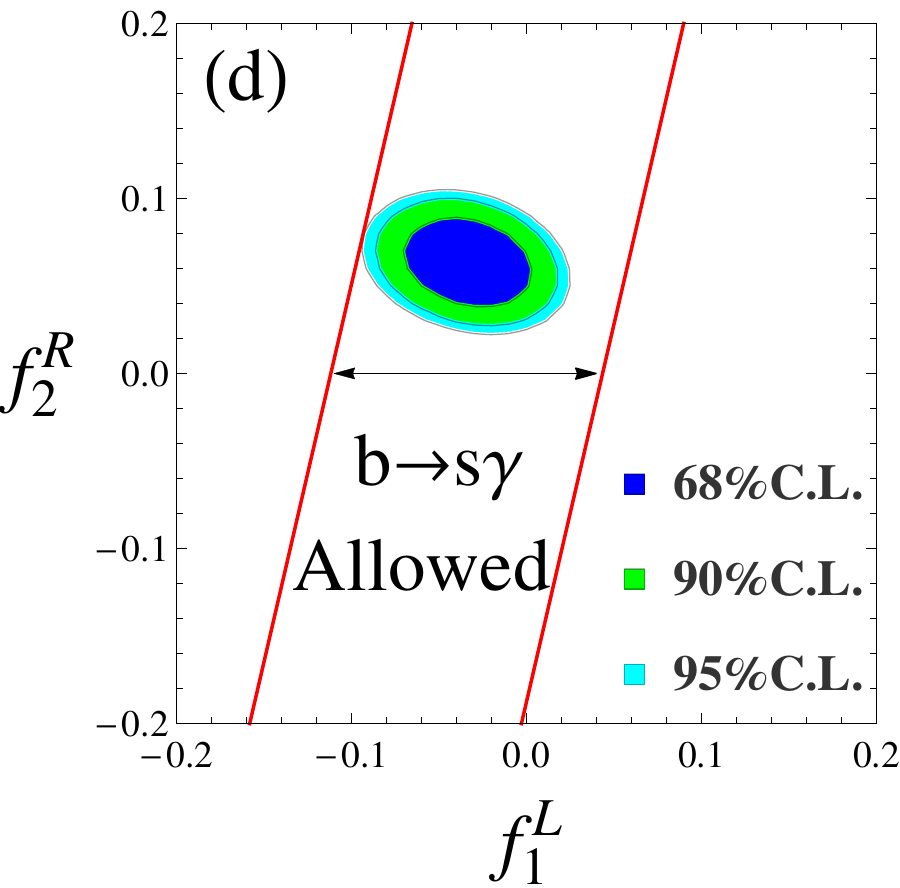}
\caption{Allowed parameter space on 68\%, 90\% and 95\% confidence levels for the effective couplings in the subspace. The sub-figures (a) and (b) correspond to the 1-dim parameter space $f_1^L$ and $f_1^R$, respectively; The contour plots (c) and (d) correspond to the cases that only $f_1^L,f_1^R$ and $f_1^L,f_2^R$ are modified, respectively. The red lines denote the constraints from $b\rightarrow s\gamma$, the detailed analyses of those constraints already exist in Ref.~\cite{Grzadkowski:2008mf}.
}
\label{Fig:scan1}
\end{figure}

So far we considered the full correlations among the four effective couplings and explored the allowed parameter space of the effective $Wtb$ couplings. However, not all but only some of the effective couplings are non-zero in many NP models.

\begin{figure}
\centering
\quad\includegraphics[width=0.3\textwidth]{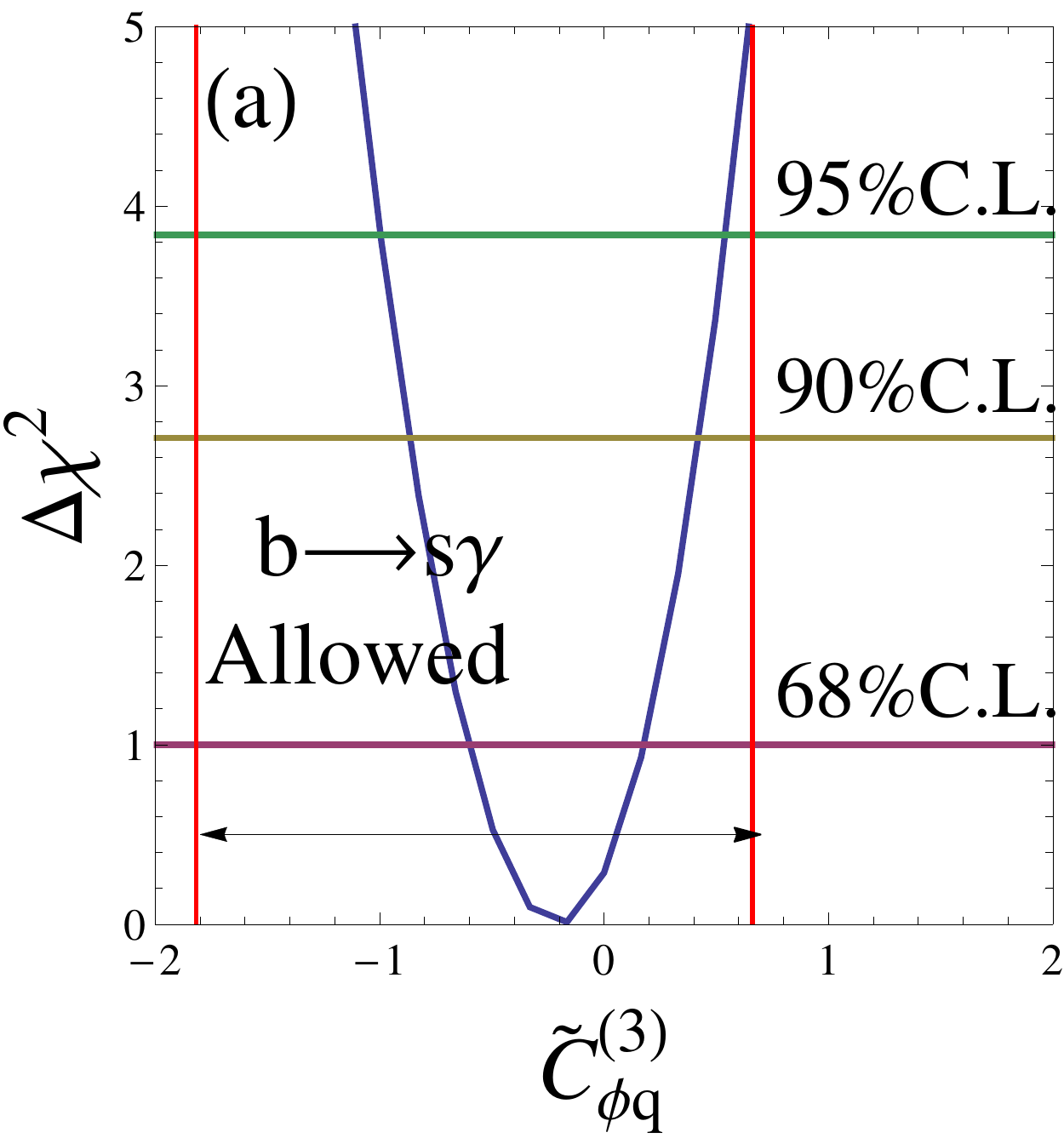}
\quad~\includegraphics[width=0.3\textwidth]{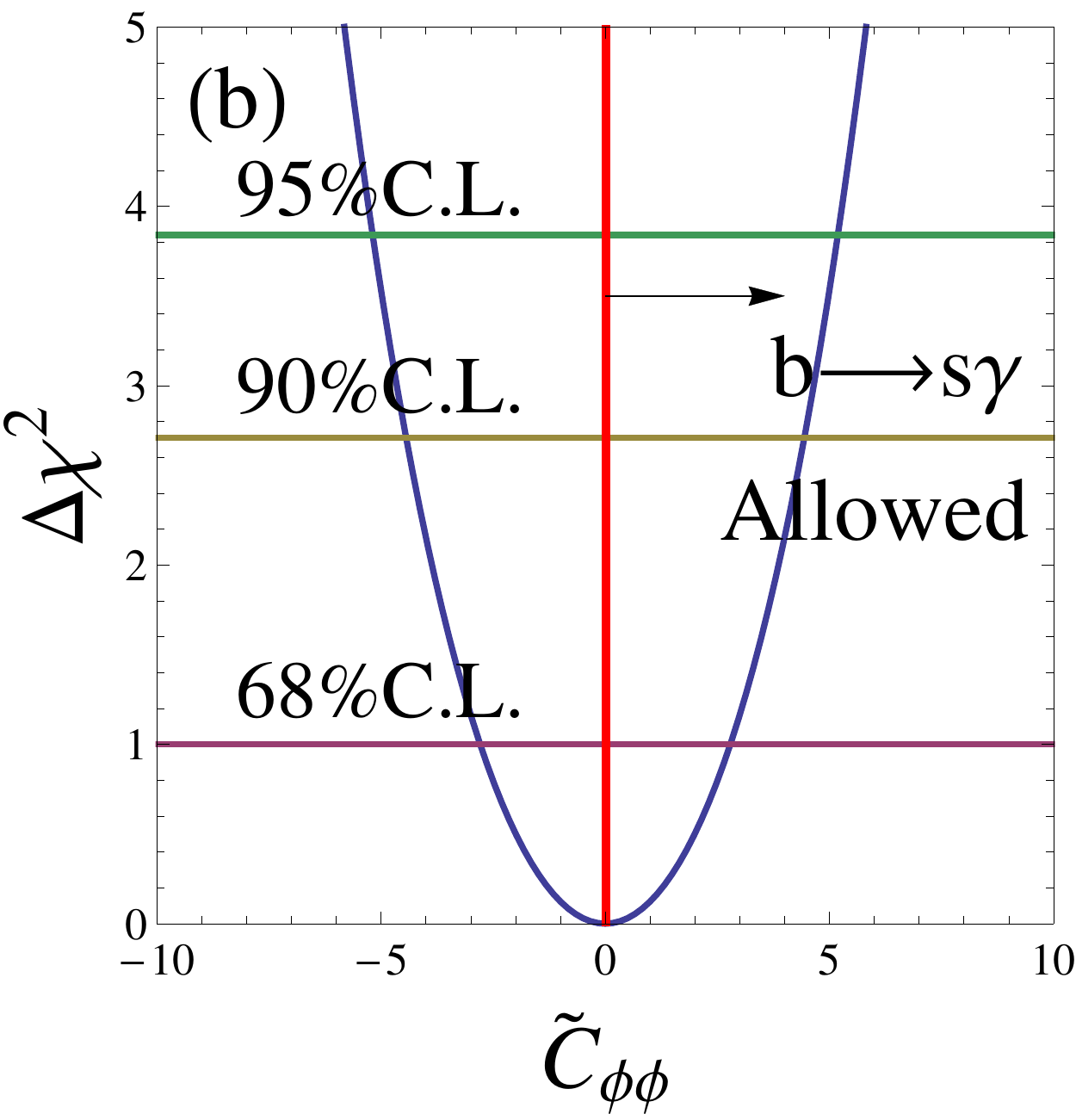}\\
\includegraphics[width=0.32\textwidth]{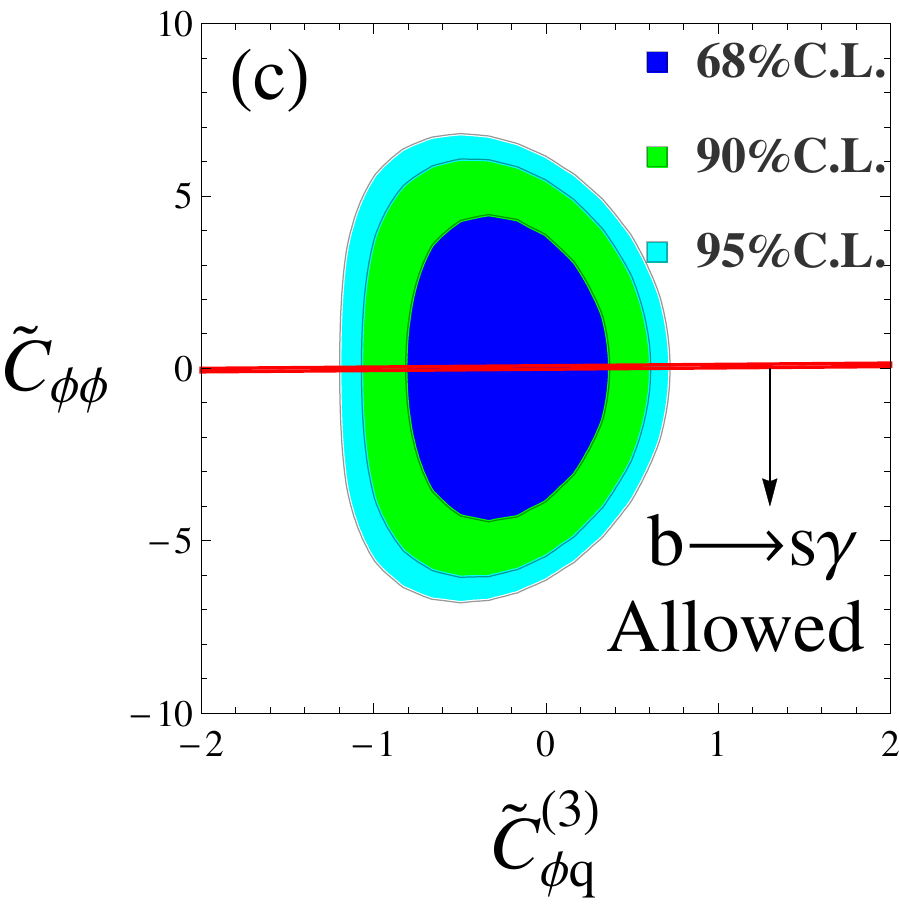}
\includegraphics[width=0.32\textwidth]{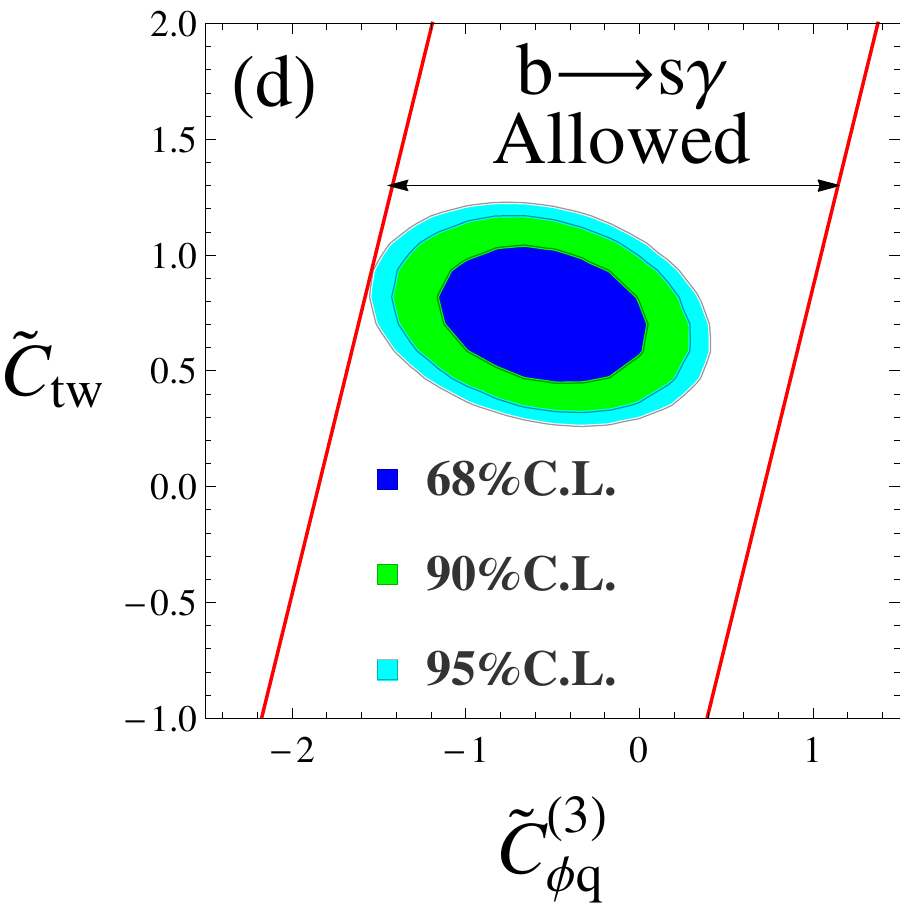}
\caption{Allowed parameter space on 68\%, 90\% and 95\% confidence levels for the parameter $\widetilde{C}_i$ in the subspace. The sub-figures (a) and (b) correspond to the 1-dim parameter space $\widetilde{C}_{\phi q}^{(3)}$ and $\widetilde{C}_{\phi\phi}$, respectively; The contour plots (c) and (d) correspond to the cases which $\widetilde{C}_{\phi q}^{(3)}$, $\widetilde{C}_{\phi\phi}$ and $\widetilde{C}_{\phi q}^{(3)}$ ,$\widetilde{C}_{tW}$ are both modified, respectively. The red lines denote the constraints from $b\rightarrow s\gamma$.
}
\label{Fig:scan2}
\end{figure}

In those cases, we limit the parameter space to subspaces and redo the global analysis in those subspaces.
Based on the NP models to be discussed, we perform the following parameter scans:
\begin{itemize}
\item We scan the 1-dim parameter space by considering either only $f_1^L$ or $f_1^R$.
For example, in the $G(221)$ models and vector-like quark models, the dominant correction is in the coupling $f_1^L$ or $f_1^R$, depending on the detail of the models (see Tables~\ref{tab:models} and \ref{tab:vectorquark}).
\item We also consider the case that both $f_1^L$ and $f_1^R$ are modified, e.g., the vector-like quark doublet ($T,B$) model and triplet cases, shown in Table~\ref{tab:vectorquark}.
\item Another case is that both $f_1^L$ and $f_2^R$ are modified. A typical example is MSSM. Ref.~\cite{Bernreuther:2008us} has shown that the
anomalous $Wtb$ couplings have the following features $f_1^L>f_2^R\gg f_1^R,f_2^L$ in the MSSM.
\end{itemize}
The results of the parameter scan are plotted in Fig.~\ref{Fig:scan1}.
In our analysis, we include the constraints from the $b\rightarrow s\gamma$~\cite{Grzadkowski:2008mf} and update the limits of those anomalous couplings using the updated experimental data~\cite{Agashe:2014kda}, see Table~\ref{tab:bounds}. From the Table~\ref{tab:bounds}, we note that the $b\rightarrow s\gamma$ impose strong constraints on $f_1^R$ and $f_2^L$,
which can be viewed as the results of $m_b$ suppression for the right-handed bottom quark in the pure left-handed $Wtb$ vertex, while the amplitude is enhanced by $m_t$ if the right-handed $Wtb$ vertex exists~\cite{Chetyrkin:1996vx,Larios:1999au,Burdman:1999fw,Grzadkowski:2008mf,Drobnak:2011aa,Drobnak:2011wj}. We also note that the central value of $f_1^L$ is negative, while $f_1^R$ is zero in that case (see Fig.~\ref{Fig:scan1}(a) and Fig.~\ref{Fig:scan1}(b)). It comes from the fact that the
cross section is  proportional to $(1+f_1^L)^2$, and  the minimal $\chi^2$ is dominant determined by the most precise experiment. In this case, the most precise experiment is the measurement of the $t$ channel cross section (see Table~\ref{measurements}). However, the central value of the experiment is smaller than the approximate NNLO SM prediction (see Table~\ref{coeffxi}). Therefore the best-fitted value of $f_1^L$ is negative. For the $f_1^R$ case, the cross section is proportional to $(1+(f_1^R)^2)$, thus the minimal $\chi^2$ is corresponding to the SM case, the central value of $f_1^R$ is zero. As shown in the Fig.~\ref{Fig:scan1}(a) and Fig.~\ref{Fig:scan1}(b), the $f_1^L$ and $f_1^R$ are constrained to be $-0.05<f_1^L<0.025$ and $|f_1^R|<0.13$ at the 90\% C.L. by the direct experiment measurements of top quark, respectively. Similarly, we translate the allowed region of the effective couplings to the coefficients $\widetilde{C}_i$ of dimension-6 operators, with the results shown in Fig.~\ref{Fig:scan2}.

\begin{table}
\centering
\caption{ The current $95\%\,$C.L. bound on the structure of the $Wtb$ vertices from $\bar{B}\rightarrow X_s \gamma$ with $\mu_0 = 160\,$GeV, where $\mu_0$ is the top quark and electroweak gauge boson decoupling scale  ~\cite{Grzadkowski:2008mf}. The branching ratio
$\mathcal{B} (\bar{B}\rightarrow X_s\gamma)_{E_{\gamma}>1.6\rm GeV}=(3.43\pm 0.21\pm 0.07)\times 10^{-4}$
 is used~\cite{Agashe:2014kda}.}
\label{tab:bounds}
\begin{tabular}{c|c|c|c|c} \hline
bound &$ f_1^L $&$   f_1^R   $&$   f_2^L   $&$  f_2^R  $\\\hline
upper &$ ~~0.04    $&$~~0.0021 $&$~~0.0011 $&$ ~~0.19$     \\
lower &$   -0.11    $&$ -0.0008 $&$ -0.0004 $&$ -0.48 $\\\hline
\end{tabular}
\end{table}

\section{The Top Couplings in New Physics Models \label{sec:couplings}}


As mentioned in Sec.~{\ref{sec:1}}, the $Wtb$ coupling can be modified by many kinds of NP models. In this section, we discuss the
constraints from the effective $Wtb$ couplings in several NP models.

\subsection{$G(221)$ models}

The $G(221)$ models~\cite{Mohapatra:1974gc, Mohapatra:1974hk, Mohapatra:1980yp,Georgi:1989ic,Georgi:1989xz,Li:1981nk,Malkawi:1996fs,He:1999vp,Chivukula:2003wj,Hsieh:2010zr,Du:2012vh,Abe:2012fb,Wang:2013jwa} represent a class of NP models with $SU(2)_1 \otimes SU(2)_2 \otimes U(1)_X$ gauge symmetry. There are two breaking patterns: $SU(2)_L \otimes SU(2)_2 \otimes U(1)_X \to SU(2)_L \otimes U(1)_Y$ (BP-I) or $SU(2)_1 \otimes SU(2)_2 \otimes U(1)_Y \to SU(2)_L\otimes U(1)_Y$ (BP-II).
In BP-I, a scalar doublet field $\Phi\sim(1,2)_{1/2}$ or a triplet field $\Sigma\sim(1,3)_1$ with a vacuum expectation value (VEV) $u$ is introduced to induce the symmetry breaking of $SU(2)_L\otimes SU(2)_2\otimes U(1)_X \to SU(2)_L \otimes U(1)_Y$  at the TeV scale, where the numbers in the parentheses are the quantum numbers of $SU(2)_1$ and $SU(2)_2$, respectively, and the numbers in the subscripts of the parentheses are the $U(1)_X$ charges. At the electroweak scale, the symmetry is further broken by a bi-doublet scalar filed $H\sim (2,\bar{2})_0$ with two VEVs $v_1$ and $v_2$. We introduce $v=\sqrt{v_1^2+v_2^2}$ and a mixing angle  $\tan\beta=v_1/v_2$ for convenience.
In BP-II, the breaking of $SU(2)_1\otimes SU(2)_2\to SU(2)_L$ is induced by a scalar bi-doublet $\Phi\sim(2,\bar{2})_0$ with one VEV $u$ at TeV scale, and the electroweak symmetry breaking is induced by a Higgs doublet $H\sim(2,1)_{1/2}$ with VEV $v$. After the symmetry breaking, the new gauge boson $W^{\prime}$ obtains mass and mixes with the SM gauge boson $W$.
For simplicity, we define a new mixing angle $\phi$,
\begin{align}
  \tan\phi &= \frac{g_x}{g_2}, &&\mathbf{(BP-I)}\\
  \tan\phi &= \frac{g_1}{g_2}, &&\mathbf{(BP-II)}
\end{align}
where $g_1$, $g_2$ and $g_x$ are the gauge couplings of $SU(2)_1$, $SU(2)_2$ and $U(1)_X$, respectively.
The gauge bosons' masses are
\begin{align}\label{G221mass}
  M_{W^{\pm}}^2 &=\frac{e^2v^2}{4s_W^2}(1-\frac{s_{2\beta}^2}{x}), &
  M_{W^{\prime\pm}}^2 &= \frac{e^2v^2}{4c_W^2s_{\phi}^2}(x+1+\frac{s_{\phi}^2s_{2\beta}^2c_W^2}{xs_W^2}),&&\mathbf{(BP-I)}\\
  M_{W^{\pm}}^2 &=\frac{e^2v^2}{4s_W^2}(1-\frac{s_{\phi}^4}{x}), &
  M_{W^{\prime\pm}}^2 &=\frac{e^2v^2}{4s_W^2s_{\phi}^2c_{\phi}^2}(x+s_{\phi}^4+\frac{s_{\phi}^6c_{\phi}^2}{x}),&&\mathbf{(BP-II)}
\end{align}
where $e$ denotes the electron charge and $x=u^2/v^2$. We also abbreviate the trigonometric functions as $c_{\phi}\equiv \cos\phi$, $s_{\phi}\equiv \sin\phi$, $s_{2\beta}\equiv\sin2\beta$, $c_W\equiv \cos\theta_w$ and $s_W\equiv \sin\theta_W$ where $\theta_W$ is the weak mixing angle of SM.

The third generation quarks play a special role in several $G(221)$ models and the $Wtb$ couplings are modified through the mixing effects between the new gauge boson $W^{\prime}$ and SM gauge boson $W$ at tree-level. In this work, we will use the left-right model~\cite{Mohapatra:1974gc, Mohapatra:1974hk, Mohapatra:1980yp}, un-unified model~\cite{Georgi:1989ic,Georgi:1989xz} and top-flavor model~\cite{Li:1981nk,Malkawi:1996fs,He:1999vp,Berger:2011xk}  as  examples to discuss the impact of the $Wtb$ measurements on those NP models. 
The charge assignments of the third generation quark fields under the $G(221)$ gauge groups and the detailed expressions of $f_1^L$ and $f_1^R$ of those NP models
are listed in Table~\ref{tab:models}.
\begin{table}
\centering
\begin{center}
\caption{The charge assignments of the third generation quark fields under the $G(221)$ gauge groups and the gauge couplings of the third generation quarks with $W$ boson in several $G(221)$ models.}
\label{tab:models}
\vspace{0.125in}
\begin{tabular}{c|c|c|c|c|c}
\hline
Model             & $SU(2)_1$       & $SU(2)_2$ & $U(1)_X$ & $f_1^L$ & $f_1^R$ \\
\hline
left-right        &
$\begin{pmatrix} t_L \\ b_L \end{pmatrix}$   &
$\begin{pmatrix} t_R \\ b_R \end{pmatrix}$   &
$\dfrac{1}{6}$   &  0   &  $\dfrac{\sin2\beta}{x}$ \\
\hline
un-unified       &
$\begin{pmatrix} t_L \\ b_L \end{pmatrix}$   &  - & $\dfrac{1}{6}$  &  $ - \dfrac{s_{\phi}^{4}}{x}$ & $0$ \\
\hline
top-flavor      & -      &$\begin{pmatrix} t_L \\ b_L \end{pmatrix}$ & $\dfrac{1}{6}$ &
$ \dfrac{s_{\phi}^{2}c_{\phi}^{2}}{x}$    & $0$ \\
\hline
\end{tabular}
\end{center}
\end{table}

\begin{figure}
\centering
\includegraphics[width=0.215\textwidth]{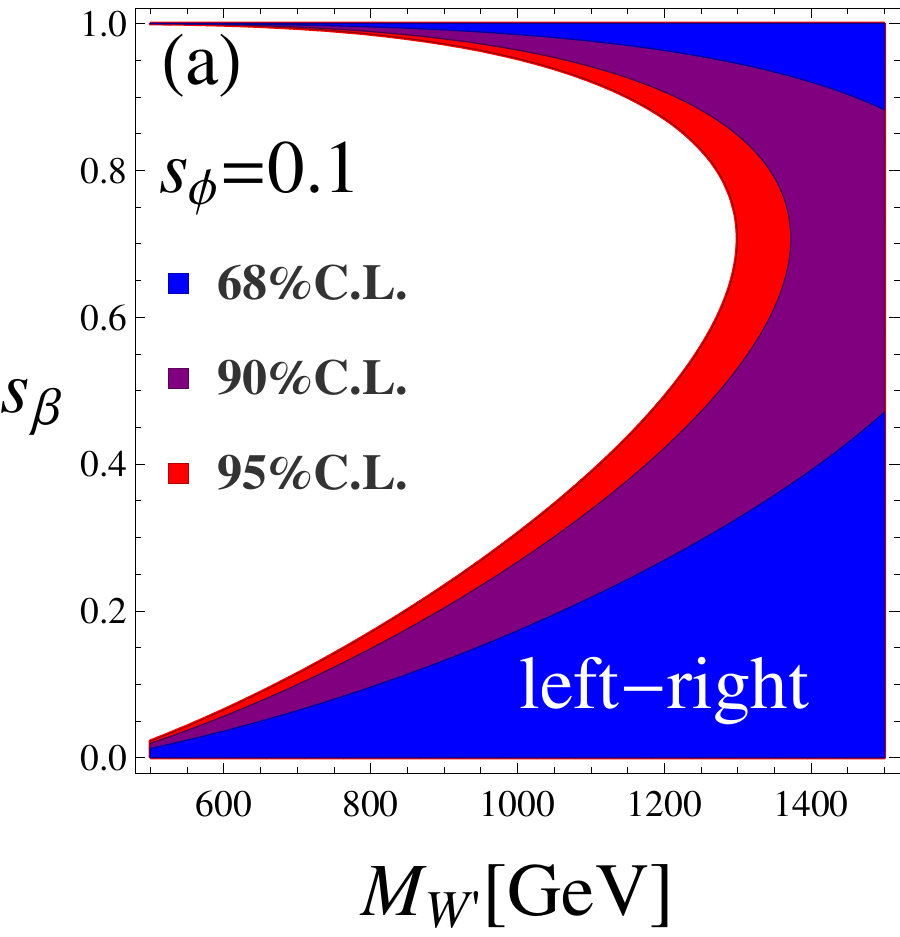}
\includegraphics[width=0.22\textwidth]{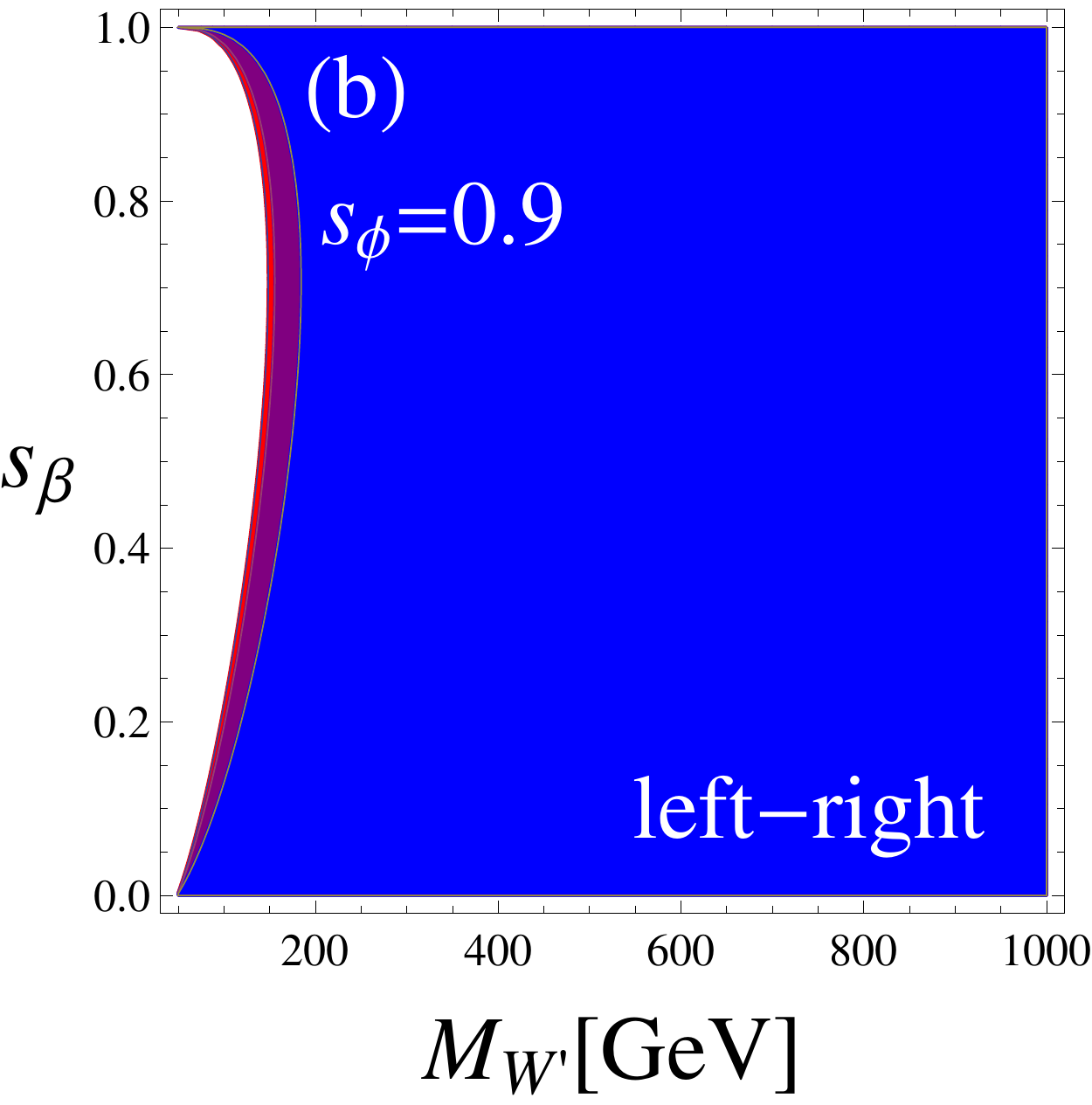}
\includegraphics[width=0.22\textwidth]{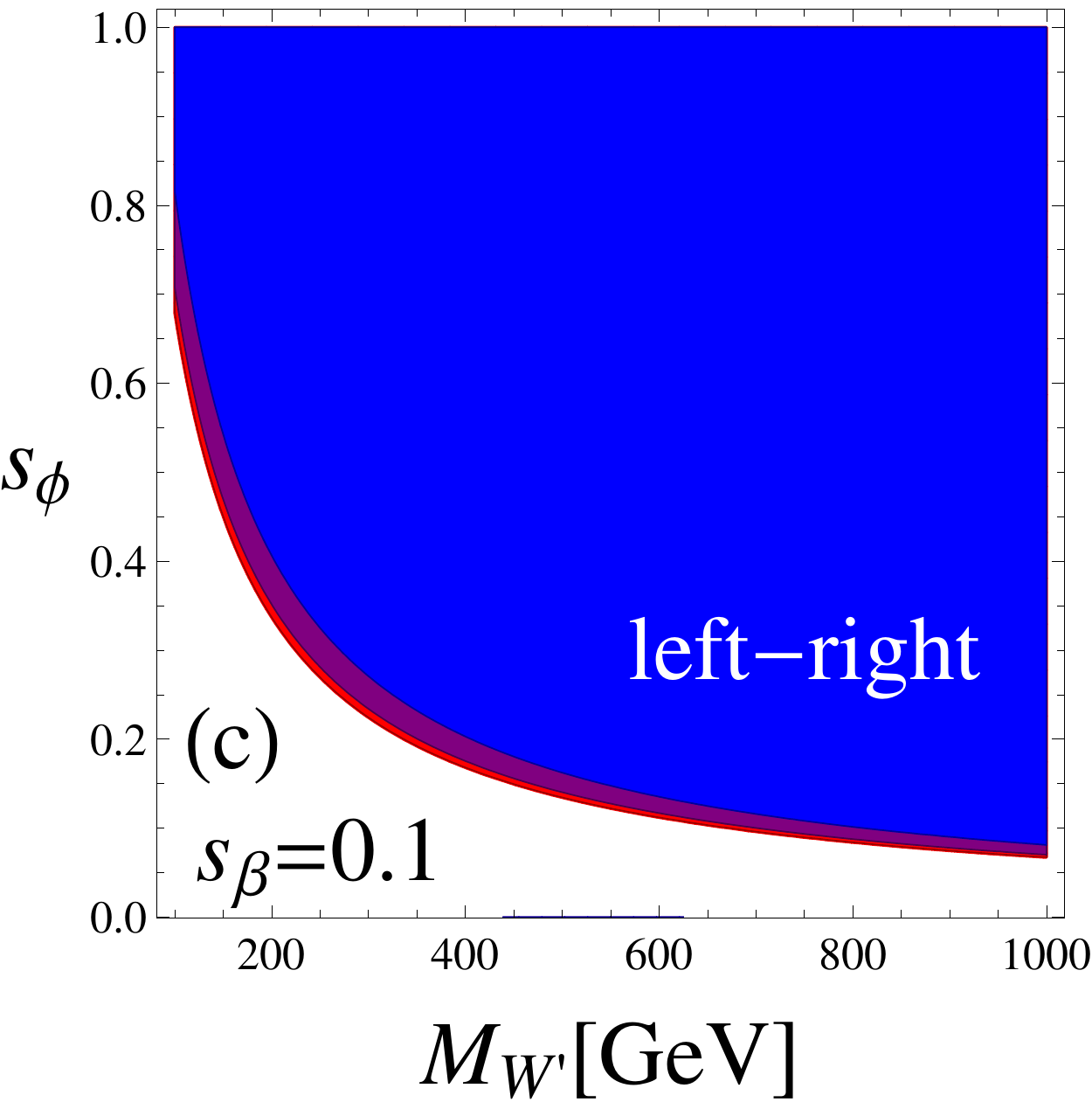}
\includegraphics[width=0.22\textwidth]{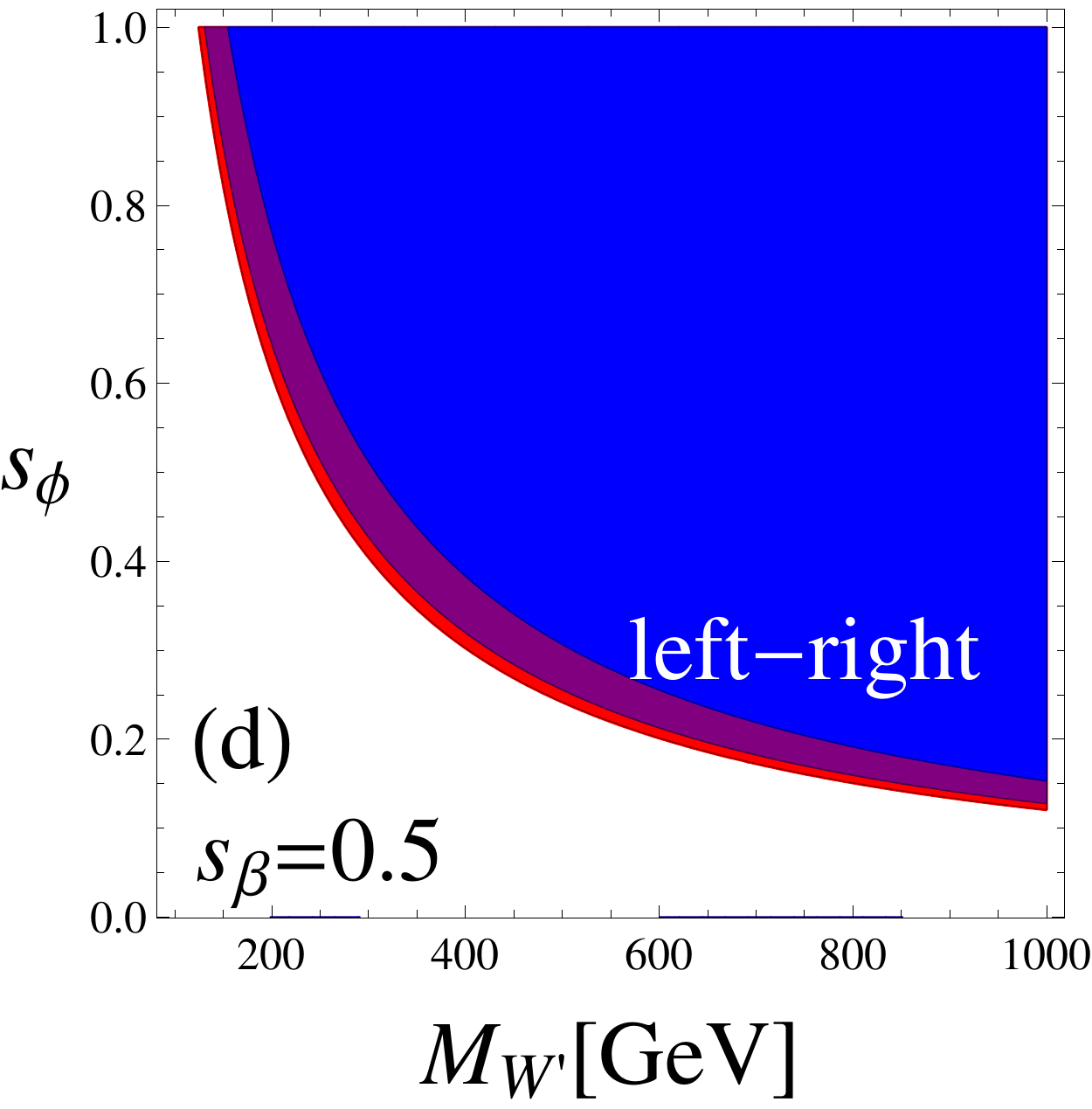}\\
\includegraphics[width=0.22\textwidth]{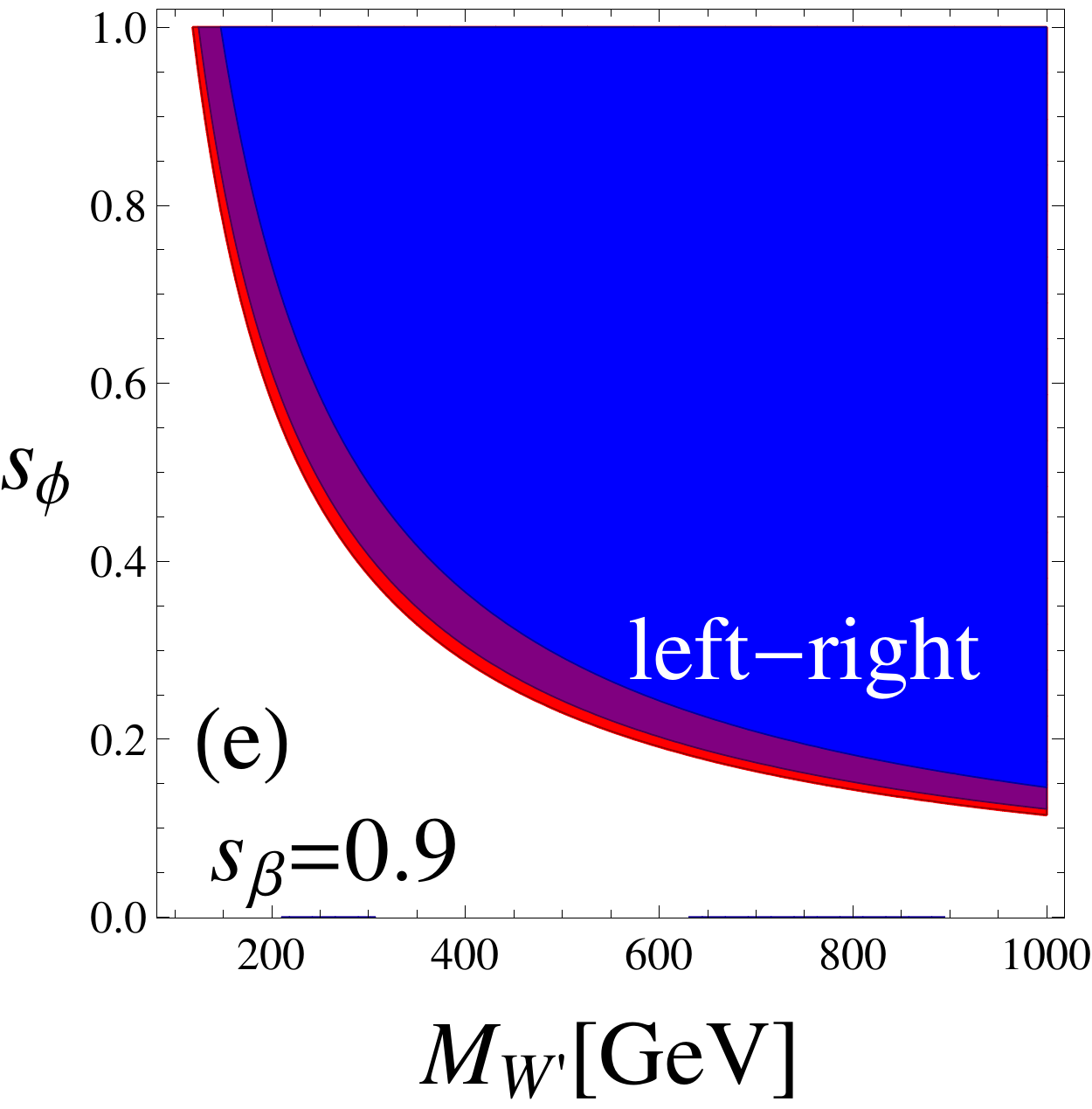}
\includegraphics[width=0.22\textwidth]{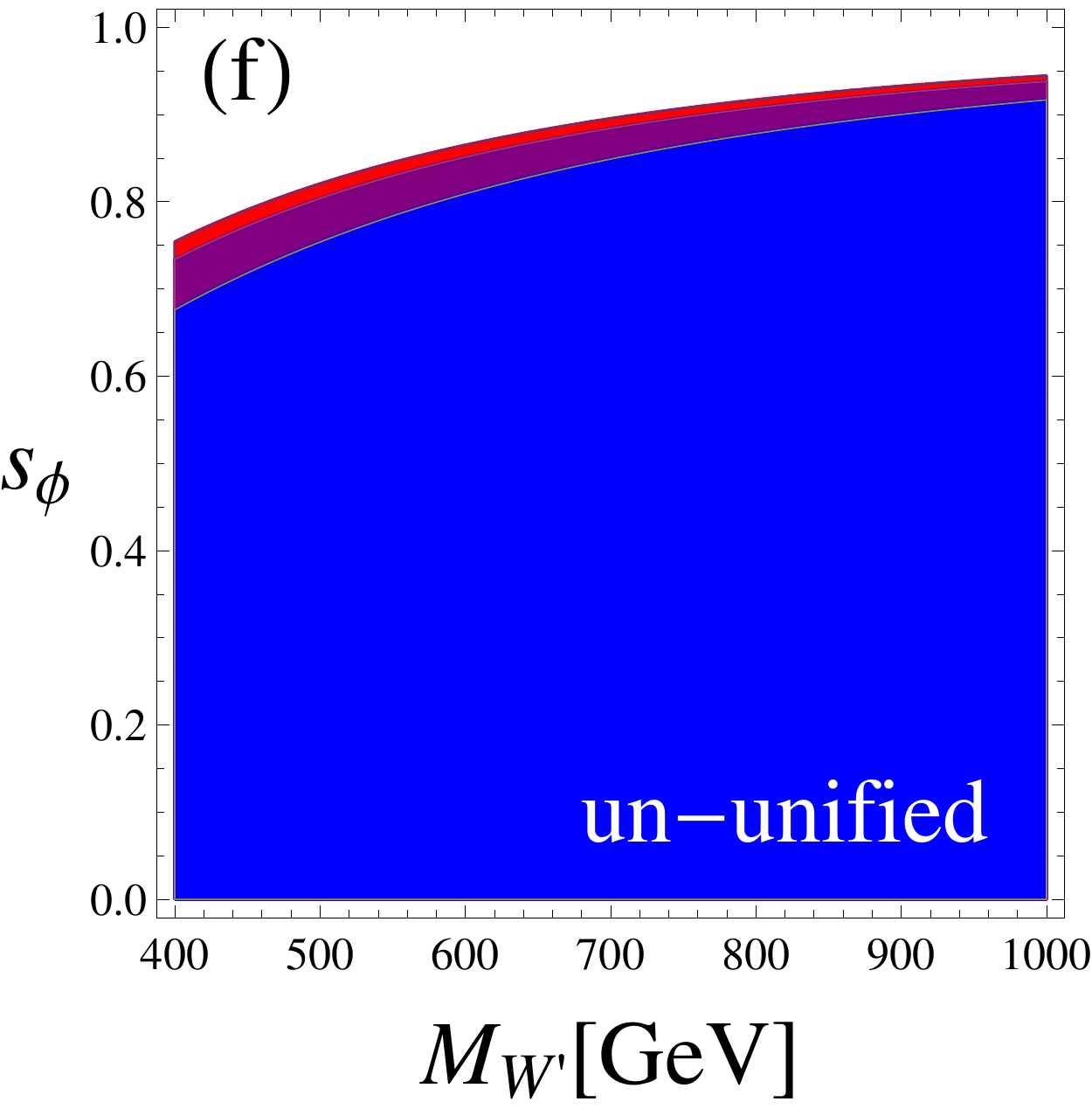}
\includegraphics[width=0.22\textwidth]{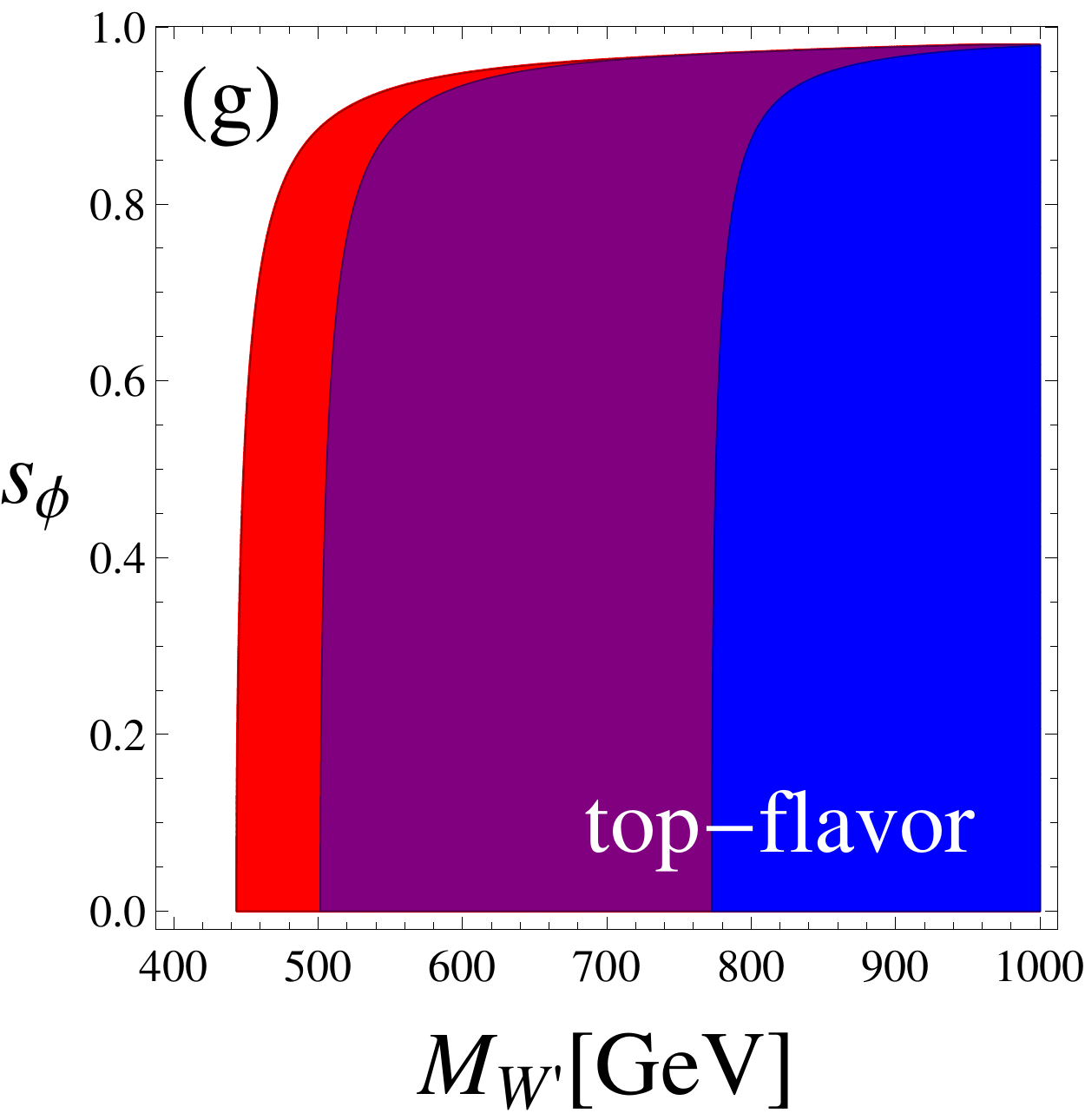}
\caption{Allowed parameter space on 68\%, 90\% and 95\% confidence levels for the several $G(221)$ models in the ($M_{W'},s_{\beta}$) or ($M_{W'},s_{\phi}$) plane. Sub-figures (a-e) correspond to the left-right model with $s_{\phi}=0.1,~0.9$ (a, b) or $s_{\beta}=0.1,~0.5,~0.9$ (c, d, e). The contour plots (f) and (g) correspond to un-unified and  top-flavor models, respectively.
}
\label{221model}
\end{figure}

We discuss the impact of the effective $Wtb$ coupling measurements on the parameter space of the $G(221)$ models.
Figure~\ref{221model} presents the allowed parameter space on 68\%, 90\% and 95\% C.L., respectively in the left-right model in BP-I and  the un-unified and top-flavor models in BP-II.
In the left-right model, we show the allowed parameter space in the $(M_{W'}, s_\beta)$ plane for  $s_\phi=0.1$ and $s_\phi=0.9$ in Fig.~\ref{221model}(a) and Fig.~\ref{221model}(b), respectively.
Figure~\ref{221model}(a) shows, for $s_\phi=0.1$, a vast parameter space of moderate $s_\beta$ and smaller $M_{W^{\prime}}$ is not allowed by the highly constrained right-handed $Wtb$ coupling, see Table~\ref{tab:models}. For a
larger $s_\phi=0.9$, the constraint for $M_{W^{\prime}}$ tends to be looser, see Fig.~\ref{221model}(b). The reason is that the gauge coupling of the SM $U(1)_Y$ is related to the gauge couplings $g_2$ and $g_x$ as: $1/g_Y^2=1/g_2^2+1/g_X^2$, thus $g_Y$ is approximately equal to $g_X$ in the limit $s_\phi\rightarrow 0$, which is corresponding to the decoupling region.
The shape of the allowed parameter space for $s_\phi=0.9$ is the same as the one of $s_\phi=0.1$ in small $M_{W^{\prime}}$ region.

Figures~\ref{221model}(c-e) display the allowed parameter space in the plane of $(M_{W'}, s_\phi)$ for $s_\beta=0.1, 0.5, 0.9$, which shows that the $M_{W^{\prime}}$ is not sensitive to $s_\beta$. It is owing to the fact  $s_\beta$ only induces the mixing between the gauge boson $W$ and $W^{\prime}$ (see Eq.\ref{G221mass}), and the primary source of the mass of $W^{\prime}$ is from the first step breaking which is proportional to the VEV $u$~\cite{Hsieh:2010zr}.

In the un-unified and top-flavor models, only the left-handed $Wtb$ coupling is modified.
The allowed parameter space in the $(M_{W^{\prime}}, s_{\phi})$ plane is shown in Fig.~\ref{221model}(f) (un-unified model) and Fig.~\ref{221model}(g) (top-flavor model). We note that the shape in the large $s_{\phi}$ region is different between the un-unified model and top-flavor model. The difference can be understood from the effective $Wtb$ couplings shown in Table~\ref{tab:models}. The effect consists of two factors: one is $W^{\prime}tb$ coupling, the other is $W-W^{\prime}$ mixing. The former is different in the two models, but the latter is the same. The gauge coupling of the heavy gauge boson $W^{\prime}$ with the top and bottom quarks is proportional to $\tan\phi$ in the un-unified model, while to $\cot\phi$ in the top-flavor model. Thus larger $s_{\phi}$ is allowed in the top flavor model compared with the un-unified model. Another important feature in the top-flavor model is that the constraint on $M_{W^{\prime}}$ is not sensitive to the parameter $s_{\phi}$ as shown in Fig.~\ref{221model}(g). In the model, $f_1^L$ is modified by the $W^\prime$ mass which is proportional to $x/(s_{\phi}^2c_{\phi}^2)$.

In order to validate our EFT prescription of the NP effects, $W^{\prime}$ should be heavy. Figure~\ref{221model} shows the constraints of $G(221)$ models from $Wtb$ measurements are  weak when new gauge bosons are heavier than 800 GeV. The constraints from low energy precision measurements and  direct searches at the Tevatron and LHC have pushed the new heavy gauge bosons to several TeV~\cite{Cao:2012ng}. Therefore, it is difficult to further constrain the parameter space by the $Wtb$ measurements.

\subsection{Vectorlike Quark Models: $T$ and $B$}

The vector-like quark (VLQ) is a common ingredient of many NP models. In order to keep the discussion general, we employ an
effective Lagrangian approach to parametrize the effects of vector-like quarks. The quantum numbers of the new VLQ with respect to the $SU(2)_L\otimes U(1)_Y$ gauge group are summarized in Table~\ref{tab:VLQ1}~\cite{delAguila:2000rc,AguilarSaavedra:2009es,Cacciapaglia:2010vn,Aguilar-Saavedra:2013qpa}.

\begin{table}
\centering
\caption{The quantum numbers of the vector-like quarks under the SM $SU(2)_L\otimes U(1)_Y$ gauge symmetry, the electric charge of the quark is obtained by $Q=T_3+Y$, where $T_3$ is the third component of the isospin, and $Y$ is the hypercharge of $U(1)_Y$.}
 \begin{tabular}{c|c|c|c|c|c|c|c}
\hline
model & $T$  & $B$  &
$\begin{pmatrix} X\\T \end{pmatrix}$ & $\begin{pmatrix} T\\B \end{pmatrix}$ &
$\begin{pmatrix} B \\ Y \end{pmatrix}$ & $\begin{pmatrix} X \\ T\\ B \end{pmatrix}$ &
$\begin{pmatrix} T \\ B \\Y \end{pmatrix}$ \tabularnewline
\hline
$SU(2)_L$ & 1 & 1 & 2 & 2 & 2  & 3 & 3  \tabularnewline
\hline
$U(1)_Y$ & $\tfrac{2}{3}$ & $-\tfrac{1}{3}$ & $\tfrac{7}{6}$  & $\tfrac{1}{6}$ & $-\tfrac{5}{6}$ & $\tfrac{2}{3}$ & $-\tfrac{1}{3}$ \tabularnewline
\hline
\end{tabular}
\label{tab:VLQ1}
\end{table}

The $Wtb$ coupling is modified by the mixing between top-quark or bottom-quark with their corresponding vector-quark partner, e.g. top-partner $T$ or bottom partner $B$. After the spontaneous symmetry breaking, the mass terms of the top-quark and top-partner $T$ in the singlet or triplet models are~\cite{Cacciapaglia:2010vn}
\begin{equation}\label{STLag}
\mathcal{L}_{mass}^t=-\frac{y_tv}{\sqrt{2}}\bar{t}_Lt_R-x_t\bar{t}_LT_R-M\bar{T}_LT_R+h.c.,
\end{equation}
where $y_t$ is the Yukawa coupling of top quark in the SM, $x_t$ represents the mixing parameter between $t$ and $T$, and $M$ denotes VLQ's mass.

In the case of doublet VLQ models, the mass terms are~\cite{Cacciapaglia:2010vn}
\begin{equation}\label{DLag}
\mathcal{L}_{mass}^t=-\frac{y_tv}{\sqrt{2}}\bar{t}_Lt_R-x_t\bar{T}_Lt_R-M\bar{T}_LT_R+h.c..
\end{equation}
The weak and mass eigenstates can be related by $2\times 2$ unitary matrices,
\bea
	\left(
	 \begin {array} {c}
	         t_{L,R} \\ T_{L,R} \\
	    \end {array}
	   \right)
	=\left(
	\begin{array}{cc}
     c^t_{L,R}  & s^t_{L,R}  \\
     -s^t_{L,R} & c^t_{L,R}
	\end{array}
	\right)
      \left(
	 \begin {array} {c}
	    t_{L,R}^{\prime}  \\ T_{L,R}^{\prime}\\
	    \end {array}
	   \right),
\eea
where $c^t_{L,R}$ and $s^t_{L,R}$ denote the cosine and sine of the mixing angles between left-handed and right-handed top quark $t_{L,R}$
with the top partner $T_{L,R}$. In this section, we use the prime in the superscript on mass eigenstates to distinguish from the weak eigenstates.
After diagonalizing the mass matrices, we can rewrite the mixing angles in according to the parameters in Eqs. \ref{STLag} and \ref{DLag}.

In the case of singlet and triplet VLQ models
\bea\label{stmix}
  s^t_L  =\frac{M x_t}{\sqrt{(M^2-m_t^2)^2+M^2x_t^2}}, \quad
  s^t_R  =\frac{m_t}{M}s^t_L.
\eea
For the doublet VLQ models
\bea\label{dmix}
  s^t_R  =\frac{M x_t}{\sqrt{(M^2-m_t^2)^2+M^2x_t^2}}, \quad
  s^t_L  =\frac{m_t}{M}s^t_R,
\eea
where $m_t$ is the top quark mass. The mass of the heavy top partner is
\bea\label{eqmass}
M_T^2=M^2\left(1+\frac{x_t^2}{M^2-m_t^2}\right).
\eea

We can obtain the similar formulae in the bottom quark sector, and use $m_b$ and $x_b$  to denote the bottom quark mass and mixing parameter between bottom quark and bottom partner $B$ hereafter. In the triplet VLQ models, $s^t_{L/R}$ is correlated with $s^b_{L/R}$ as the mixing  parameters $x_t$ and $x_b$ are linearly related to each other. For example, in the $(X,T,B)$ model~\cite{Cacciapaglia:2010vn},
\begin{align}
\mathcal{L}_{(X,T,B)} &=-\frac{y_t v}{\sqrt{2}}\bar{t}_L t_R-x_t\bar{t}_L T_R- x_b\bar{b}_L B_R
 -M(\bar{T}_L T_R+\bar{B}_L B_R+\bar{X}_L X_R)+h.c.,
 \label{XTB}
\end{align}
with $x_b=\sqrt{2}x_t$,
and in the $(T,B,Y)$ model~\cite{Cacciapaglia:2010vn},
\begin{align}
\mathcal{L}_{(T,B,Y)}&=-\frac{y_t v}{\sqrt{2}}\bar{t}_L t_R-x_t\bar{t}_L T_R+x_b\bar{b}_L B_R
 -M(\bar{T}_L T_R+\bar{B}_L B_R+\bar{Y}_L Y_R)+h.c.,
 \label{TBY}
\end{align}
with $x_t=\sqrt{2}x_b$.
After diagonalizing the mass matrices, we obtain
the couplings of the gauge boson fields to the third generation quarks,
\begin{eqnarray}
\mathcal{L}_{Zbb} &= & \frac{g}{2 c_W} \bar b^{\prime} \gamma^\mu \left( -f_{b}^L P_L - f_{b}^R P_R + \frac{2}{3} s_W^2 \right) b^{\prime} \Zm \,.
\label{eq:ll}
\end{eqnarray}
The couplings  $f_1^{L,R}$ and $f_{b}^{L,R}$ for all 7 models are listed in Table~\ref{tab:vectorquark}, see also Ref.~\cite{Aguilar-Saavedra:2013qpa}.

\begin{table}
\centering
\caption{The $Wtb$ and $Zbb$ couplings in 7 models, where $\clu$($\slu$) and $\cld$($\sld$) denote the cosine (sine) of the mixing angles of left-handed top quark and bottom quark with their heavy partners, respectively. Similarly, $s_R^t$ and $s_R^b$ are the sine of the mixing angles of right-handed top quark and bottom quark with their heavy partners, respectively.}
\label{tab:vectorquark}
\scalebox{0.82}[0.82]{%
\begin{tabular}{c|c|c|c|c|c|c|c}
\hline
model & \ts  & \bs  & \xt & \tb & \by & \xtb & \tby \tabularnewline
\hline
$f_1^L$ & $\clu-1$ &$\cld-1$ & $\clu-1$ & $\clu \cld +\slu\sld-1$ & $\cld-1$ & $\clu\cld+\sqt\slu\sld-1$  & $\clu \cld +\sqt \slu \sld-1$   \tabularnewline
\hline
$f_1^R$ & 0 & 0 & 0 & $\sru \srd $ & 0 & $\sqt \sru \srd$ &  $\sqt \sru \srd$  \tabularnewline
\hline
$f_{b}^L$ & 1 & $(c_L^b)^2$ & 1 & 1 & $(c_L^b)^2-(s_L^b)^2$ & $1+(s_L^b)^2$ & $(c_L^b)^2$ \tabularnewline
\hline
$f_{b}^R$ & 0 & 0 & 0 & $(s_R^b)^2$ & -$(s_L^b)^2$ & $2(s_R^b)^2$ & 0 \tabularnewline
\hline
\end{tabular}}

\end{table}

\subsubsection{The mixing angles}
\begin{figure}
\centering
\includegraphics[width=0.3\textwidth]{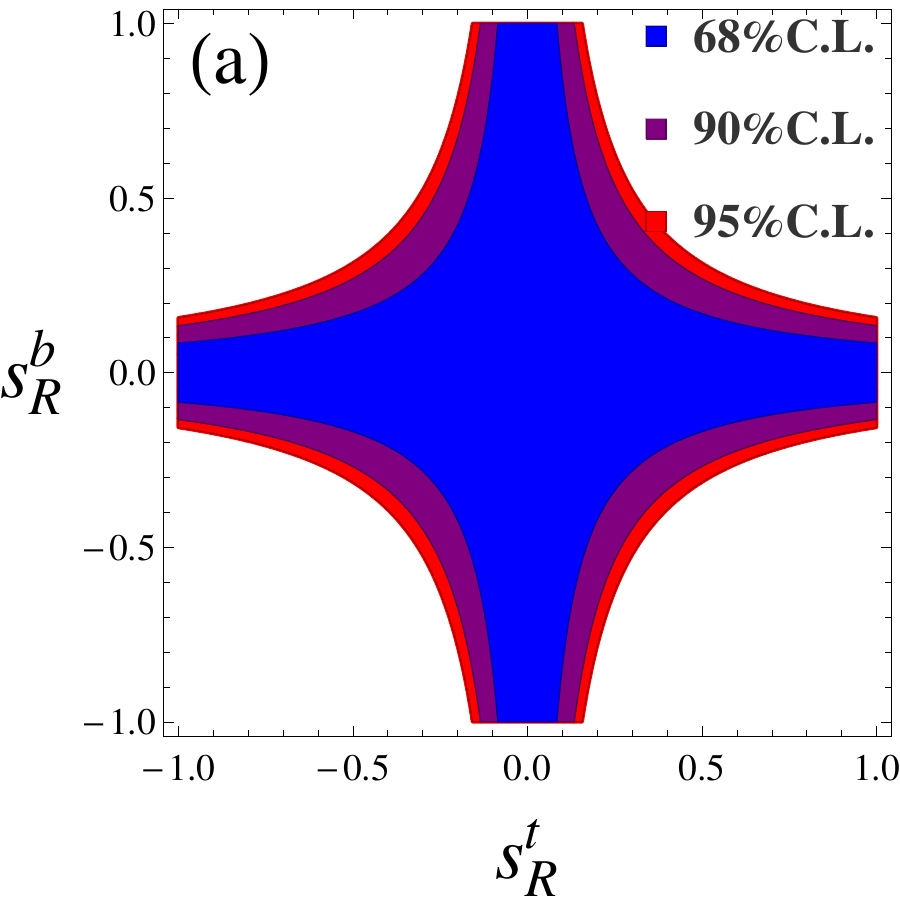}
\includegraphics[width=0.3\textwidth]{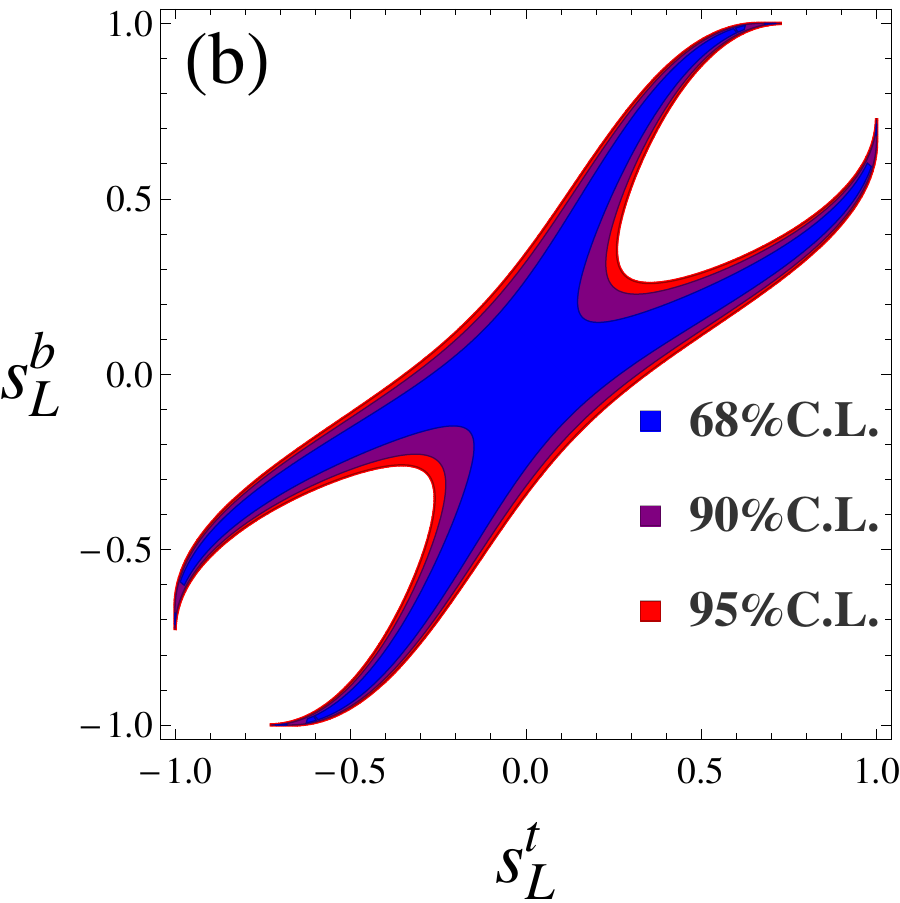}
\includegraphics[width=0.29\textwidth]{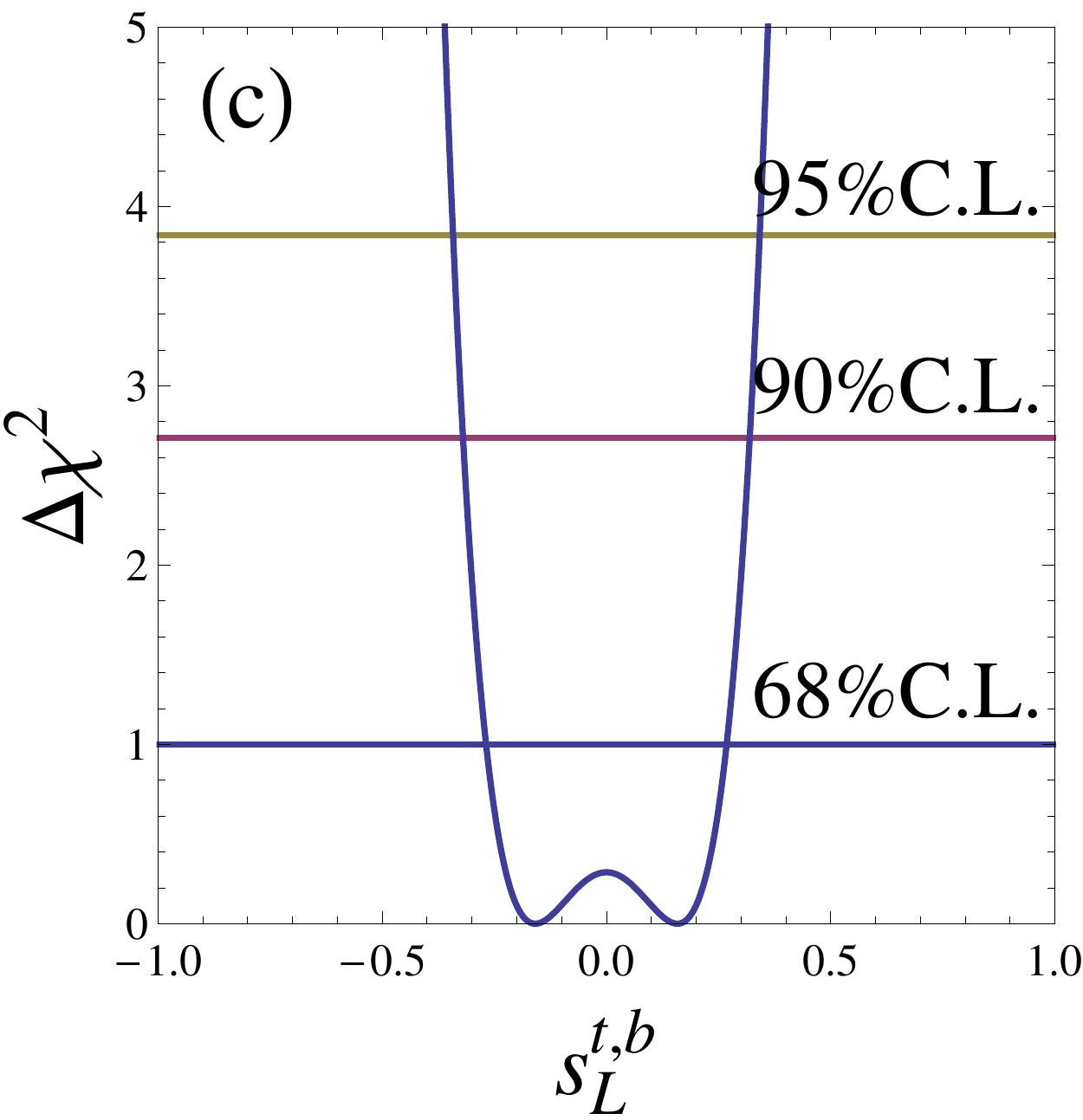}
\caption{Allowed parameter space on 68\%, 90\% and 95\% confidence levels for the VLQ models: (a) Doublet $(T,B)$ model; (b) Triplet models; (c) Singlet, Doublet $(X,T)$, $(B,Y)$ models.
}
\label{VLQ}
\end{figure}

Using the results of  Sec.~{\ref{sec:fitting}}, we translate the allowed region of the effective $Wtb$
couplings to the parameter space of VLQ models on the 68\%, 90\% and 95\% C.L., respectively;
see Fig.~\ref{VLQ}. In the triplet and singlet models, VLQs mainly couple to the left-handed top or bottom quark, while in the doublet models to the right-handed top or bottom quark. As a result, $s^{t/b}_L \gg s^{t/b}_R$ in the triplet model, and  $s^{t/b}_R \gg s^{t/b}_L$ in the doublet model. It is, therefore, convenient to neglect the smaller mixing angles in our parameter scan.
Figure~\ref{VLQ}(a) shows the allowed parameter space of $(s_R^t,s_R^b)$ in the doublet $(T,B)$ model.
It yields the hyperbola contour region on the plane, which is determined by the right-handed $Wtb$ coupling $f_1^R=s_R^t s_R^b$. Figure~\ref{VLQ}(b) displays the contour on the plane of $(s_L^t,s_L^b)$ in triplet models, and the allowed parameter space is symmetric about $\slu\leftrightarrow\sld$ and $\slu\leftrightarrow-\sld$. Such a behavior can be understood from the deviation of the left-handed $Wtb$ coupling between the triplet models and the SM
\begin{equation}
 f_1^L=\sqrt{1-(\slu)^2}\sqrt{1-(\sld)^2}+\sqrt{2}\slu\sld-1.
\end{equation}
Furthermore, the upper limit of $f_1^L$ determines the shape of the parameter contour in the
top-right and bottom-left regions, and the lower limit of $f_1^L$ determines the boundary of the parameter space in the top-left and bottom-right regions.
Note that only one mixing angle exists in the singlet models $T$ and $B$, and in the doublet models $(X,T)$ and $(B,Y)$.
Figure~\ref{VLQ}(c) shows the allowed parameter space of the sole mixing angle, $s_L^t$ or $s_L^b$, which indicates $|s_L^{t,b}| < 0.34$ at 95\% C.L..

\subsubsection{Mixing angles versus $M_{T,B}$}
We note that the constraints from the $Wtb$ measurements on the parameter space of the VLQ models are weak.
However, the left-handed top quark and bottom quark form a $SU(2)_L$ weak doublet,
the left-handed $Wt_Lb_L$ coupling is always related to the left-handed $Zt_Lt_L$ and $Zb_Lb_L$ couplings.
The $R_b$ and $A_{FB}^b$ measurements at the LEP-II~\cite{Abdallah:2008ab} impose a severe constraint on the $Zb_Lb_L$, which yields a correlation between $Zt_Lt_L$ and $Wt_Lb_L$~\cite{Agashe:2006at,Berger:2009hi}. To fully constrain the allowed parameter space of the VLQ models, it is necessary to include the $Zbb$ and $Ztt$ couplings in the analysis.  The complete study will be presented elsewhere.
In this work, we will allow a variation of -0.2\% and +1\% for left-handed $Zb_Lb_L$ coupling, while in the right-handed case, -5\% and +20\% is used~\cite{Cacciapaglia:2010vn}.

Another important constraint comes from the Peskin-Takeuchi parameters $\hat{S}$, $\hat{T}$ and $\hat{U}$~\cite{Peskin:1991sw}. The contribution of arbitrary number of vector-like singlet and doublet quarks to the $\hat{S}$, $\hat{T}$ and $\hat{U}$ parameters have been calculated in Ref.~\cite{Lavoura:1992np}, and is generalised to arbitrary couplings in Ref.~\cite{Anastasiou:2009rv}. In our work, we calculate the $\hat{T}$ parameter of all possible VLQ models. Our analytical results are consistent with Ref.~\cite{Cai:2012ji}. The definition of the $\hat{T}$ parameter is
\begin{equation}
\alpha \hat{T}= \frac{\Pi_{WW}(0)}{m_W^2}-\frac{\Pi_{ZZ}(0)}{m_Z^2},
\end{equation}
where the notation $\Pi_{WW}(0)$  and $\Pi_{ZZ}(0)$ denotes the vacuum polarization amplitudes of $W$ loop and $Z$ loop at zero momentum, respectively, $\alpha$ is the fine-structure constant and $m_{W/Z}$ are the mass of $W$ and $Z$ bosons. In the NP model, the contribution of the $\hat{U}$-parameter is usually very small and can be neglected. Fixing $\hat{U}=0$, the $\hat{T}$ parameter is obtained~\cite{Baak:2014ora}
\begin{equation}
\Delta \hat{T}=\hat{T}-\hat{T}_{\rm SM}=0.10\pm 0.07,
\end{equation}
where the reference of top quark and Higgs boson mass are $m_t=173$GeV and $m_H=125$GeV.

\begin{figure}
\centering
\includegraphics[width=0.32\textwidth]{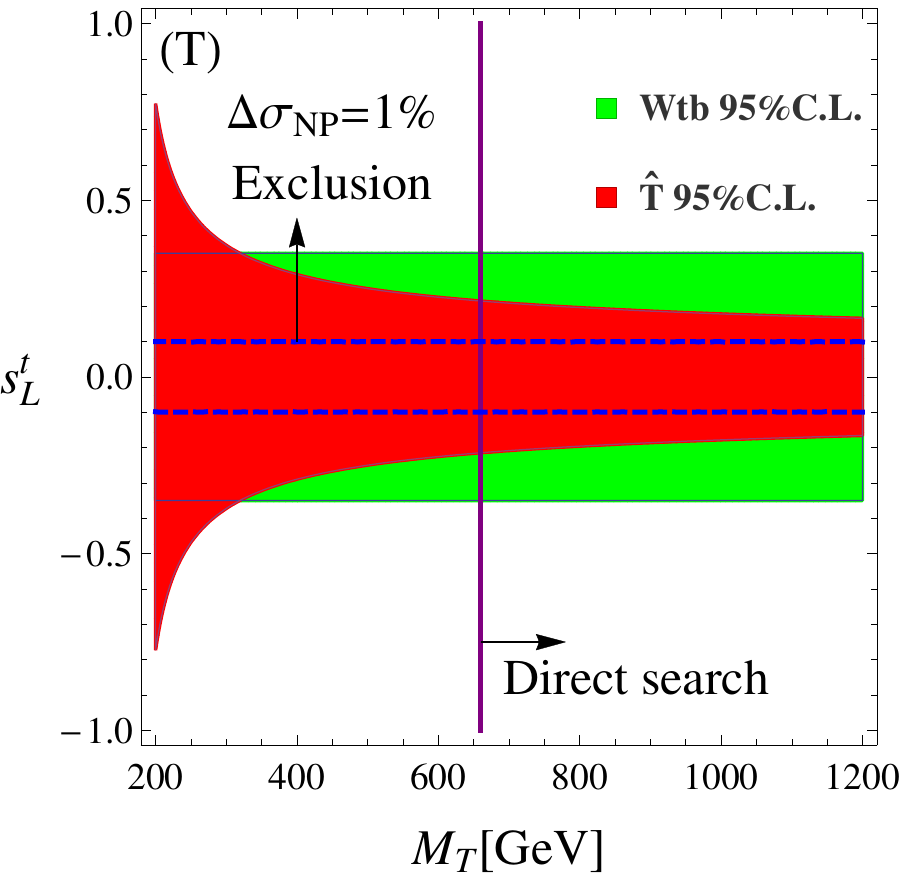}
\includegraphics[width=0.32\textwidth]{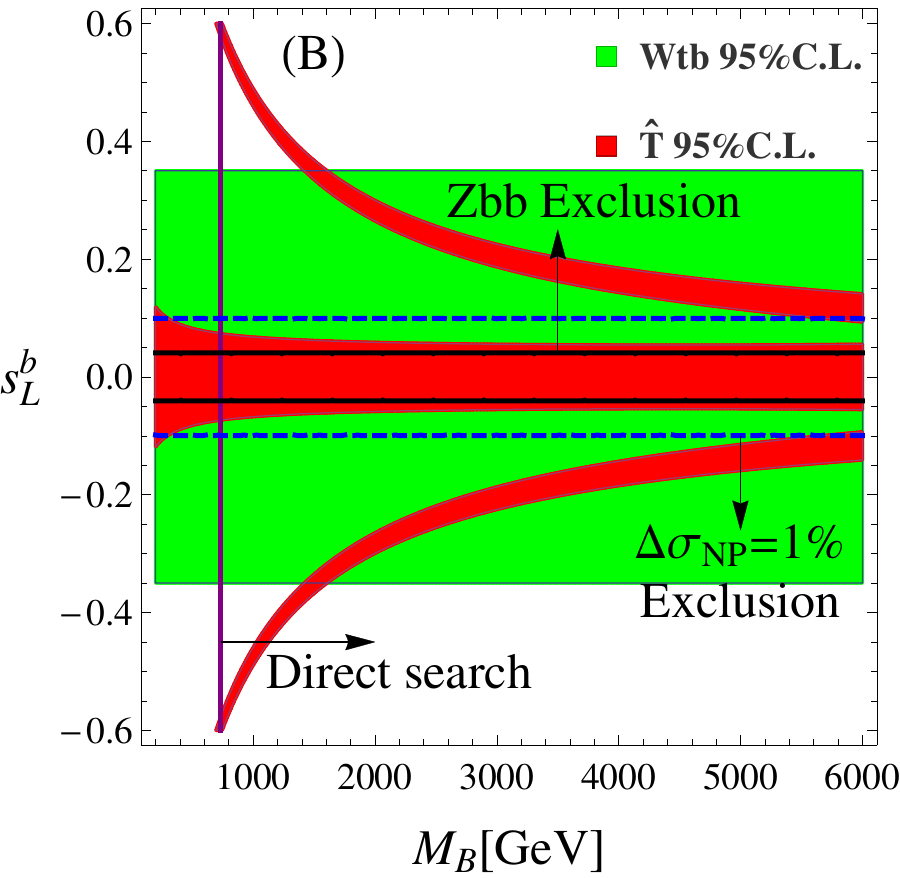}
\caption{Allowed parameter space on 95\% C.L. of the singlet $T$ model in the plane of ($M_T, s_L^t$) and of the singlet $B$ model in the plane of ($M_B, s_L^b$) after including indirect and direct constraints:
$Wtb$ coupling (green region), $\hat{T}$ parameter (red region), $Zbb$ coupling (black line), direct search at the LHC (purple line).
}
\label{VLQB}
\end{figure}

\begin{figure}
\centering
\includegraphics[width=0.32\textwidth]{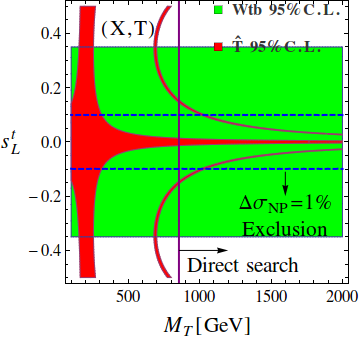}
\includegraphics[width=0.32\textwidth]{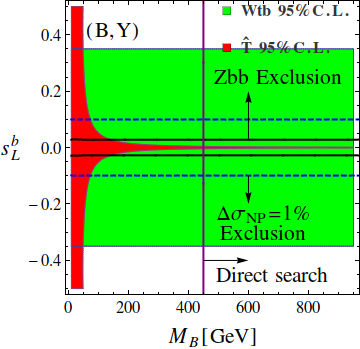}
\caption{Allowed parameter space on 95\% C.L. of the $(X,T)$ doublet model in the plane of ($M_T, s_L^t$) and of the $(B, Y)$ doublet model in the plane of ($M_B, s_L^b$) after including indirect and direct constraints.
The convention is the same as in Fig.~\ref{VLQB}.
}
\label{VLQB2}
\end{figure}

\begin{figure}
\centering
\includegraphics[width=0.32\textwidth]{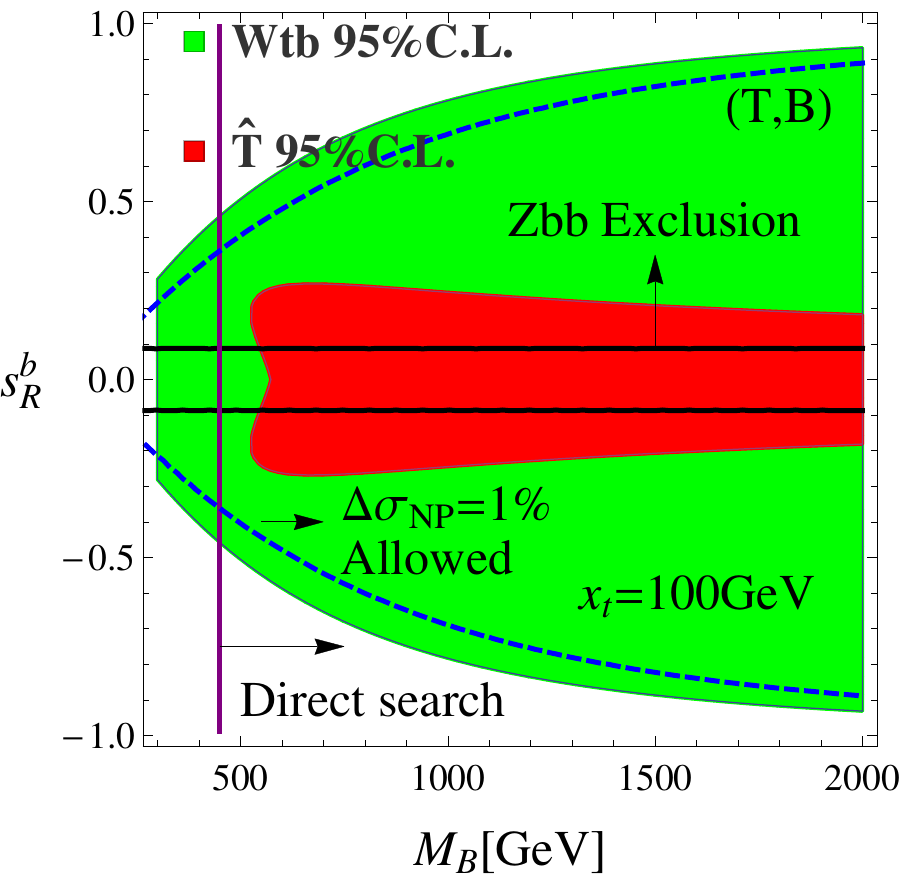}
\includegraphics[width=0.32\textwidth]{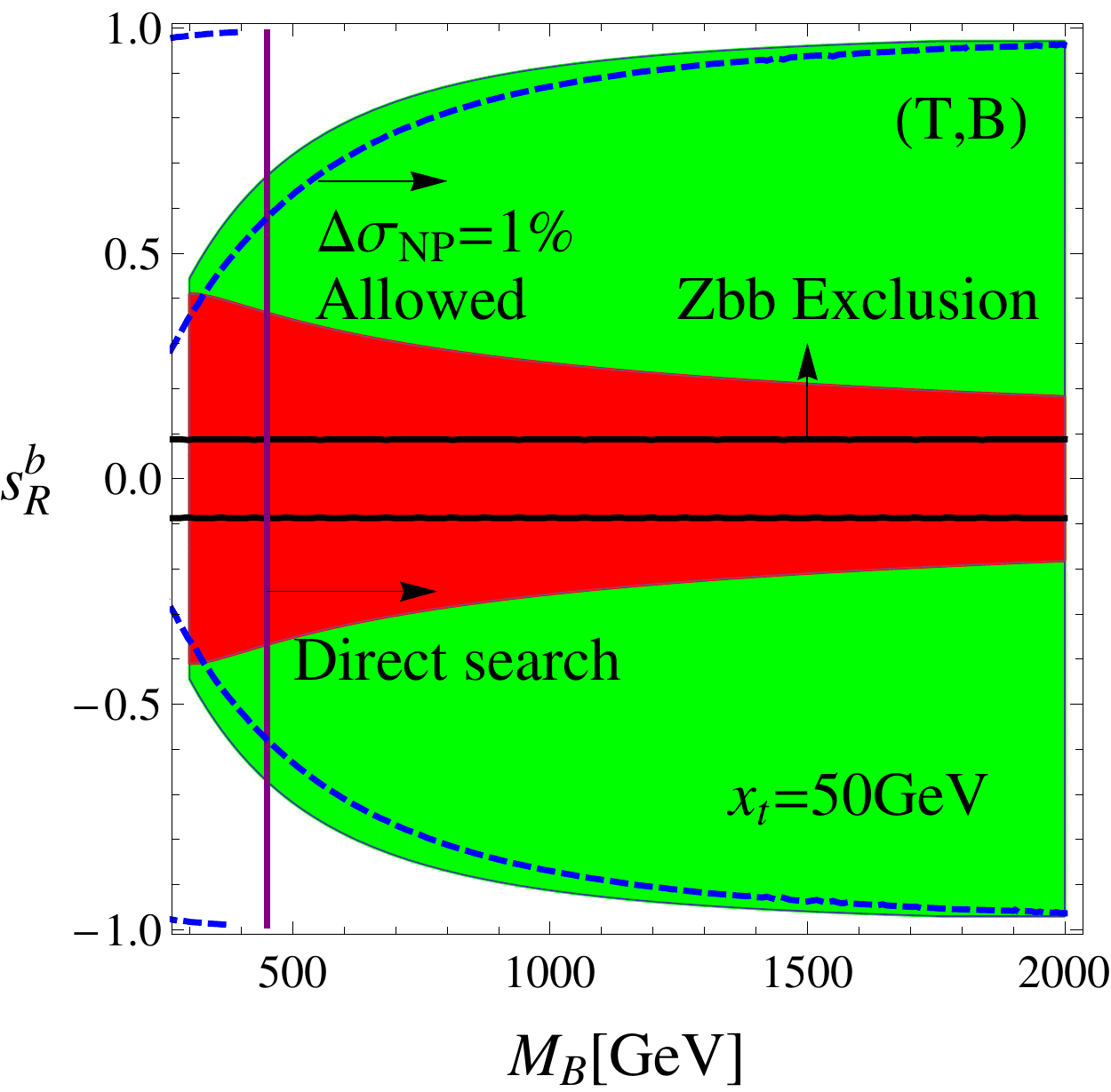}
\includegraphics[width=0.32\textwidth]{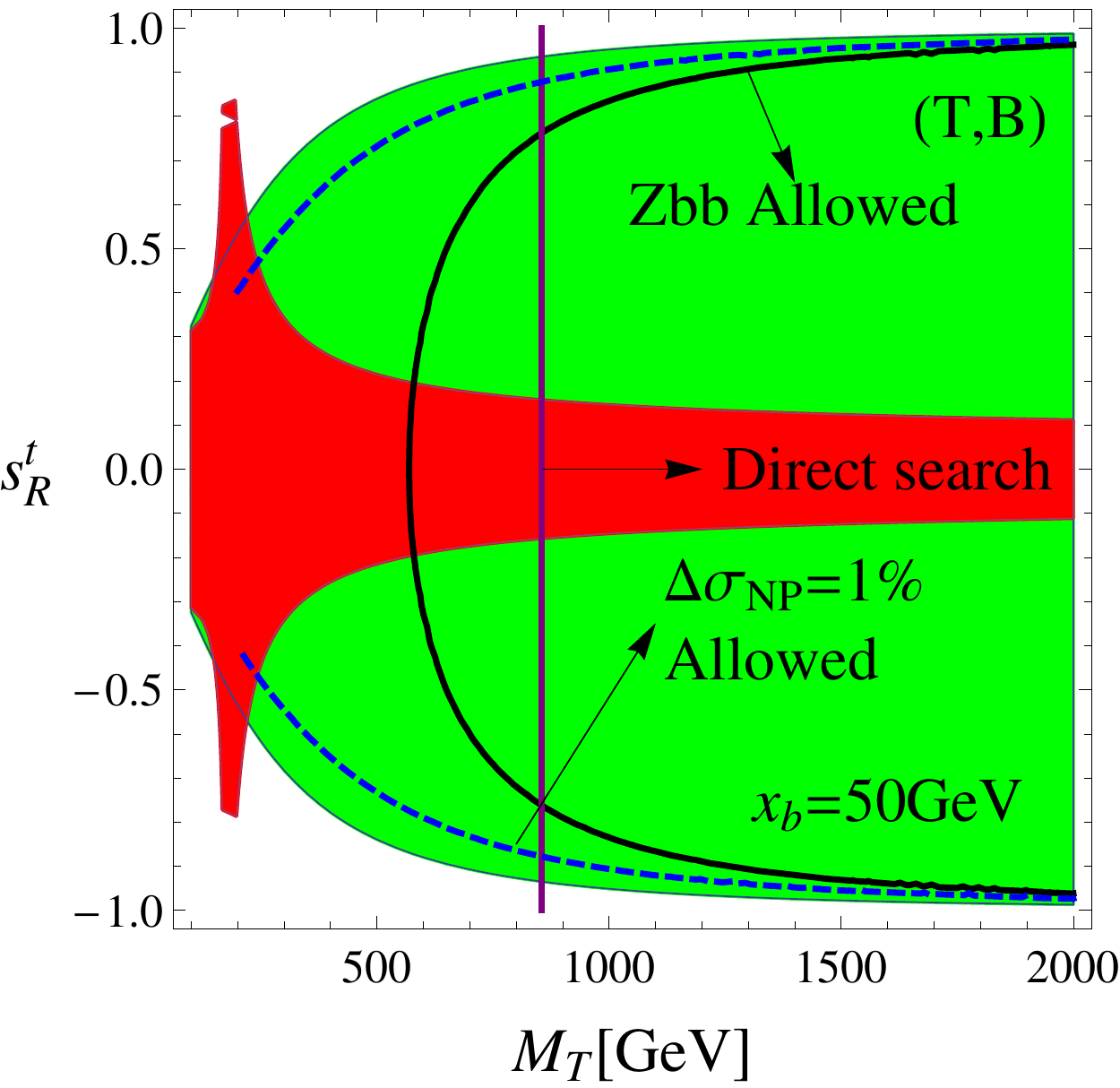}
\caption{Allowed parameter space on 95\% C.L. of the SM-like $(T,B)$ doublet model in the plane of ($M_B, s_R^b$) and ($M_T, s_R^t$) after including indirect and direct constraints. In the plane ($M_B, s_R^b$), we fix the top quark sector's mixing parameter $x_t=100, 50$ GeV, and in the ($M_T, s_R^t$) plane, the bottom quark sector's mixing parameter $x_b$ is fixed to 50 GeV. The convention is the same as in Fig.~\ref{VLQB}.
}
\label{VLQB3}
\end{figure}

\begin{figure}
\centering
\includegraphics[width=0.32\textwidth]{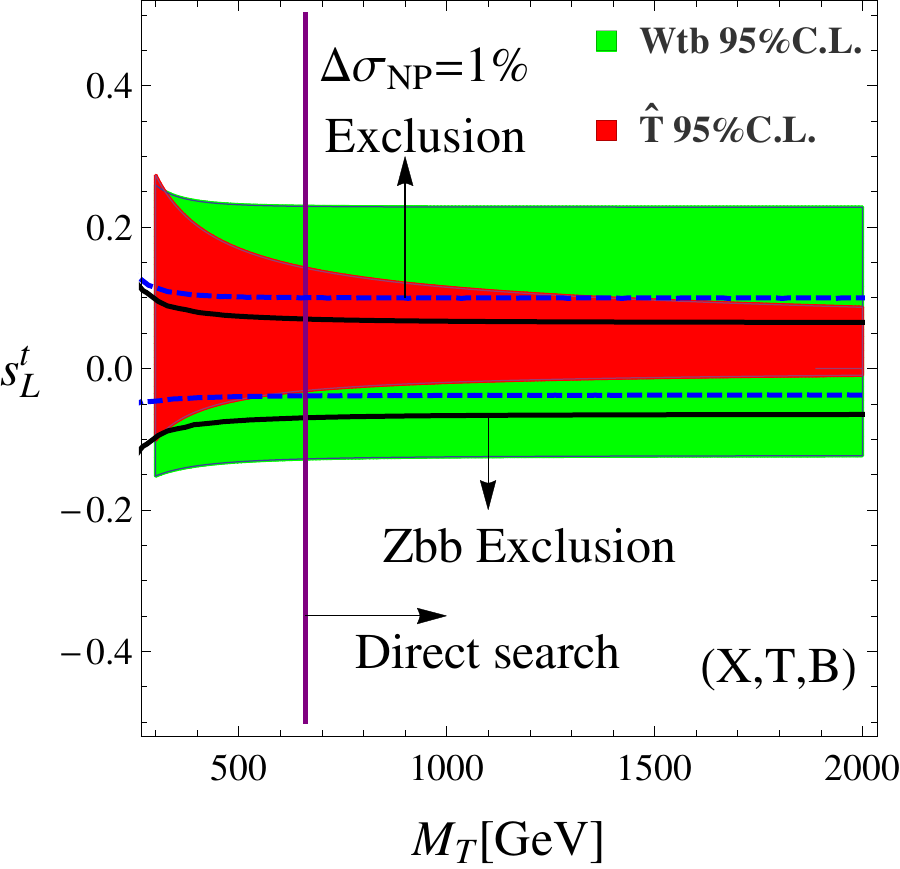}
\includegraphics[width=0.32\textwidth]{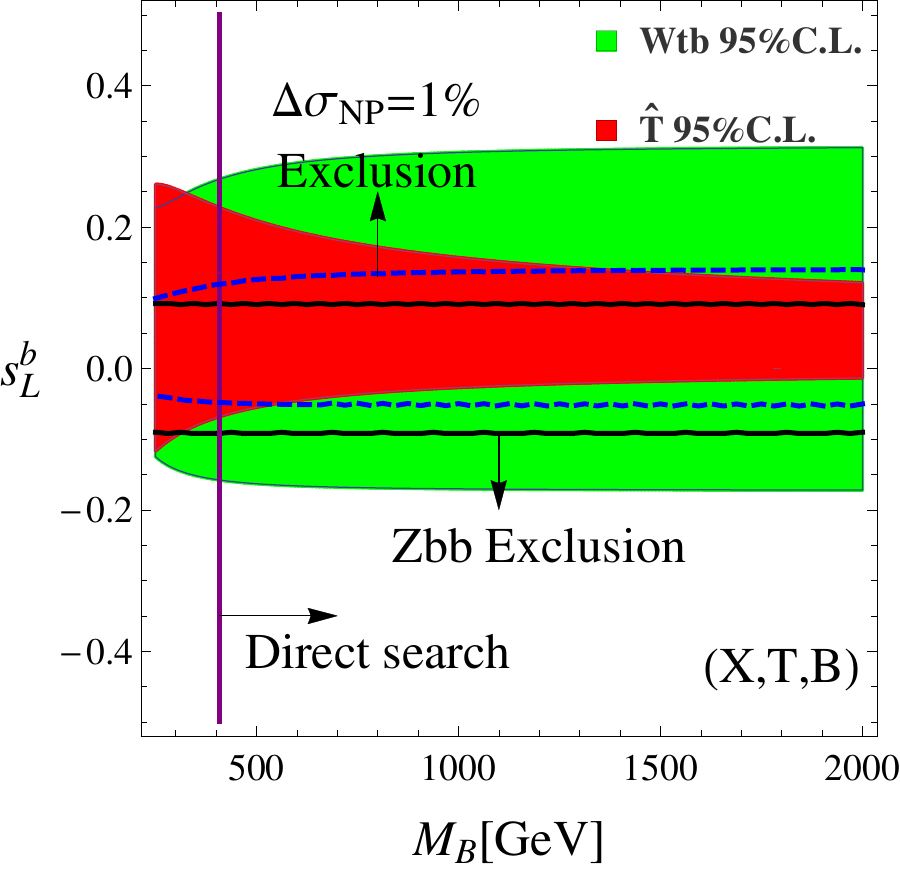}\\
\includegraphics[width=0.32\textwidth]{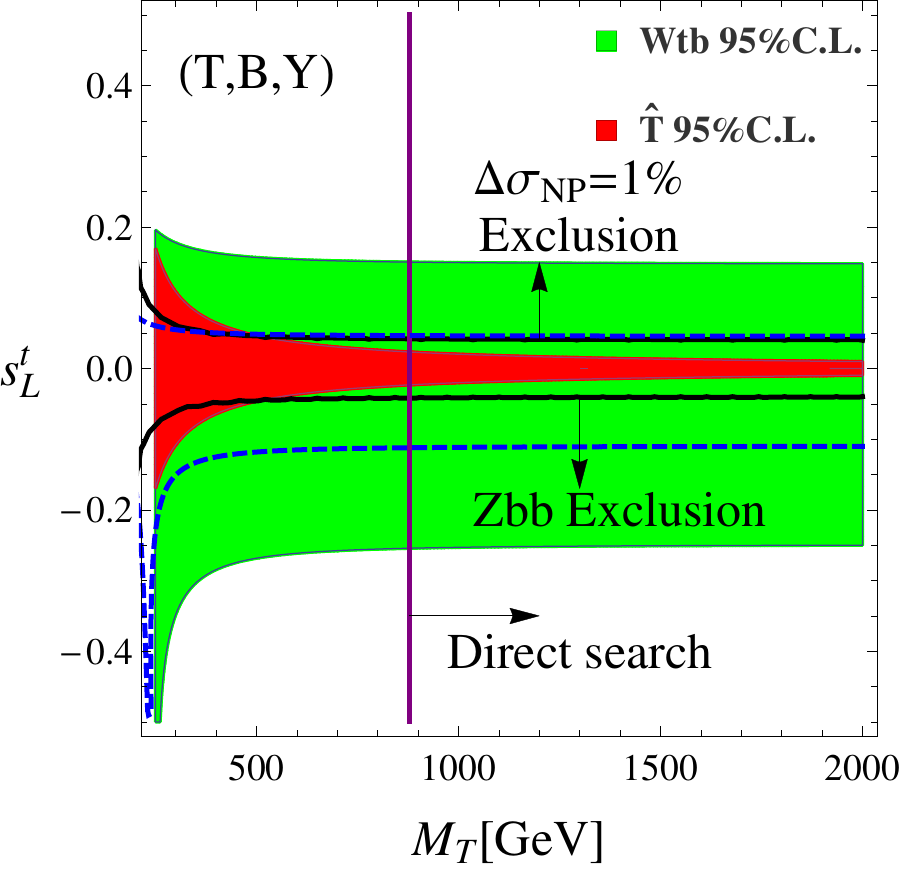}
\includegraphics[width=0.32\textwidth]{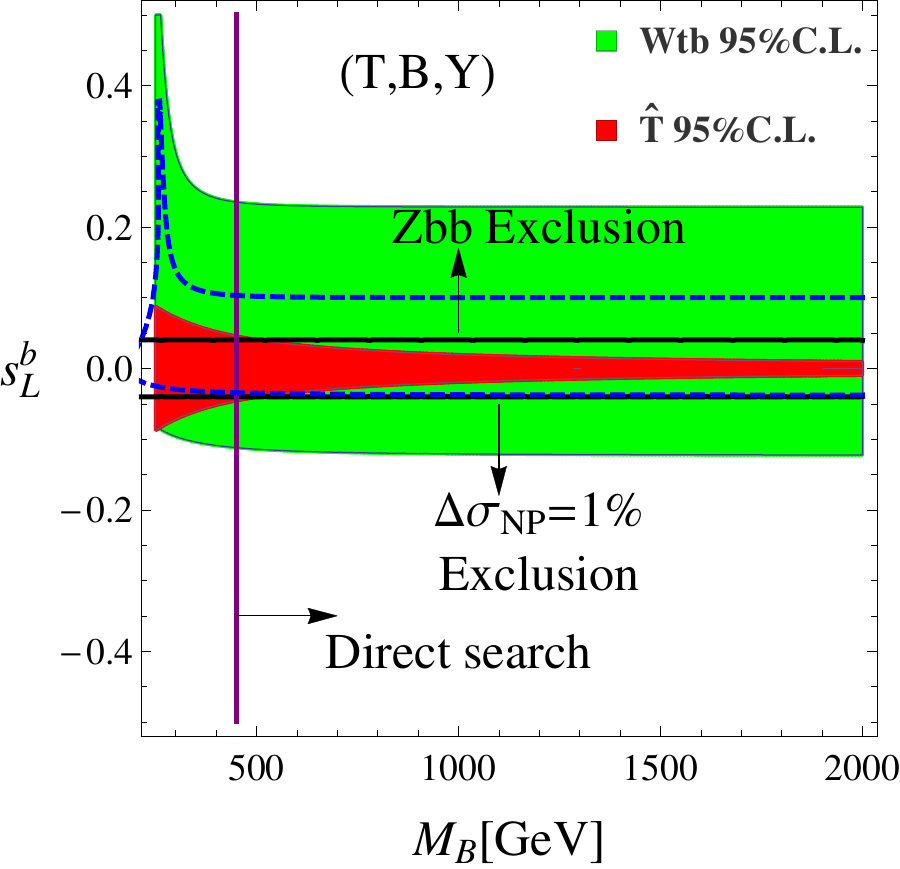}
\caption{Allowed parameter space on 95\% C.L. of the $(X,T,B)$ triplet model in the ($M_T, s_L^t$) plane and of the $(T,B,Y)$ triplet model in the ($M_B, s_L^b$) plane after including indirect and direct constraints. The convention is the same as in Fig.~\ref{VLQB}.
}
\label{VLQB4}
\end{figure}

We present the allowed regions of the VLQ models in Figs.~\ref{VLQB}-\ref{VLQB4}.
In order to better understand the impact of various bounds, we separate the constraints into different categories: the $Wtb$ coupling constraint at 95\% C.L. (green region), $\hat{T}$ parameter constraint at 95\% C.L. (red region) and $Zbb$ coupling constraint (black line). One should keep in mind that top and bottom quark partners must be heavy in order to validate our EFT prescription of the NP effects.  Currently, both the ATLAS and CMS collaborations searched the various VLQs and imposed bounds on the heavy quark's mass~\cite{Khachatryan:2015axa,ATLAS:2015VLQ}.
The current mass limit of top partners are $m_T>660$ GeV for singlet $(T)$ and triplet $(X,T,B)$, $m_T>855$ GeV for doublet $(X,T)$ and doublet $(T,B)$, and $m_T>878$ GeV for triplet $(T,B,Y)$.
Bounds on the bottom quark partner's mass are $m_B>735$ GeV for singlet $(B)$, $m_B>450$ GeV for doublets $(T,B)$, $(B,Y)$ and triplet $(T,B,Y)$, and $m_B>408$ GeV for triplet $(X,T,B)$ (see the purple lines).

We plot in Figs.~\ref{VLQB}-\ref{VLQB4} the effects of the $Wtb$ measurements on the parameter space of the VLQ models. We note the following common features when comparing different experimental constraints. There is no direct constraint on the heavy VLQ's mass from the $Wtb$ measurements as $f_1^L$ and $f_1^R$ depend only on the quark mixing angles (see Table~\ref{tab:vectorquark}). For the most parameter space of the VLQ models, the bounds from the $Wtb$ coupling measurements are weak.
We also consider the measurements of the $Zbb$ coupling if a bottom-quark partner is present. The $Zbb$ coupling is measured very precisely at the LEP II~\cite{Abdallah:2008ab} such that it leads to a much tighter bound than the $Wtb$ coupling measurements at the Tevatron and LHC.

Figure~\ref{VLQB} shows the allowed region of the singlet $T$-quark model (left) and the singlet $B$-quark model (right).  In the singlet $T$ model,
the light $T$-quark contributes largely to the $\hat{T}$-parameter, thus a large quark mixing angle is needed to respect the $\hat{T}$-parameter constraints~\cite{Xiao:2014kba}, and the bound from the $Wtb$  measurements plays an important role for a  light $T$-quark, say $M_T\lsim 300~{\rm GeV}$,
\begin{equation}\label{obT}
\dfrac{\Delta\hat{T}_T}{\hat{T}_{\rm SM}^F} = (s_t^L)^2\left[-\left(1+(c_t^L)^2\right)+(s_t^L)^2\dfrac{M_T^2}{m_t^2}+(c_t^L)^2\dfrac{2M_T^2}{M_T^2-m_t^2}\ln\dfrac{M_T^2}{m_t^2}\right],
\end{equation}
where $\hat{T}_{\rm SM}^F\equiv 3/(16\pi s_W^2c_W^2)(m_t^2/m_Z^2)$ denotes the contributions from the third generation quarks of SM.
 For a heavy $T$ quark, the $\hat{T}$-parameter constraint dominates over the $Wtb$ coupling,
Fig.~\ref{VLQB} also shows the allowed parameter space of the singlet $B$ model. We note that, different from the singlet $T$ model,  a two-fold contour of $\hat{T}$ parameter occurs. The reason is that the $\Delta\hat{T}_T>0$ in the singlet $T$ model, and there is no constraint from the lower limit of $\hat{T}$-parameter. That result is obvious in the heavy mass region $M_T\gg m_t$. In the small mass region ($M_T\sim m_t$), Eq.~\ref{obT} can be written as
\begin{equation}
\dfrac{\Delta\hat{T}_T}{\hat{T}_{\rm SM}^F} = (s_t^L)^2\left(\frac{M_T^2}{m_t^2}-1\right)\left(1+(c_t^L)^2\right),
\end{equation}
which shows the $\Delta\hat{T}_T$ is positive if $M_T >m_t$. But the $\Delta\hat{T}$ is not always positive in the singlet $B$, thus both the upper and lower limits of $\hat{T}$-parameter give a constraint on the parameter space of the singlet $B$ ~\cite{Lavoura:1992np},
\begin{equation}
\dfrac{\Delta\hat{T}_B}{\hat{T}_{\rm SM}^F} = (s_b^L)^2\left[(s_b^L)^2\dfrac{M_B^2}{m_t^2}-\dfrac{2M_B^2}{M_B^2-m_t^2}\ln\dfrac{M_B^2}{m_t^2}\right].
\end{equation}
We also note that the two-fold contour of the $\hat{T}$-parameter in the singlet $B$ model tends to  overlap each other when $M_B>5.5$ TeV due to the decoupling of the heavy VLQ.

Figure~\ref{VLQB2} shows the allowed parameter space of  the doublet $(X,T)$ model (left) and the doublet $(B,Y)$ model (right), respectively. The bound on $s_L$ is dominated by the $\hat{T}$ parameter in the heavy VLQ's mass region. It mainly stems from the fact that the top partner or bottom partner of doublet models mainly has right-handed coupling to SM particles, and the left-handed mixing angle is highly suppressed by factor $m_{(t,b)}/M$ in the heavy mass region, see Eq.~\ref{dmix}.

The $(X,T)$ quark doublet contributes to the $\hat{T}$-parameter as following,
\beq
\frac{\hat{T}_{(X,T)}}{\hat{T}_{\rm SM}^F} =\frac{2}{m_t^2}\left[ M_X^2 + M_T^2 -\frac{2M_X^2 M_T^2}{M_X^2-M_T^2}\ln\left(\frac{M_X^2}{M_T^2}\right)
+2 M_T M_X\left(\frac{M_T^2+M_X^2}{M_T^2-M_X^2}\ln\frac{M_T^2}{M_X^2}-2\right)\right].
\eeq
The quark mixing yields a mass splitting between the two vector-quarks in the same doublet which breaks the $SU(2)$ symmetry. Such a breaking effects lead to a non-zero contribution to the $\hat{T}$-parameter.

Figure~\ref{VLQB3} shows the allowed region of doublet $(T,B)$ model. We note that the $Zbb$ constraint is different between the ($s^b_R,M_B$) and ($s^t_R,M_T$) plane. The reason is that the dangerous bound of the $Zbb$ coupling is sensitive to the mixing angle of the bottom quark sector, whereas for the top quark sector, the constraint is indirect and comes from the allowed range of the mass parameter $M$. However, the mixing parameter $x_t$ is not fixed in the ($s^t_R,M_T$) plane, therefore the constraint is very weak in ($s^t_R,M_T$) plane, see Eq.~\ref{dmix}.  The constraint from $\hat{T}$ parameter is sensitive to the parameter $x_t$, and the constraint is weaker for the smaller $x_t$.  This is because $x_t$ represents the mixing between the top quark and top partner, and the smaller $x_t$ is corresponding to the decoupling limit. Note that, even in the case of $s_R^b=0$, the $\hat{T}$-parameter constraint demands $M_B$ to be larger than several hundred GeV for $x_t=100$ GeV; see the left figure in Fig.~\ref{VLQB3}. It arises from the non-decoupling effect of the top-quark partner.

The $(X,T,B)$ quark triplet contributes to the $\hat{T}$-parameter as following,
\begin{align}
\frac{\hat{T}_{(X,T,B)}}{\hat{T}_{\rm SM}^F} &=  \frac{4}{m_t^2}\left[
M_T^2 + M_B^2 -\frac{2M_T^2 M_B^2}{M_T^2-M_B^2}\ln\frac{M_T^2}{M_B^2}
+2 M_T M_B\left(\frac{M_T^2+M_B^2}{M_T^2-M_B^2}\ln\frac{M_T^2}{M_B^2}-2\right)\right.\nn\\
&\quad\quad\quad \left. + M_T^2 + M_X^2 -\frac{2M_T^2 M_X^2}{M_T^2-M_X^2}\ln\frac{M_T^2}{M_X^2}
+2 M_T M_X\left(\frac{M_T^2+M_X^2}{M_T^2-M_X^2}\ln\frac{M_T^2}{M_X^2}-2\right)\right] .
\end{align}
The allowed regions of triplet $(X,T,B)$ and $(T,B,Y)$ model are show in Fig.~\ref{VLQB4}. It displays that the allowed parameter space of $s_L^t$ is smaller than the $s_L^b$ in the heavy mass region. The reason is that the mixing parameter of bottom quark sector  ($\sqrt{2}x_t$) is larger than the top quark sector ($x_t$), see Eq.~\ref{XTB}. The similar result holds for triplet $(T,B,Y)$ model except that $s_L^b<s_L^t$ in the heavy mass region and the sign of the mixing parameter is opposite ($x_b\rightarrow-x_b,x_t\rightarrow \sqrt{2}x_b$); see Eq.~\ref{TBY}.

\subsection{Little Higgs Models}

Little Higgs models are proposed to solve the hierarchy problem~\cite{ArkaniHamed:2002qy}. The Higgs boson in the Little Higgs models is a pseudo-Goldstone boson arising from the global symmetry breaking at TeV scale and is light due to the collective symmetry breaking mechanism. The one-loop divergence to the Higgs boson mass is cancelled between top-quark and SM gauge bosons and their partners.
The Littlest Higgs model is proposed in Ref.~\cite{ArkaniHamed:2002qy} which is based on the $SU(5)/SO(5)$ nonlinear sigma model, with a locally gauged subgroup $G_1\otimes G_2=[SU(2)_1\otimes U(1)_1]\times[SU(2)_2\otimes U(1)_2]$.

The global symmetry $SU(5)$ is spontaneously broken down to the subgroup $SO(5)$ at the scale of $f$. At the same time, the gauge symmetry $G_1\otimes G_2$ is broken to the diagonal subgroup $SU(2)_L\otimes U(1)_Y$ which is identified as
the SM electroweak symmetry.

In the Littlest Higgs model, vector-like $T$-quark and new heavy gauge boson $W^{\prime}$ mix with SM particles and modify the $Wtb$ vertex~\cite{Han:2003wu},
\begin{eqnarray}\label{LHcoupling}
\mathcal{L}_{Wtb} = \frac{g}{\sqrt{2}} \left[c_L-\frac{v^2}{2f^2}  c_\phi^2 (c^2_\phi - s^2_\phi) \right]\bar{t} \gamma^\mu P_L b W_\mu^++h.c.,
\end{eqnarray}
where $c_\phi = g_1/\sqrt{g_1^2 + g_2^2}$ with $g_1$ and $g_2$ are the gauge couplings of the $SU(2)_1$ and $SU(2)_2$, respectively, and $c_L$ is the cosine of the mixing angle of the top-quark and $T$-quark.

\begin{figure}
\centering
\includegraphics[width=0.32\textwidth]{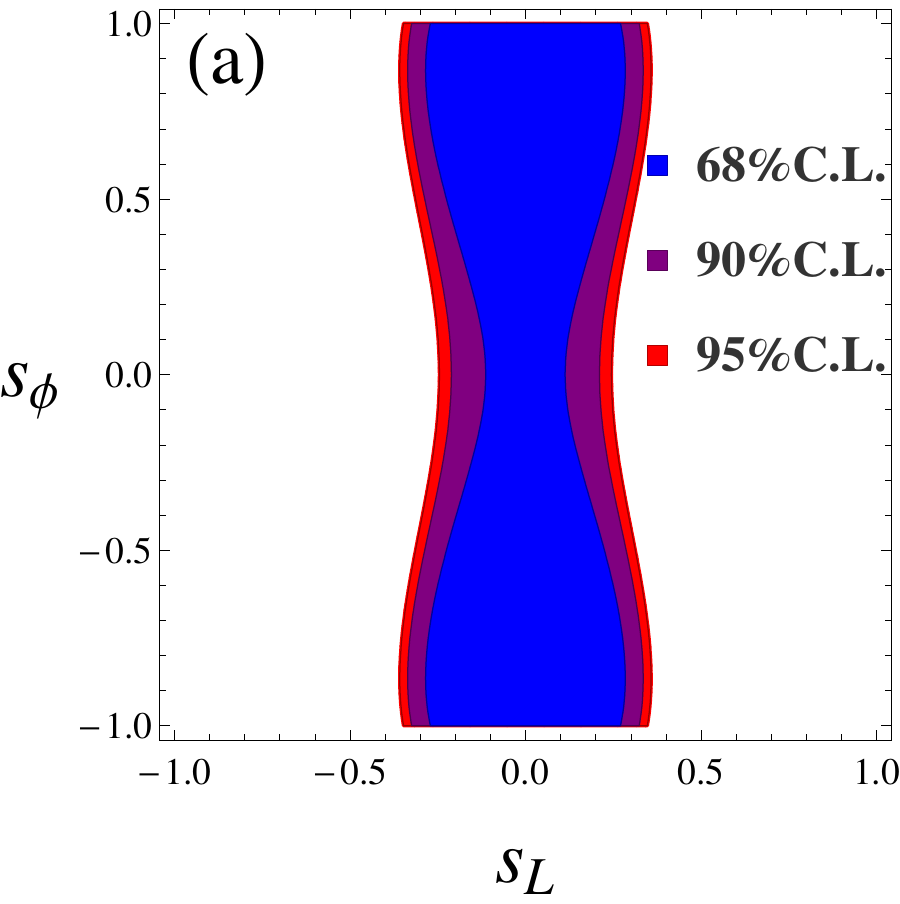}
\includegraphics[width=0.31\textwidth]{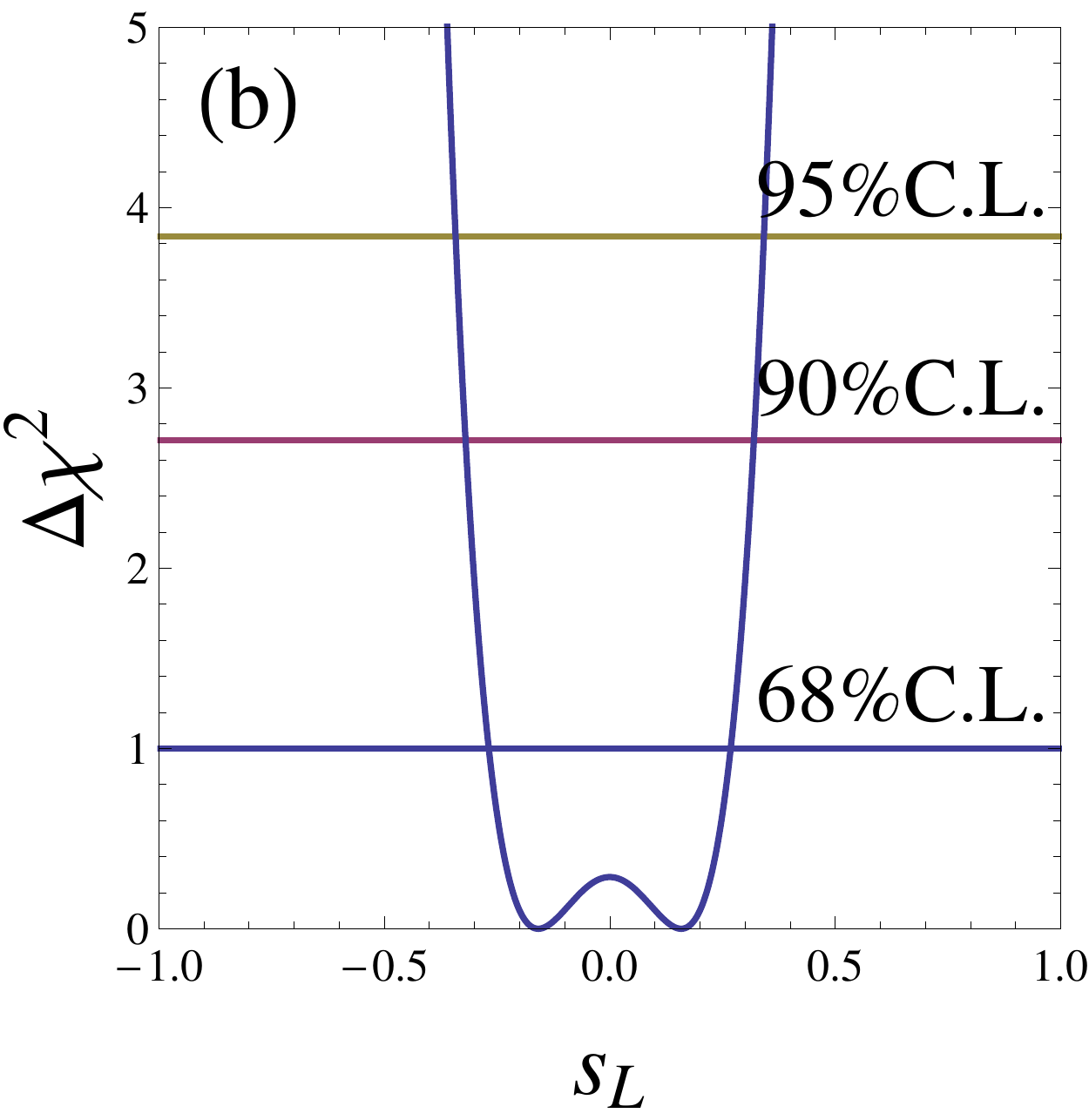}
\caption{Allowed parameter space on 68\%, 90\% and 95\% confidence levels of the Littlest Higgs model (a) and  the Littlest Higgs with T parity model (b). $s_L$ is the sine of the mixing angle of the top-quark and $T$-quark.
}
\label{LH}
\end{figure}

After considering the $Wtb$ measurements, we plot the allowed parameter space of the Littlest Higgs model in Fig.~\ref{LH}, which shows
the allowed parameter space at 68\% (blue region)\,,90\% (purple region) and 95\% (red region) C.L., respectively for $f=1$ TeV.
It shows that the constraint on the mixing angle $s_\phi$ is weak but the constraint on $s_L$ is much tighter. That can be understood from Eq.\ref{LHcoupling} that the contribution of $s_\phi$ to the $Wtb$ coupling is suppressed by the scale $f$, while $s_L$ direct modifies the $Wtb$ coupling.

The $W^{\prime}$ mixing with the $W$-boson at the tree-level modifies the oblique $\hat{T}$-parameter significantly. In order to respect the electroweak precision tests (EWPT), the scale $f$ has to be above several TeV. The current lower bound of $f$ in the Littlest Higgs model at 95\% C.L. is 5.1 TeV~\cite{Reuter:2013zja}. Refs.~\cite{Cheng:2003ju,Cheng:2004yc,Low:2004xc} introduce a new discrete symmetry to forbid the tree-level mixing between $W$ and $W^{\prime}$ and relax $f$ down to hundreds of GeV~\cite{Reuter:2013iya}.
A simple case is Littlest Higgs with T-parity (LHT) model, in which the SM particles are even and the new heavy gauge bosons and scalars are odd under the T-parity. The only T-parity even non-SM particle is the top partner $T$ and it will contribute to the $Wtb$ coupling through its mixing with top quark.
In the LHT model, the $Wtb$ coupling is~\cite{Belyaev:2006jh},
\begin{eqnarray}
\mathcal{L}_{Wtb} &=& \frac{g}{\sqrt{2}} c_L \bar{t}\gamma^\mu P_Lb W_\mu^++h.c..
\end{eqnarray}
Figure~\ref{LH}(b) shows that the effective $Wtb$ coupling measurements require $|s_L|<0.34$ at 95\% C.L..

In order to further constrain the mixing angle, we include the $Zbb$ coupling and oblique parameter constraints.
To the leading order and in the limit $M_T\gg m_t\gg m_W$, the variation of $Zb_Lb_L$ is given by~\cite{Hubisz:2005tx},
\begin{equation}\label{zbbLHT}
 \delta g_L^{b}=\frac{g^3}{32\pi^2 c_W}\frac{m_t^4}{m_W^2M_T^2}R^2\log{\frac{M_T^2}{m_t^2}},
\end{equation}
where $M_T$ denotes $T$-quark's mass, $R=\lambda_1/\lambda_2=s_L M_T/m_t$. The correction to the right-handed $Zb_Rb_R$ vertex is negligible in this case.
The electroweak precision constraints on the LHT model have been calculated in Ref.~\cite{Hubisz:2005tx}, which shows that the $\hat{T}$ parameter induced by the $T$-quark loop is much large than the $\hat{S}$ and $\hat{U}$ parameters for the same model parameters. Therefore we will only include the $\hat{T}$ parameter bound in our analysis. The contribution from the heavy T-quark is,
\begin{equation}\label{TPLHT}
\dfrac{\Delta \hat{T}_F}{\hat{T}_{\rm SM}^F}= s_L^2\left[\frac{s_L^2}{x_t}-1-c_L^2-\frac{2c_L^2}{1-x_t}\log{x_t}\right],
\end{equation}
where $x_t=m_t^2/M_T^2$.

\begin{figure}
\centering
\includegraphics[width=0.32\textwidth]{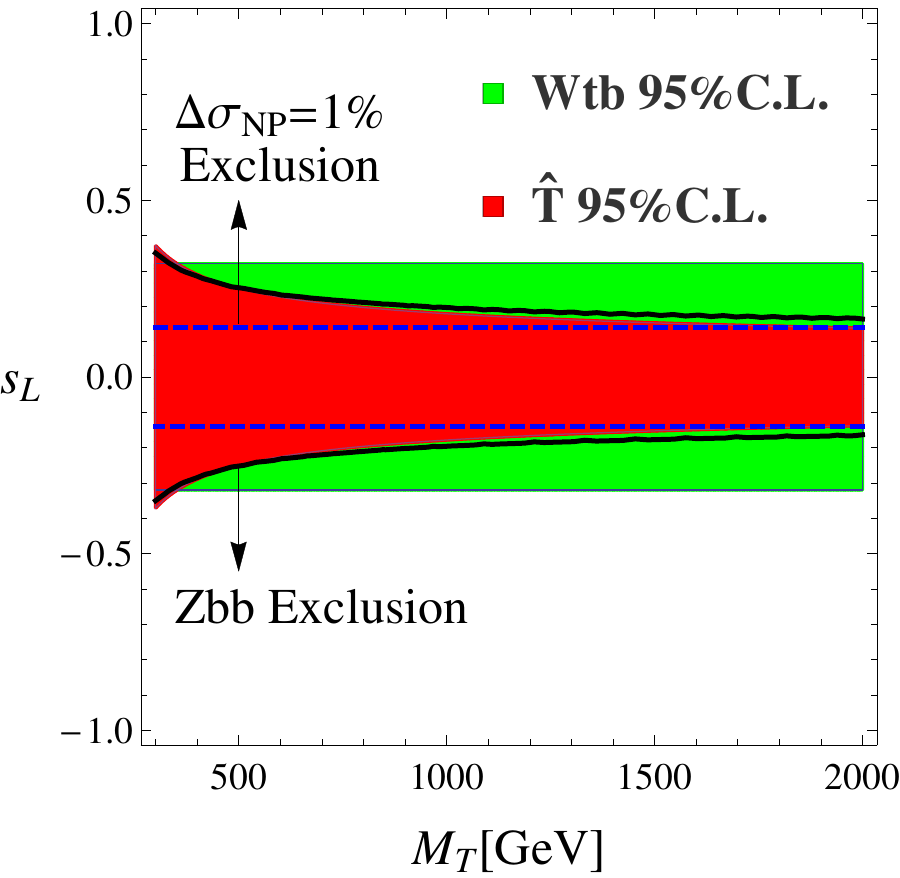}
\caption{Allowed parameter space on 95\% C.L. of the Littlest Higgs with T parity model with $f=1$ TeV after including indirect and direct
constraints. The convention is the same as in Fig.~\ref{VLQB}.}
\label{LHTB}
\end{figure}

After considering the $Wtb$, $Zbb$ and oblique $\hat{T}$-parameter experimental data, we plot in Fig.~\ref{LHTB} the allowed parameter space of LHT model at 95\% C.L. with $f=1$ TeV in the $(s_L,M_T)$ plane.

We separate the constraints into different categories: the $Wtb$ coupling constraint at 95\% C.L. (green region), $\hat{T}$-parameter constraint at 95\% C.L. (red region) and $Zbb$ coupling constraint (black line).

From Fig.~\ref{LHTB}, we note that the bounds from $\hat{T}$-parameter and $Zbb$ coupling are almost identical and are tighter than the bound from the $Wtb$ coupling.
All of the constraints are not sensitive to the heavy $T$-quark's mass.

\section{Conclusion and Discussion}
\label{sec:conclusion}

We study the top quark effective couplings using a set of higher dimensional operators made out of the SM fields. The leading contributions from NP can be captured by the dimension-six operators that are related to the top quark anomalous couplings $f_{1,2}^{L,R}$.
Using the recent data of $t$-channel single top production $\sigma_t$, $tW$ associated production $\sigma_{tW}$, $s$-channel single top production $\sigma_s$ and $W$-helicity fractions $F_0$, $F_L$ and $F_R$ collected at the 7 TeV, 8 TeV and 13 TeV LHC as well as at the Tevatron, we perform a global fit to impose constraints on anomalous couplings $f_{1,2}^{L,R}$. The current data is sensitive to the top quark effective couplings and yields strong correlations among the top quark anomalous couplings.

We introduce $x_{0}$, $x_p$, $x_m$ and $x_5$ (see Eq.~\ref{xnotation}) to study the correlations among the top quark effective couplings.
The variables $x_i$'s are sensitive to the correlations among the top quark anomalous couplings; for example, $x_m$ probes the correlation between $f_1^L$ and $f_2^R$, $x_p$ tests the correlation between $f_1^R$ and $f_2^L$,  $x_5$ is sensitive to the relation between $f_2^L$ and $f_2^R$, and $x_0$ knows about all the four anomalous couplings. Precisely measuring $x_i$'s can probe the correlations among the top anomalous couplings, which may shed light on new physics models.
We note that
\begin{itemize}
\item[(i)] Improving the measurements of $\sigma_t$ and $\sigma_{tW}$ is important for constraining $x_0$, which can be translated into the correlation of $(f_1^R,f_2^R)$ and $(f_2^L,f_2^R)$; see Fig.~\ref{Fig:scan3}(e) and Fig.~\ref{Fig:scan3}(f).
\item[(ii)] The $x_m$ is sensitive to all the four experiments. As $f_1^L$ and $f_2^R$ are anticorrelated in $x_m$, reducing the uncertainty of $x_m$ would tighten the correlation between $f_1^L$ and $f_2^R$; see Fig.~\ref{Fig:scan3}(c).
\item[(iii)] The $x_p$ is directly linked to the right-handed $W$-helicity fraction $F_R$, which is inferred from $F_L$ and $F_0$ measurements. The strong anti-correlation between $f_1^R$ and $f_2^L$ is sensitive to the $F_0$ and $F_L$ measurements; see Fig.~\ref{Fig:scan3}(d).
\item[(iv)] The $x_5$ is sensitive to the precision of $\sigma_t$, $\sigma_{tW}$ and $F_0$ measurements.
\end{itemize}
At the 13 or 14~TeV LHC, the single top production cross sections increase and further affect the dependence of $x_i$'s on the single top production and $W$-helicity fraction measurements. For example,
the coefficients of $\sigma_t$ and $\sigma_{tW}$  decrease  when the collider energy increases from 8~TeV to 13 or 14~TeV. On the contrary, the $W$-helicity fractions are measured in the top quark decay process, the coefficients of  $F_0$ and $F_L$  remain almost the same as those at the 8~TeV LHC. Therefore, at the 13 or 14~TeV machine, the $W$-helicity fraction measurements play a more important role in determining $x_i$'s. The $x_i$'s can be better measured at the 13 or 14~TeV machine. More specifically, the $F_0$ measurement is important for the precision of $x_i$'s.

After exploring the allowed parameter space of those top quark effective couplings, we discuss their impact on the following three new physics models: the $G(221)=SU(2)_1\otimes SU(2)_2\otimes U(1)_X$ , vector-like quark models and littlest Higgs models with and without $T$-pairity. These NP models modify the $W$-$t$-$b$ coupling through gauge boson mixing or quark mixing if such a mixing is not forbidden by symmetry. For example, the $W^\prime$ in the so-called $G(221)$ model can mix with the SM $W$-boson to affect the $Wtb$ coupling;  in the vector quark models  new heavy quarks mix with the SM top quark or bottom quark to shift the $Wtb$ coupling; in the Littlest Higgs model both gauge boson mixing and quark mixing are present. We translate our model-independent constraints of the top quark effective couplings into the parameter space of each new physics model.

In the $G(221)$ models we consider three typical models: the left-right model, the un-unified model and the top-flavor model. The structure of the $Wtb$ coupling highly depends on the quantum number of top and bottom quarks under the $G(221)$ group and the symmetry breaking of $G(221) \to SU(2)_L\otimes U(1)_Y$.
We show that the $Wtb$ coupling is sensitive to $\sin\phi$ which describes the ratio of gauge couplings in all $G(221)$ models except the top-flavour model when $M_{W^\prime}$ is less than several hundred GeV. In the left-right model we define $\tan\beta$ as the ratio of two vacuum expectation values. We note that the $Wtb$ coupling is not sensitive to $\sin\beta$.
The current experimental limits can be evaded if  $M_{W^\prime}$ is larger than 2~TeV.

We considered seven kinds of vector quarks in this work:  (i) weak singlet model: $(T)$ quark and $(B)$ quark;  (ii) weak doublet model: $(X,T)$, $(T, B)$ and $(B,Y)$; (iii) weak triplet models:  $(X,T,B)$ and $(T,B,Y)$ model. The structure of the $Wtb$ coupling highly depends on the weak quantum number of vector quarks. We show that the $Wtb$ coupling is sensitive to the mixing angle $s_L^t$ or $s_L^b$ in all the vector quark model except the $(T, B)$ doublet  model.  To fully constrain the allowed parameter space of vector quark models, we include the constraints of oblique $\hat{T}$-parameter and $Zbb$ couplings. We note that the $\hat{T}$-parameter and $Zbb$ couplings impose much tighter constraints on the mixing angles than effective $Wtb$ couplings obtained from $\sigma_{s/t/tW}$ and $F_{0/L/R}$.

The $Wtb$ coupling in the Littlest Higgs model is modified by both quark and gauge boson mixing. The Littlest Higgs model is severely constrained by the $W$-$W^\prime$ mixing at the tree-level. We then consider the Littlest Higgs model with $T$-parity which forbids the tree-level mixing of $W$- and $W^\prime$-bosons. The $Wtb$ coupling is modified by the mixing of top-quark and top-quark partner which gives rise to constraints similar to the vector-like quark models.

The forthcoming LHC Run-II will collect more top-quark pair and single top-quark events, which will improve the measurements of single top production and $W$-helicity fraction. We expect tighter limits will be made on the effective top-quark couplings and the dimension-six operators when new data is available. That will help us to probe new physics beyond the standard model.

\begin{acknowledgments}
The work of QHC, BY and CZ is supported in part by the National Science Foundation of China under Grand No. 11275009, 11675002 and 11635001. The research of JHY is supported in part by the National Science Foundation under Grant Numbers PHY-1315983 and PHY-1316033 and DOE Grant DE- SC0011095.
\end{acknowledgments}

\bibliographystyle{apsrev}
\bibliography{reference}

\begin{thebibliography}{120}
\expandafter\ifx\csname natexlab\endcsname\relax\def\natexlab#1{#1}\fi
\expandafter\ifx\csname bibnamefont\endcsname\relax
  \def\bibnamefont#1{#1}\fi
\expandafter\ifx\csname bibfnamefont\endcsname\relax
  \def\bibfnamefont#1{#1}\fi
\expandafter\ifx\csname citenamefont\endcsname\relax
  \def\citenamefont#1{#1}\fi
\expandafter\ifx\csname url\endcsname\relax
  \def\url#1{\texttt{#1}}\fi
\expandafter\ifx\csname urlprefix\endcsname\relax\def\urlprefix{URL }\fi
\providecommand{\bibinfo}[2]{#2}
\providecommand{\eprint}[2][]{\url{#2}}

\bibitem[{\citenamefont{Aad et~al.}(2012)}]{Aad:2012tfa}
\bibinfo{author}{\bibfnamefont{G.}~\bibnamefont{Aad}} \bibnamefont{et~al.}
  (\bibinfo{collaboration}{ATLAS Collaboration}), \bibinfo{journal}{Phys.Lett.}
  \textbf{\bibinfo{volume}{B716}}, \bibinfo{pages}{1} (\bibinfo{year}{2012}),
  \eprint{1207.7214}.

\bibitem[{\citenamefont{Chatrchyan et~al.}(2012)}]{Chatrchyan:2012ufa}
\bibinfo{author}{\bibfnamefont{S.}~\bibnamefont{Chatrchyan}}
  \bibnamefont{et~al.} (\bibinfo{collaboration}{CMS Collaboration}),
  \bibinfo{journal}{Phys.Lett.} \textbf{\bibinfo{volume}{B716}},
  \bibinfo{pages}{30} (\bibinfo{year}{2012}), \eprint{1207.7235}.

\bibitem[{\citenamefont{Abe et~al.}(1995)}]{Abe:1995hr}
\bibinfo{author}{\bibfnamefont{F.}~\bibnamefont{Abe}} \bibnamefont{et~al.}
  (\bibinfo{collaboration}{CDF Collaboration}),
  \bibinfo{journal}{Phys.Rev.Lett.} \textbf{\bibinfo{volume}{74}},
  \bibinfo{pages}{2626} (\bibinfo{year}{1995}), \eprint{hep-ex/9503002}.

\bibitem[{\citenamefont{Abachi et~al.}(1995)}]{Abachi:1995iq}
\bibinfo{author}{\bibfnamefont{S.}~\bibnamefont{Abachi}} \bibnamefont{et~al.}
  (\bibinfo{collaboration}{D0 Collaboration}),
  \bibinfo{journal}{Phys.Rev.Lett.} \textbf{\bibinfo{volume}{74}},
  \bibinfo{pages}{2632} (\bibinfo{year}{1995}), \eprint{hep-ex/9503003}.

\bibitem[{\citenamefont{Hsieh et~al.}(2010)\citenamefont{Hsieh, Schmitz, Yu,
  and Yuan}}]{Hsieh:2010zr}
\bibinfo{author}{\bibfnamefont{K.}~\bibnamefont{Hsieh}},
  \bibinfo{author}{\bibfnamefont{K.}~\bibnamefont{Schmitz}},
  \bibinfo{author}{\bibfnamefont{J.-H.} \bibnamefont{Yu}}, \bibnamefont{and}
  \bibinfo{author}{\bibfnamefont{C.-P.} \bibnamefont{Yuan}},
  \bibinfo{journal}{Phys.Rev.} \textbf{\bibinfo{volume}{D82}},
  \bibinfo{pages}{035011} (\bibinfo{year}{2010}), \eprint{1003.3482}.

\bibitem[{\citenamefont{Cao et~al.}(2012)\citenamefont{Cao, Li, Yu, and
  Yuan}}]{Cao:2012ng}
\bibinfo{author}{\bibfnamefont{Q.-H.} \bibnamefont{Cao}},
  \bibinfo{author}{\bibfnamefont{Z.}~\bibnamefont{Li}},
  \bibinfo{author}{\bibfnamefont{J.-H.} \bibnamefont{Yu}}, \bibnamefont{and}
  \bibinfo{author}{\bibfnamefont{C.}~\bibnamefont{Yuan}},
  \bibinfo{journal}{Phys.Rev.} \textbf{\bibinfo{volume}{D86}},
  \bibinfo{pages}{095010} (\bibinfo{year}{2012}), \eprint{1205.3769}.

\bibitem[{\citenamefont{del Aguila et~al.}(2000)\citenamefont{del Aguila,
  Perez-Victoria, and Santiago}}]{delAguila:2000rc}
\bibinfo{author}{\bibfnamefont{F.}~\bibnamefont{del Aguila}},
  \bibinfo{author}{\bibfnamefont{M.}~\bibnamefont{Perez-Victoria}},
  \bibnamefont{and} \bibinfo{author}{\bibfnamefont{J.}~\bibnamefont{Santiago}},
  \bibinfo{journal}{JHEP} \textbf{\bibinfo{volume}{09}}, \bibinfo{pages}{011}
  (\bibinfo{year}{2000}), \eprint{hep-ph/0007316}.

\bibitem[{\citenamefont{Aguilar-Saavedra}(2009{\natexlab{a}})}]{AguilarSaavedra:2009es}
\bibinfo{author}{\bibfnamefont{J.~A.} \bibnamefont{Aguilar-Saavedra}},
  \bibinfo{journal}{JHEP} \textbf{\bibinfo{volume}{11}}, \bibinfo{pages}{030}
  (\bibinfo{year}{2009}{\natexlab{a}}), \eprint{0907.3155}.

\bibitem[{\citenamefont{Cacciapaglia et~al.}(2010)\citenamefont{Cacciapaglia,
  Deandrea, Harada, and Okada}}]{Cacciapaglia:2010vn}
\bibinfo{author}{\bibfnamefont{G.}~\bibnamefont{Cacciapaglia}},
  \bibinfo{author}{\bibfnamefont{A.}~\bibnamefont{Deandrea}},
  \bibinfo{author}{\bibfnamefont{D.}~\bibnamefont{Harada}}, \bibnamefont{and}
  \bibinfo{author}{\bibfnamefont{Y.}~\bibnamefont{Okada}},
  \bibinfo{journal}{JHEP} \textbf{\bibinfo{volume}{1011}}, \bibinfo{pages}{159}
  (\bibinfo{year}{2010}), \eprint{1007.2933}.

\bibitem[{\citenamefont{Aguilar-Saavedra
  et~al.}(2013)\citenamefont{Aguilar-Saavedra, Benbrik, Heinemeyer, and
  P��rez-Victoria}}]{Aguilar-Saavedra:2013qpa}
\bibinfo{author}{\bibfnamefont{J.}~\bibnamefont{Aguilar-Saavedra}},
  \bibinfo{author}{\bibfnamefont{R.}~\bibnamefont{Benbrik}},
  \bibinfo{author}{\bibfnamefont{S.}~\bibnamefont{Heinemeyer}},
  \bibnamefont{and}
  \bibinfo{author}{\bibfnamefont{M.}~\bibnamefont{P��rez-Victoria}},
  \bibinfo{journal}{Phys.Rev.} \textbf{\bibinfo{volume}{D88}},
  \bibinfo{pages}{094010} (\bibinfo{year}{2013}), \eprint{1306.0572}.

\bibitem[{\citenamefont{Belyaev et~al.}(2006)\citenamefont{Belyaev, Chen, Tobe,
  and Yuan}}]{Belyaev:2006jh}
\bibinfo{author}{\bibfnamefont{A.}~\bibnamefont{Belyaev}},
  \bibinfo{author}{\bibfnamefont{C.-R.} \bibnamefont{Chen}},
  \bibinfo{author}{\bibfnamefont{K.}~\bibnamefont{Tobe}}, \bibnamefont{and}
  \bibinfo{author}{\bibfnamefont{C.-P.} \bibnamefont{Yuan}},
  \bibinfo{journal}{Phys.Rev.} \textbf{\bibinfo{volume}{D74}},
  \bibinfo{pages}{115020} (\bibinfo{year}{2006}), \eprint{hep-ph/0609179}.

\bibitem[{\citenamefont{Han et~al.}(2003)\citenamefont{Han, Logan, McElrath,
  and Wang}}]{Han:2003wu}
\bibinfo{author}{\bibfnamefont{T.}~\bibnamefont{Han}},
  \bibinfo{author}{\bibfnamefont{H.~E.} \bibnamefont{Logan}},
  \bibinfo{author}{\bibfnamefont{B.}~\bibnamefont{McElrath}}, \bibnamefont{and}
  \bibinfo{author}{\bibfnamefont{L.-T.} \bibnamefont{Wang}},
  \bibinfo{journal}{Phys.Rev.} \textbf{\bibinfo{volume}{D67}},
  \bibinfo{pages}{095004} (\bibinfo{year}{2003}), \eprint{hep-ph/0301040}.

\bibitem[{\citenamefont{Penunuri and Larios}(2009)}]{Penunuri:2008pb}
\bibinfo{author}{\bibfnamefont{F.}~\bibnamefont{Penunuri}} \bibnamefont{and}
  \bibinfo{author}{\bibfnamefont{F.}~\bibnamefont{Larios}},
  \bibinfo{journal}{Phys.Rev.} \textbf{\bibinfo{volume}{D79}},
  \bibinfo{pages}{015013} (\bibinfo{year}{2009}), \eprint{0810.4545}.

\bibitem[{\citenamefont{Contino et~al.}(2007)\citenamefont{Contino, Kramer,
  Son, and Sundrum}}]{Contino:2006nn}
\bibinfo{author}{\bibfnamefont{R.}~\bibnamefont{Contino}},
  \bibinfo{author}{\bibfnamefont{T.}~\bibnamefont{Kramer}},
  \bibinfo{author}{\bibfnamefont{M.}~\bibnamefont{Son}}, \bibnamefont{and}
  \bibinfo{author}{\bibfnamefont{R.}~\bibnamefont{Sundrum}},
  \bibinfo{journal}{JHEP} \textbf{\bibinfo{volume}{0705}}, \bibinfo{pages}{074}
  (\bibinfo{year}{2007}), \eprint{hep-ph/0612180}.

\bibitem[{\citenamefont{Dabelstein et~al.}(1995)\citenamefont{Dabelstein,
  Hollik, Junger, Jimenez, and Sola}}]{Dabelstein:1995jt}
\bibinfo{author}{\bibfnamefont{A.}~\bibnamefont{Dabelstein}},
  \bibinfo{author}{\bibfnamefont{W.}~\bibnamefont{Hollik}},
  \bibinfo{author}{\bibfnamefont{C.}~\bibnamefont{Junger}},
  \bibinfo{author}{\bibfnamefont{R.~A.} \bibnamefont{Jimenez}},
  \bibnamefont{and} \bibinfo{author}{\bibfnamefont{J.}~\bibnamefont{Sola}},
  \bibinfo{journal}{Nucl.Phys.} \textbf{\bibinfo{volume}{B454}},
  \bibinfo{pages}{75} (\bibinfo{year}{1995}), \eprint{hep-ph/9503398}.

\bibitem[{\citenamefont{Cao et~al.}(2003)\citenamefont{Cao, Oakes, Wang, and
  Yang}}]{Cao:2003yk}
\bibinfo{author}{\bibfnamefont{J.-j.} \bibnamefont{Cao}},
  \bibinfo{author}{\bibfnamefont{R.~J.} \bibnamefont{Oakes}},
  \bibinfo{author}{\bibfnamefont{F.}~\bibnamefont{Wang}}, \bibnamefont{and}
  \bibinfo{author}{\bibfnamefont{J.~M.} \bibnamefont{Yang}},
  \bibinfo{journal}{Phys.Rev.} \textbf{\bibinfo{volume}{D68}},
  \bibinfo{pages}{054019} (\bibinfo{year}{2003}), \eprint{hep-ph/0306278}.

\bibitem[{\citenamefont{Grzadkowski and Hollik}(1992)}]{Grzadkowski:1991nj}
\bibinfo{author}{\bibfnamefont{B.}~\bibnamefont{Grzadkowski}} \bibnamefont{and}
  \bibinfo{author}{\bibfnamefont{W.}~\bibnamefont{Hollik}},
  \bibinfo{journal}{Nucl.Phys.} \textbf{\bibinfo{volume}{B384}},
  \bibinfo{pages}{101} (\bibinfo{year}{1992}).

\bibitem[{\citenamefont{Czarnecki et~al.}(2010)\citenamefont{Czarnecki, Korner,
  and Piclum}}]{Czarnecki:2010gb}
\bibinfo{author}{\bibfnamefont{A.}~\bibnamefont{Czarnecki}},
  \bibinfo{author}{\bibfnamefont{J.~G.} \bibnamefont{Korner}},
  \bibnamefont{and} \bibinfo{author}{\bibfnamefont{J.~H.}
  \bibnamefont{Piclum}}, \bibinfo{journal}{Phys.Rev.}
  \textbf{\bibinfo{volume}{D81}}, \bibinfo{pages}{111503}
  (\bibinfo{year}{2010}), \eprint{1005.2625}.

\bibitem[{CMS Collaboration({\natexlab{a}})}]{CMS:2013pfa}
CMS Collaboration (\bibinfo{year}{2013}{\natexlab{a}}),
  \eprint{CMS-PAS-TOP-13-008}.

\bibitem[{ATLAS Collaboration()}]{ATLAS:2013tla}
ATLAS Collaboration (\bibinfo{year}{2013}), \eprint{ATLAS-CONF-2013-033,
  ATLAS-COM-CONF-2013-004}.

\bibitem[{\citenamefont{Kidonakis}(2011)}]{Kidonakis:2011wy}
\bibinfo{author}{\bibfnamefont{N.}~\bibnamefont{Kidonakis}},
  \bibinfo{journal}{Phys.Rev.} \textbf{\bibinfo{volume}{D83}},
  \bibinfo{pages}{091503} (\bibinfo{year}{2011}), \eprint{1103.2792}.

\bibitem[{\citenamefont{Kidonakis}(2012)}]{Kidonakis:2012db}
\bibinfo{author}{\bibfnamefont{N.}~\bibnamefont{Kidonakis}}, pp.
  \bibinfo{pages}{831--834} (\bibinfo{year}{2012}), \eprint{1205.3453}.

\bibitem[{\citenamefont{Kidonakis}(2010{\natexlab{a}})}]{Kidonakis:2010ux}
\bibinfo{author}{\bibfnamefont{N.}~\bibnamefont{Kidonakis}},
  \bibinfo{journal}{Phys.Rev.} \textbf{\bibinfo{volume}{D82}},
  \bibinfo{pages}{054018} (\bibinfo{year}{2010}{\natexlab{a}}),
  \eprint{1005.4451}.

\bibitem[{\citenamefont{Kidonakis}(2010{\natexlab{b}})}]{Kidonakis:2010tc}
\bibinfo{author}{\bibfnamefont{N.}~\bibnamefont{Kidonakis}},
  \bibinfo{journal}{Phys.Rev.} \textbf{\bibinfo{volume}{D81}},
  \bibinfo{pages}{054028} (\bibinfo{year}{2010}{\natexlab{b}}),
  \eprint{1001.5034}.

\bibitem[{\citenamefont{Kidonakis}(2016)}]{Kidonakis:2016smr}
\bibinfo{author}{\bibfnamefont{N.}~\bibnamefont{Kidonakis}}
  (\bibinfo{year}{2016}), \eprint{1609.07404},
  \urlprefix\url{https://inspirehep.net/record/1487920/files/arXiv:1609.07404.pdf}.

\bibitem[{\citenamefont{Kane et~al.}(1992)\citenamefont{Kane, Ladinsky, and
  Yuan}}]{Kane:1991bg}
\bibinfo{author}{\bibfnamefont{G.~L.} \bibnamefont{Kane}},
  \bibinfo{author}{\bibfnamefont{G.}~\bibnamefont{Ladinsky}}, \bibnamefont{and}
  \bibinfo{author}{\bibfnamefont{C.}~\bibnamefont{Yuan}},
  \bibinfo{journal}{Phys.Rev.} \textbf{\bibinfo{volume}{D45}},
  \bibinfo{pages}{124} (\bibinfo{year}{1992}).

\bibitem[{\citenamefont{Malkawi and Yuan}(1994)}]{Malkawi:1994tg}
\bibinfo{author}{\bibfnamefont{E.}~\bibnamefont{Malkawi}} \bibnamefont{and}
  \bibinfo{author}{\bibfnamefont{C.}~\bibnamefont{Yuan}},
  \bibinfo{journal}{Phys.Rev.} \textbf{\bibinfo{volume}{D50}},
  \bibinfo{pages}{4462} (\bibinfo{year}{1994}), \eprint{hep-ph/9405322}.

\bibitem[{\citenamefont{Carlson et~al.}(1994)\citenamefont{Carlson, Malkawi,
  and Yuan}}]{Carlson:1994bg}
\bibinfo{author}{\bibfnamefont{D.~O.} \bibnamefont{Carlson}},
  \bibinfo{author}{\bibfnamefont{E.}~\bibnamefont{Malkawi}}, \bibnamefont{and}
  \bibinfo{author}{\bibfnamefont{C.}~\bibnamefont{Yuan}},
  \bibinfo{journal}{Phys.Lett.} \textbf{\bibinfo{volume}{B337}},
  \bibinfo{pages}{145} (\bibinfo{year}{1994}), \eprint{hep-ph/9405277}.

\bibitem[{\citenamefont{Whisnant et~al.}(1997)\citenamefont{Whisnant, Yang,
  Young, and Zhang}}]{Whisnant:1997qu}
\bibinfo{author}{\bibfnamefont{K.}~\bibnamefont{Whisnant}},
  \bibinfo{author}{\bibfnamefont{J.-M.} \bibnamefont{Yang}},
  \bibinfo{author}{\bibfnamefont{B.-L.} \bibnamefont{Young}}, \bibnamefont{and}
  \bibinfo{author}{\bibfnamefont{X.}~\bibnamefont{Zhang}},
  \bibinfo{journal}{Phys.Rev.} \textbf{\bibinfo{volume}{D56}},
  \bibinfo{pages}{467} (\bibinfo{year}{1997}), \eprint{hep-ph/9702305}.

\bibitem[{\citenamefont{Yang and Young}(1997)}]{Yang:1997iv}
\bibinfo{author}{\bibfnamefont{J.~M.} \bibnamefont{Yang}} \bibnamefont{and}
  \bibinfo{author}{\bibfnamefont{B.-L.} \bibnamefont{Young}},
  \bibinfo{journal}{Phys.Rev.} \textbf{\bibinfo{volume}{D56}},
  \bibinfo{pages}{5907} (\bibinfo{year}{1997}), \eprint{hep-ph/9703463}.

\bibitem[{\citenamefont{Cao et~al.}(1998)\citenamefont{Cao, Wang, Yang, Young,
  and Zhang}}]{Cao:1998at}
\bibinfo{author}{\bibfnamefont{J.-J.} \bibnamefont{Cao}},
  \bibinfo{author}{\bibfnamefont{J.-X.} \bibnamefont{Wang}},
  \bibinfo{author}{\bibfnamefont{J.~M.} \bibnamefont{Yang}},
  \bibinfo{author}{\bibfnamefont{B.-L.} \bibnamefont{Young}}, \bibnamefont{and}
  \bibinfo{author}{\bibfnamefont{X.-m.} \bibnamefont{Zhang}},
  \bibinfo{journal}{Phys.Rev.} \textbf{\bibinfo{volume}{D58}},
  \bibinfo{pages}{094004} (\bibinfo{year}{1998}), \eprint{hep-ph/9804343}.

\bibitem[{\citenamefont{Hikasa et~al.}(1998)\citenamefont{Hikasa, Whisnant,
  Yang, and Young}}]{Hikasa:1998wx}
\bibinfo{author}{\bibfnamefont{K.-i.} \bibnamefont{Hikasa}},
  \bibinfo{author}{\bibfnamefont{K.}~\bibnamefont{Whisnant}},
  \bibinfo{author}{\bibfnamefont{J.~M.} \bibnamefont{Yang}}, \bibnamefont{and}
  \bibinfo{author}{\bibfnamefont{B.-L.} \bibnamefont{Young}},
  \bibinfo{journal}{Phys.Rev.} \textbf{\bibinfo{volume}{D58}},
  \bibinfo{pages}{114003} (\bibinfo{year}{1998}), \eprint{hep-ph/9806401}.

\bibitem[{\citenamefont{Larios et~al.}(1999)\citenamefont{Larios, Perez, and
  Yuan}}]{Larios:1999au}
\bibinfo{author}{\bibfnamefont{F.}~\bibnamefont{Larios}},
  \bibinfo{author}{\bibfnamefont{M.}~\bibnamefont{Perez}}, \bibnamefont{and}
  \bibinfo{author}{\bibfnamefont{C.}~\bibnamefont{Yuan}},
  \bibinfo{journal}{Phys.Lett.} \textbf{\bibinfo{volume}{B457}},
  \bibinfo{pages}{334} (\bibinfo{year}{1999}), \eprint{hep-ph/9903394}.

\bibitem[{\citenamefont{Lin et~al.}(2002)\citenamefont{Lin, Han, Huang, Wang,
  and Zhang}}]{Lin:2001yq}
\bibinfo{author}{\bibfnamefont{Z.}~\bibnamefont{Lin}},
  \bibinfo{author}{\bibfnamefont{T.}~\bibnamefont{Han}},
  \bibinfo{author}{\bibfnamefont{T.}~\bibnamefont{Huang}},
  \bibinfo{author}{\bibfnamefont{J.}~\bibnamefont{Wang}}, \bibnamefont{and}
  \bibinfo{author}{\bibfnamefont{X.}~\bibnamefont{Zhang}},
  \bibinfo{journal}{Phys.Rev.} \textbf{\bibinfo{volume}{D65}},
  \bibinfo{pages}{014008} (\bibinfo{year}{2002}), \eprint{hep-ph/0106344}.

\bibitem[{\citenamefont{Espriu and Manzano}(2002)}]{Espriu:2001vj}
\bibinfo{author}{\bibfnamefont{D.}~\bibnamefont{Espriu}} \bibnamefont{and}
  \bibinfo{author}{\bibfnamefont{J.}~\bibnamefont{Manzano}},
  \bibinfo{journal}{Phys.Rev.} \textbf{\bibinfo{volume}{D65}},
  \bibinfo{pages}{073005} (\bibinfo{year}{2002}), \eprint{hep-ph/0107112}.

\bibitem[{\citenamefont{Chen et~al.}(2005)\citenamefont{Chen, Larios, and
  Yuan}}]{Chen:2005vr}
\bibinfo{author}{\bibfnamefont{C.-R.} \bibnamefont{Chen}},
  \bibinfo{author}{\bibfnamefont{F.}~\bibnamefont{Larios}}, \bibnamefont{and}
  \bibinfo{author}{\bibfnamefont{C.-P.} \bibnamefont{Yuan}},
  \bibinfo{journal}{Phys.Lett.} \textbf{\bibinfo{volume}{B631}},
  \bibinfo{pages}{126} (\bibinfo{year}{2005}), \eprint{hep-ph/0503040}.

\bibitem[{\citenamefont{Batra and Tait}(2006)}]{Batra:2006iq}
\bibinfo{author}{\bibfnamefont{P.}~\bibnamefont{Batra}} \bibnamefont{and}
  \bibinfo{author}{\bibfnamefont{T.~M.} \bibnamefont{Tait}},
  \bibinfo{journal}{Phys.Rev.} \textbf{\bibinfo{volume}{D74}},
  \bibinfo{pages}{054021} (\bibinfo{year}{2006}), \eprint{hep-ph/0606068}.

\bibitem[{\citenamefont{Cao et~al.}(2007)\citenamefont{Cao, Wudka, and
  Yuan}}]{Cao:2007ea}
\bibinfo{author}{\bibfnamefont{Q.-H.} \bibnamefont{Cao}},
  \bibinfo{author}{\bibfnamefont{J.}~\bibnamefont{Wudka}}, \bibnamefont{and}
  \bibinfo{author}{\bibfnamefont{C.-P.} \bibnamefont{Yuan}},
  \bibinfo{journal}{Phys.Lett.} \textbf{\bibinfo{volume}{B658}},
  \bibinfo{pages}{50} (\bibinfo{year}{2007}), \eprint{0704.2809}.

\bibitem[{\citenamefont{Aguilar-Saavedra}(2008)}]{AguilarSaavedra:2008gt}
\bibinfo{author}{\bibfnamefont{J.~A.} \bibnamefont{Aguilar-Saavedra}},
  \bibinfo{journal}{Nucl. Phys.} \textbf{\bibinfo{volume}{B804}},
  \bibinfo{pages}{160} (\bibinfo{year}{2008}), \eprint{0803.3810}.

\bibitem[{\citenamefont{Berger et~al.}(2009)\citenamefont{Berger, Cao, and
  Low}}]{Berger:2009hi}
\bibinfo{author}{\bibfnamefont{E.~L.} \bibnamefont{Berger}},
  \bibinfo{author}{\bibfnamefont{Q.-H.} \bibnamefont{Cao}}, \bibnamefont{and}
  \bibinfo{author}{\bibfnamefont{I.}~\bibnamefont{Low}},
  \bibinfo{journal}{Phys.Rev.} \textbf{\bibinfo{volume}{D80}},
  \bibinfo{pages}{074020} (\bibinfo{year}{2009}), \eprint{0907.2191}.

\bibitem[{\citenamefont{Zhang and Willenbrock}(2011)}]{Zhang:2010dr}
\bibinfo{author}{\bibfnamefont{C.}~\bibnamefont{Zhang}} \bibnamefont{and}
  \bibinfo{author}{\bibfnamefont{S.}~\bibnamefont{Willenbrock}},
  \bibinfo{journal}{Phys.Rev.} \textbf{\bibinfo{volume}{D83}},
  \bibinfo{pages}{034006} (\bibinfo{year}{2011}), \eprint{1008.3869}.

\bibitem[{\citenamefont{Aguilar-Saavedra and
  Bernabeu}(2010)}]{AguilarSaavedra:2010nx}
\bibinfo{author}{\bibfnamefont{J.~A.} \bibnamefont{Aguilar-Saavedra}}
  \bibnamefont{and} \bibinfo{author}{\bibfnamefont{J.}~\bibnamefont{Bernabeu}},
  \bibinfo{journal}{Nucl. Phys.} \textbf{\bibinfo{volume}{B840}},
  \bibinfo{pages}{349} (\bibinfo{year}{2010}), \eprint{1005.5382}.

\bibitem[{\citenamefont{Rindani and Sharma}(2011)}]{Rindani:2011pk}
\bibinfo{author}{\bibfnamefont{S.~D.} \bibnamefont{Rindani}} \bibnamefont{and}
  \bibinfo{author}{\bibfnamefont{P.}~\bibnamefont{Sharma}},
  \bibinfo{journal}{JHEP} \textbf{\bibinfo{volume}{1111}}, \bibinfo{pages}{082}
  (\bibinfo{year}{2011}), \eprint{1107.2597}.

\bibitem[{\citenamefont{Rindani and Sharma}(2012)}]{Rindani:2011gt}
\bibinfo{author}{\bibfnamefont{S.~D.} \bibnamefont{Rindani}} \bibnamefont{and}
  \bibinfo{author}{\bibfnamefont{P.}~\bibnamefont{Sharma}},
  \bibinfo{journal}{Phys.Lett.} \textbf{\bibinfo{volume}{B712}},
  \bibinfo{pages}{413} (\bibinfo{year}{2012}), \eprint{1108.4165}.

\bibitem[{\citenamefont{Bach and Ohl}(2012)}]{Bach:2012fb}
\bibinfo{author}{\bibfnamefont{F.}~\bibnamefont{Bach}} \bibnamefont{and}
  \bibinfo{author}{\bibfnamefont{T.}~\bibnamefont{Ohl}},
  \bibinfo{journal}{Phys.Rev.} \textbf{\bibinfo{volume}{D86}},
  \bibinfo{pages}{114026} (\bibinfo{year}{2012}), \eprint{1209.4564}.

\bibitem[{\citenamefont{Fabbrichesi et~al.}(2014)\citenamefont{Fabbrichesi,
  Pinamonti, and Tonero}}]{Fabbrichesi:2014wva}
\bibinfo{author}{\bibfnamefont{M.}~\bibnamefont{Fabbrichesi}},
  \bibinfo{author}{\bibfnamefont{M.}~\bibnamefont{Pinamonti}},
  \bibnamefont{and} \bibinfo{author}{\bibfnamefont{A.}~\bibnamefont{Tonero}}
  (\bibinfo{year}{2014}), \eprint{1406.5393}.

\bibitem[{\citenamefont{Bernardo et~al.}(2014)\citenamefont{Bernardo, Castro,
  Fiolhais, Gon?alves, Guerra et~al.}}]{Bernardo:2014vha}
\bibinfo{author}{\bibfnamefont{C.}~\bibnamefont{Bernardo}},
  \bibinfo{author}{\bibfnamefont{N.}~\bibnamefont{Castro}},
  \bibinfo{author}{\bibfnamefont{M.~C.~N.} \bibnamefont{Fiolhais}},
  \bibinfo{author}{\bibfnamefont{H.}~\bibnamefont{Gon?alves}},
  \bibinfo{author}{\bibfnamefont{A.~G.~C.} \bibnamefont{Guerra}},
  \bibnamefont{et~al.} (\bibinfo{year}{2014}), \eprint{1408.7063}.

\bibitem[{\citenamefont{Sarmiento-Alvarado
  et~al.}(2014)\citenamefont{Sarmiento-Alvarado, Bouzas, and
  Larios}}]{Sarmiento-Alvarado:2014eha}
\bibinfo{author}{\bibfnamefont{I.}~\bibnamefont{Sarmiento-Alvarado}},
  \bibinfo{author}{\bibfnamefont{A.~O.} \bibnamefont{Bouzas}},
  \bibnamefont{and} \bibinfo{author}{\bibfnamefont{F.}~\bibnamefont{Larios}}
  (\bibinfo{year}{2014}), \eprint{1412.6679}.

\bibitem[{\citenamefont{Bach and Ohl}(2014)}]{Bach:2014zca}
\bibinfo{author}{\bibfnamefont{F.}~\bibnamefont{Bach}} \bibnamefont{and}
  \bibinfo{author}{\bibfnamefont{T.}~\bibnamefont{Ohl}},
  \bibinfo{journal}{Phys.Rev.} \textbf{\bibinfo{volume}{D90}},
  \bibinfo{pages}{074022} (\bibinfo{year}{2014}), \eprint{1407.2546}.

\bibitem[{\citenamefont{Buchmuller and Wyler}(1986)}]{Buchmuller:1985jz}
\bibinfo{author}{\bibfnamefont{W.}~\bibnamefont{Buchmuller}} \bibnamefont{and}
  \bibinfo{author}{\bibfnamefont{D.}~\bibnamefont{Wyler}},
  \bibinfo{journal}{Nucl.Phys.} \textbf{\bibinfo{volume}{B268}},
  \bibinfo{pages}{621} (\bibinfo{year}{1986}).

\bibitem[{\citenamefont{Peccei and Zhang}(1990)}]{Peccei:1989kr}
\bibinfo{author}{\bibfnamefont{R.}~\bibnamefont{Peccei}} \bibnamefont{and}
  \bibinfo{author}{\bibfnamefont{X.}~\bibnamefont{Zhang}},
  \bibinfo{journal}{Nucl.Phys.} \textbf{\bibinfo{volume}{B337}},
  \bibinfo{pages}{269} (\bibinfo{year}{1990}).

\bibitem[{\citenamefont{Georgi}(1993)}]{Georgi:1994qn}
\bibinfo{author}{\bibfnamefont{H.}~\bibnamefont{Georgi}},
  \bibinfo{journal}{Ann.Rev.Nucl.Part.Sci.} \textbf{\bibinfo{volume}{43}},
  \bibinfo{pages}{209} (\bibinfo{year}{1993}).

\bibitem[{\citenamefont{Larios and Yuan}(1997)}]{Larios:1996ib}
\bibinfo{author}{\bibfnamefont{F.}~\bibnamefont{Larios}} \bibnamefont{and}
  \bibinfo{author}{\bibfnamefont{C.}~\bibnamefont{Yuan}},
  \bibinfo{journal}{Phys.Rev.} \textbf{\bibinfo{volume}{D55}},
  \bibinfo{pages}{7218} (\bibinfo{year}{1997}), \eprint{hep-ph/9606397}.

\bibitem[{\citenamefont{Tait and Yuan}(2000)}]{Tait:2000sh}
\bibinfo{author}{\bibfnamefont{T.~M.} \bibnamefont{Tait}} \bibnamefont{and}
  \bibinfo{author}{\bibfnamefont{C.-P.} \bibnamefont{Yuan}},
  \bibinfo{journal}{Phys.Rev.} \textbf{\bibinfo{volume}{D63}},
  \bibinfo{pages}{014018} (\bibinfo{year}{2000}), \eprint{hep-ph/0007298}.

\bibitem[{\citenamefont{Aguilar-Saavedra}(2009{\natexlab{b}})}]{AguilarSaavedra:2008zc}
\bibinfo{author}{\bibfnamefont{J.}~\bibnamefont{Aguilar-Saavedra}},
  \bibinfo{journal}{Nucl.Phys.} \textbf{\bibinfo{volume}{B812}},
  \bibinfo{pages}{181} (\bibinfo{year}{2009}{\natexlab{b}}),
  \eprint{0811.3842}.

\bibitem[{\citenamefont{Drobnak et~al.}(2010)\citenamefont{Drobnak, Fajfer, and
  Kamenik}}]{Drobnak:2010ej}
\bibinfo{author}{\bibfnamefont{J.}~\bibnamefont{Drobnak}},
  \bibinfo{author}{\bibfnamefont{S.}~\bibnamefont{Fajfer}}, \bibnamefont{and}
  \bibinfo{author}{\bibfnamefont{J.~F.} \bibnamefont{Kamenik}},
  \bibinfo{journal}{Phys.Rev.} \textbf{\bibinfo{volume}{D82}},
  \bibinfo{pages}{114008} (\bibinfo{year}{2010}), \eprint{1010.2402}.

\bibitem[{\citenamefont{Degrande et~al.}(2015)\citenamefont{Degrande, Maltoni,
  Wang, and Zhang}}]{Degrande:2014tta}
\bibinfo{author}{\bibfnamefont{C.}~\bibnamefont{Degrande}},
  \bibinfo{author}{\bibfnamefont{F.}~\bibnamefont{Maltoni}},
  \bibinfo{author}{\bibfnamefont{J.}~\bibnamefont{Wang}}, \bibnamefont{and}
  \bibinfo{author}{\bibfnamefont{C.}~\bibnamefont{Zhang}},
  \bibinfo{journal}{Phys.Rev.} \textbf{\bibinfo{volume}{D91}},
  \bibinfo{pages}{034024} (\bibinfo{year}{2015}), \eprint{1412.5594}.

\bibitem[{\citenamefont{Arzt et~al.}(1995)\citenamefont{Arzt, Einhorn, and
  Wudka}}]{Arzt:1994gp}
\bibinfo{author}{\bibfnamefont{C.}~\bibnamefont{Arzt}},
  \bibinfo{author}{\bibfnamefont{M.}~\bibnamefont{Einhorn}}, \bibnamefont{and}
  \bibinfo{author}{\bibfnamefont{J.}~\bibnamefont{Wudka}},
  \bibinfo{journal}{Nucl.Phys.} \textbf{\bibinfo{volume}{B433}},
  \bibinfo{pages}{41} (\bibinfo{year}{1995}), \eprint{hep-ph/9405214}.

\bibitem[{\citenamefont{Li et~al.}(1993)\citenamefont{Li, Yang, and
  Hu}}]{Li:1992ga}
\bibinfo{author}{\bibfnamefont{C.-S.} \bibnamefont{Li}},
  \bibinfo{author}{\bibfnamefont{J.-M.} \bibnamefont{Yang}}, \bibnamefont{and}
  \bibinfo{author}{\bibfnamefont{B.-Q.} \bibnamefont{Hu}},
  \bibinfo{journal}{Phys.Rev.} \textbf{\bibinfo{volume}{D48}},
  \bibinfo{pages}{5425} (\bibinfo{year}{1993}).

\bibitem[{\citenamefont{Arzt}(1995)}]{Arzt:1993gz}
\bibinfo{author}{\bibfnamefont{C.}~\bibnamefont{Arzt}},
  \bibinfo{journal}{Phys.Lett.} \textbf{\bibinfo{volume}{B342}},
  \bibinfo{pages}{189} (\bibinfo{year}{1995}), \eprint{hep-ph/9304230}.

\bibitem[{The ATLAS collaboration({\natexlab{a}})}]{ATLAS:2013ula}
The ATLAS collaboration (\bibinfo{year}{2013}{\natexlab{a}}),
  \eprint{ATLAS-CONF-2013-032}.

\bibitem[{\citenamefont{Pumplin et~al.}(2002)\citenamefont{Pumplin, Stump,
  Huston, Lai, Nadolsky et~al.}}]{Pumplin:2002vw}
\bibinfo{author}{\bibfnamefont{J.}~\bibnamefont{Pumplin}},
  \bibinfo{author}{\bibfnamefont{D.}~\bibnamefont{Stump}},
  \bibinfo{author}{\bibfnamefont{J.}~\bibnamefont{Huston}},
  \bibinfo{author}{\bibfnamefont{H.}~\bibnamefont{Lai}},
  \bibinfo{author}{\bibfnamefont{P.~M.} \bibnamefont{Nadolsky}},
  \bibnamefont{et~al.}, \bibinfo{journal}{JHEP}
  \textbf{\bibinfo{volume}{0207}}, \bibinfo{pages}{012} (\bibinfo{year}{2002}),
  \eprint{hep-ph/0201195}.

\bibitem[{\citenamefont{Stelzer et~al.}(1997)\citenamefont{Stelzer, Sullivan,
  and Willenbrock}}]{Stelzer:1997ns}
\bibinfo{author}{\bibfnamefont{T.}~\bibnamefont{Stelzer}},
  \bibinfo{author}{\bibfnamefont{Z.}~\bibnamefont{Sullivan}}, \bibnamefont{and}
  \bibinfo{author}{\bibfnamefont{S.}~\bibnamefont{Willenbrock}},
  \bibinfo{journal}{Phys. Rev.} \textbf{\bibinfo{volume}{D56}},
  \bibinfo{pages}{5919} (\bibinfo{year}{1997}), \eprint{hep-ph/9705398}.

\bibitem[{\citenamefont{Zhu}(2002)}]{Zhu:2002uj}
\bibinfo{author}{\bibfnamefont{S.}~\bibnamefont{Zhu}},
  \bibinfo{journal}{Phys.Lett.} \textbf{\bibinfo{volume}{B524}},
  \bibinfo{pages}{283} (\bibinfo{year}{2002}).

\bibitem[{\citenamefont{Harris et~al.}(2002)\citenamefont{Harris, Laenen, Phaf,
  Sullivan, and Weinzierl}}]{Harris:2002md}
\bibinfo{author}{\bibfnamefont{B.}~\bibnamefont{Harris}},
  \bibinfo{author}{\bibfnamefont{E.}~\bibnamefont{Laenen}},
  \bibinfo{author}{\bibfnamefont{L.}~\bibnamefont{Phaf}},
  \bibinfo{author}{\bibfnamefont{Z.}~\bibnamefont{Sullivan}}, \bibnamefont{and}
  \bibinfo{author}{\bibfnamefont{S.}~\bibnamefont{Weinzierl}},
  \bibinfo{journal}{Phys.Rev.} \textbf{\bibinfo{volume}{D66}},
  \bibinfo{pages}{054024} (\bibinfo{year}{2002}), \eprint{hep-ph/0207055}.

\bibitem[{\citenamefont{Campbell et~al.}(2004)\citenamefont{Campbell, Ellis,
  and Tramontano}}]{Campbell:2004ch}
\bibinfo{author}{\bibfnamefont{J.~M.} \bibnamefont{Campbell}},
  \bibinfo{author}{\bibfnamefont{R.~K.} \bibnamefont{Ellis}}, \bibnamefont{and}
  \bibinfo{author}{\bibfnamefont{F.}~\bibnamefont{Tramontano}},
  \bibinfo{journal}{Phys.Rev.} \textbf{\bibinfo{volume}{D70}},
  \bibinfo{pages}{094012} (\bibinfo{year}{2004}), \eprint{hep-ph/0408158}.

\bibitem[{\citenamefont{Cao and Yuan}(2005)}]{Cao:2004ky}
\bibinfo{author}{\bibfnamefont{Q.-H.} \bibnamefont{Cao}} \bibnamefont{and}
  \bibinfo{author}{\bibfnamefont{C.-P.} \bibnamefont{Yuan}},
  \bibinfo{journal}{Phys.Rev.} \textbf{\bibinfo{volume}{D71}},
  \bibinfo{pages}{054022} (\bibinfo{year}{2005}), \eprint{hep-ph/0408180}.

\bibitem[{\citenamefont{Cao et~al.}(2005{\natexlab{a}})\citenamefont{Cao,
  Schwienhorst, and Yuan}}]{Cao:2004ap}
\bibinfo{author}{\bibfnamefont{Q.-H.} \bibnamefont{Cao}},
  \bibinfo{author}{\bibfnamefont{R.}~\bibnamefont{Schwienhorst}},
  \bibnamefont{and} \bibinfo{author}{\bibfnamefont{C.-P.} \bibnamefont{Yuan}},
  \bibinfo{journal}{Phys.Rev.} \textbf{\bibinfo{volume}{D71}},
  \bibinfo{pages}{054023} (\bibinfo{year}{2005}{\natexlab{a}}),
  \eprint{hep-ph/0409040}.

\bibitem[{\citenamefont{Cao et~al.}(2005{\natexlab{b}})\citenamefont{Cao,
  Schwienhorst, Benitez, Brock, and Yuan}}]{Cao:2005pq}
\bibinfo{author}{\bibfnamefont{Q.-H.} \bibnamefont{Cao}},
  \bibinfo{author}{\bibfnamefont{R.}~\bibnamefont{Schwienhorst}},
  \bibinfo{author}{\bibfnamefont{J.~A.} \bibnamefont{Benitez}},
  \bibinfo{author}{\bibfnamefont{R.}~\bibnamefont{Brock}}, \bibnamefont{and}
  \bibinfo{author}{\bibfnamefont{C.-P.} \bibnamefont{Yuan}},
  \bibinfo{journal}{Phys.Rev.} \textbf{\bibinfo{volume}{D72}},
  \bibinfo{pages}{094027} (\bibinfo{year}{2005}{\natexlab{b}}),
  \eprint{hep-ph/0504230}.

\bibitem[{\citenamefont{Campbell et~al.}(2009)\citenamefont{Campbell, Frederix,
  Maltoni, and Tramontano}}]{Campbell:2009ss}
\bibinfo{author}{\bibfnamefont{J.~M.} \bibnamefont{Campbell}},
  \bibinfo{author}{\bibfnamefont{R.}~\bibnamefont{Frederix}},
  \bibinfo{author}{\bibfnamefont{F.}~\bibnamefont{Maltoni}}, \bibnamefont{and}
  \bibinfo{author}{\bibfnamefont{F.}~\bibnamefont{Tramontano}},
  \bibinfo{journal}{Phys.Rev.Lett.} \textbf{\bibinfo{volume}{102}},
  \bibinfo{pages}{182003} (\bibinfo{year}{2009}), \eprint{0903.0005}.

\bibitem[{\citenamefont{Heim et~al.}(2010)\citenamefont{Heim, Cao,
  Schwienhorst, and Yuan}}]{Heim:2009ku}
\bibinfo{author}{\bibfnamefont{S.}~\bibnamefont{Heim}},
  \bibinfo{author}{\bibfnamefont{Q.-H.} \bibnamefont{Cao}},
  \bibinfo{author}{\bibfnamefont{R.}~\bibnamefont{Schwienhorst}},
  \bibnamefont{and} \bibinfo{author}{\bibfnamefont{C.-P.} \bibnamefont{Yuan}},
  \bibinfo{journal}{Phys.Rev.} \textbf{\bibinfo{volume}{D81}},
  \bibinfo{pages}{034005} (\bibinfo{year}{2010}), \eprint{0911.0620}.

\bibitem[{\citenamefont{Schwienhorst et~al.}(2011)\citenamefont{Schwienhorst,
  Yuan, Mueller, and Cao}}]{Schwienhorst:2010je}
\bibinfo{author}{\bibfnamefont{R.}~\bibnamefont{Schwienhorst}},
  \bibinfo{author}{\bibfnamefont{C.-P.} \bibnamefont{Yuan}},
  \bibinfo{author}{\bibfnamefont{C.}~\bibnamefont{Mueller}}, \bibnamefont{and}
  \bibinfo{author}{\bibfnamefont{Q.-H.} \bibnamefont{Cao}},
  \bibinfo{journal}{Phys.Rev.} \textbf{\bibinfo{volume}{D83}},
  \bibinfo{pages}{034019} (\bibinfo{year}{2011}), \eprint{1012.5132}.

\bibitem[{\citenamefont{Wang et~al.}(2013{\natexlab{a}})\citenamefont{Wang, Li,
  and Zhu}}]{Wang:2012dc}
\bibinfo{author}{\bibfnamefont{J.}~\bibnamefont{Wang}},
  \bibinfo{author}{\bibfnamefont{C.~S.} \bibnamefont{Li}}, \bibnamefont{and}
  \bibinfo{author}{\bibfnamefont{H.~X.} \bibnamefont{Zhu}},
  \bibinfo{journal}{Phys.Rev.} \textbf{\bibinfo{volume}{D87}},
  \bibinfo{pages}{034030} (\bibinfo{year}{2013}{\natexlab{a}}),
  \eprint{1210.7698}.

\bibitem[{\citenamefont{Kidonakis}(2014)}]{Kidonakis:2012rm}
\bibinfo{author}{\bibfnamefont{N.}~\bibnamefont{Kidonakis}},
  \bibinfo{journal}{Phys.Part.Nucl.} \textbf{\bibinfo{volume}{45}},
  \bibinfo{pages}{714} (\bibinfo{year}{2014}), \eprint{1210.7813}.

\bibitem[{\citenamefont{Aaltonen et~al.}(2014)}]{CDF:2014uma}
\bibinfo{author}{\bibfnamefont{T.~A.} \bibnamefont{Aaltonen}}
  \bibnamefont{et~al.} (\bibinfo{collaboration}{CDF Collaboration, D0
  Collaboration}), \bibinfo{journal}{Phys.Rev.Lett.}
  \textbf{\bibinfo{volume}{112}}, \bibinfo{pages}{231803}
  (\bibinfo{year}{2014}), \eprint{1402.5126}.

\bibitem[{\citenamefont{collaboration}(2015)}]{ATLAS:2015047}
\bibinfo{author}{\bibfnamefont{T.~A.} \bibnamefont{collaboration}}
  (\bibinfo{year}{2015}), \eprint{ATLAS-CONF-2015-047}.

\bibitem[{\citenamefont{Khachatryan et~al.}(2014)}]{Khachatryan:2014iya}
\bibinfo{author}{\bibfnamefont{V.}~\bibnamefont{Khachatryan}}
  \bibnamefont{et~al.} (\bibinfo{collaboration}{CMS Collaboration}),
  \bibinfo{journal}{JHEP} \textbf{\bibinfo{volume}{1406}}, \bibinfo{pages}{090}
  (\bibinfo{year}{2014}), \eprint{1403.7366}.

\bibitem[{ATLAS collaboration()}]{ATLAS-CONF-2014-007}
ATLAS collaboration (\bibinfo{year}{2014}), \eprint{ATLAS-CONF-2014-007,
  ATLAS-COM-CONF-2014-008}.

\bibitem[{\citenamefont{Collaboration}(2016)}]{CMS:2016ayb}
\bibinfo{author}{\bibfnamefont{C.}~\bibnamefont{Collaboration}}
  (\bibinfo{collaboration}{CMS}) (\bibinfo{year}{2016}),
  \eprint{CMS-PAS-TOP-16-003}.

\bibitem[{\citenamefont{Aaboud et~al.}(2016)}]{Aaboud:2016ymp}
\bibinfo{author}{\bibfnamefont{M.}~\bibnamefont{Aaboud}} \bibnamefont{et~al.}
  (\bibinfo{collaboration}{ATLAS}) (\bibinfo{year}{2016}), \eprint{1609.03920}.

\bibitem[{CMS Collaboration({\natexlab{b}})}]{CMS:2014efa}
CMS Collaboration (\bibinfo{year}{2014}{\natexlab{b}}),
  \eprint{CMS-PAS-TOP-14-009}.

\bibitem[{\citenamefont{collaboration}(2016)}]{ATLAS:2016lte}
\bibinfo{author}{\bibfnamefont{T.~A.} \bibnamefont{collaboration}}
  (\bibinfo{collaboration}{ATLAS}) (\bibinfo{year}{2016}).

\bibitem[{\citenamefont{James and Roos}(1975)}]{James:1975dr}
\bibinfo{author}{\bibfnamefont{F.}~\bibnamefont{James}} \bibnamefont{and}
  \bibinfo{author}{\bibfnamefont{M.}~\bibnamefont{Roos}},
  \bibinfo{journal}{Comput.Phys.Commun.} \textbf{\bibinfo{volume}{10}},
  \bibinfo{pages}{343} (\bibinfo{year}{1975}).

\bibitem[{\citenamefont{Grzadkowski and Misiak}(2008)}]{Grzadkowski:2008mf}
\bibinfo{author}{\bibfnamefont{B.}~\bibnamefont{Grzadkowski}} \bibnamefont{and}
  \bibinfo{author}{\bibfnamefont{M.}~\bibnamefont{Misiak}},
  \bibinfo{journal}{Phys.Rev.} \textbf{\bibinfo{volume}{D78}},
  \bibinfo{pages}{077501} (\bibinfo{year}{2008}), \eprint{0802.1413}.

\bibitem[{\citenamefont{Bernreuther et~al.}(2009)\citenamefont{Bernreuther,
  Gonzalez, and Wiebusch}}]{Bernreuther:2008us}
\bibinfo{author}{\bibfnamefont{W.}~\bibnamefont{Bernreuther}},
  \bibinfo{author}{\bibfnamefont{P.}~\bibnamefont{Gonzalez}}, \bibnamefont{and}
  \bibinfo{author}{\bibfnamefont{M.}~\bibnamefont{Wiebusch}},
  \bibinfo{journal}{Eur.Phys.J.} \textbf{\bibinfo{volume}{C60}},
  \bibinfo{pages}{197} (\bibinfo{year}{2009}), \eprint{0812.1643}.

\bibitem[{\citenamefont{Olive et~al.}(2014)}]{Agashe:2014kda}
\bibinfo{author}{\bibfnamefont{K.}~\bibnamefont{Olive}} \bibnamefont{et~al.}
  (\bibinfo{collaboration}{Particle Data Group}), \bibinfo{journal}{Chin.Phys.}
  \textbf{\bibinfo{volume}{C38}}, \bibinfo{pages}{090001}
  (\bibinfo{year}{2014}).

\bibitem[{\citenamefont{Chetyrkin et~al.}(1997)\citenamefont{Chetyrkin, Misiak,
  and Munz}}]{Chetyrkin:1996vx}
\bibinfo{author}{\bibfnamefont{K.~G.} \bibnamefont{Chetyrkin}},
  \bibinfo{author}{\bibfnamefont{M.}~\bibnamefont{Misiak}}, \bibnamefont{and}
  \bibinfo{author}{\bibfnamefont{M.}~\bibnamefont{Munz}},
  \bibinfo{journal}{Phys.Lett.} \textbf{\bibinfo{volume}{B400}},
  \bibinfo{pages}{206} (\bibinfo{year}{1997}), \eprint{hep-ph/9612313}.

\bibitem[{\citenamefont{Burdman et~al.}(2000)\citenamefont{Burdman,
  Gonzalez-Garcia, and Novaes}}]{Burdman:1999fw}
\bibinfo{author}{\bibfnamefont{G.}~\bibnamefont{Burdman}},
  \bibinfo{author}{\bibfnamefont{M.}~\bibnamefont{Gonzalez-Garcia}},
  \bibnamefont{and} \bibinfo{author}{\bibfnamefont{S.}~\bibnamefont{Novaes}},
  \bibinfo{journal}{Phys.Rev.} \textbf{\bibinfo{volume}{D61}},
  \bibinfo{pages}{114016} (\bibinfo{year}{2000}), \eprint{hep-ph/9906329}.

\bibitem[{\citenamefont{Drobnak et~al.}(2012)\citenamefont{Drobnak, Fajfer, and
  Kamenik}}]{Drobnak:2011aa}
\bibinfo{author}{\bibfnamefont{J.}~\bibnamefont{Drobnak}},
  \bibinfo{author}{\bibfnamefont{S.}~\bibnamefont{Fajfer}}, \bibnamefont{and}
  \bibinfo{author}{\bibfnamefont{J.~F.} \bibnamefont{Kamenik}},
  \bibinfo{journal}{Nucl.Phys.} \textbf{\bibinfo{volume}{B855}},
  \bibinfo{pages}{82} (\bibinfo{year}{2012}), \eprint{1109.2357}.

\bibitem[{\citenamefont{Drobnak et~al.}(2011)\citenamefont{Drobnak, Fajfer, and
  Kamenik}}]{Drobnak:2011wj}
\bibinfo{author}{\bibfnamefont{J.}~\bibnamefont{Drobnak}},
  \bibinfo{author}{\bibfnamefont{S.}~\bibnamefont{Fajfer}}, \bibnamefont{and}
  \bibinfo{author}{\bibfnamefont{J.~F.} \bibnamefont{Kamenik}},
  \bibinfo{journal}{Phys.Lett.} \textbf{\bibinfo{volume}{B701}},
  \bibinfo{pages}{234} (\bibinfo{year}{2011}), \eprint{1102.4347}.

\bibitem[{\citenamefont{Mohapatra and
  Pati}(1975{\natexlab{a}})}]{Mohapatra:1974gc}
\bibinfo{author}{\bibfnamefont{R.}~\bibnamefont{Mohapatra}} \bibnamefont{and}
  \bibinfo{author}{\bibfnamefont{J.~C.} \bibnamefont{Pati}},
  \bibinfo{journal}{Phys.Rev.} \textbf{\bibinfo{volume}{D11}},
  \bibinfo{pages}{2558} (\bibinfo{year}{1975}{\natexlab{a}}).

\bibitem[{\citenamefont{Mohapatra and
  Pati}(1975{\natexlab{b}})}]{Mohapatra:1974hk}
\bibinfo{author}{\bibfnamefont{R.~N.} \bibnamefont{Mohapatra}}
  \bibnamefont{and} \bibinfo{author}{\bibfnamefont{J.~C.} \bibnamefont{Pati}},
  \bibinfo{journal}{Phys.Rev.} \textbf{\bibinfo{volume}{D11}},
  \bibinfo{pages}{566} (\bibinfo{year}{1975}{\natexlab{b}}).

\bibitem[{\citenamefont{Mohapatra and Senjanovic}(1981)}]{Mohapatra:1980yp}
\bibinfo{author}{\bibfnamefont{R.~N.} \bibnamefont{Mohapatra}}
  \bibnamefont{and}
  \bibinfo{author}{\bibfnamefont{G.}~\bibnamefont{Senjanovic}},
  \bibinfo{journal}{Phys.Rev.} \textbf{\bibinfo{volume}{D23}},
  \bibinfo{pages}{165} (\bibinfo{year}{1981}).

\bibitem[{\citenamefont{Georgi et~al.}(1989)\citenamefont{Georgi, Jenkins, and
  Simmons}}]{Georgi:1989ic}
\bibinfo{author}{\bibfnamefont{H.}~\bibnamefont{Georgi}},
  \bibinfo{author}{\bibfnamefont{E.~E.} \bibnamefont{Jenkins}},
  \bibnamefont{and} \bibinfo{author}{\bibfnamefont{E.~H.}
  \bibnamefont{Simmons}}, \bibinfo{journal}{Phys.Rev.Lett.}
  \textbf{\bibinfo{volume}{62}}, \bibinfo{pages}{2789} (\bibinfo{year}{1989}).

\bibitem[{\citenamefont{Georgi et~al.}(1990)\citenamefont{Georgi, Jenkins, and
  Simmons}}]{Georgi:1989xz}
\bibinfo{author}{\bibfnamefont{H.}~\bibnamefont{Georgi}},
  \bibinfo{author}{\bibfnamefont{E.~E.} \bibnamefont{Jenkins}},
  \bibnamefont{and} \bibinfo{author}{\bibfnamefont{E.~H.}
  \bibnamefont{Simmons}}, \bibinfo{journal}{Nucl.Phys.}
  \textbf{\bibinfo{volume}{B331}}, \bibinfo{pages}{541} (\bibinfo{year}{1990}).

\bibitem[{\citenamefont{Li and Ma}(1981)}]{Li:1981nk}
\bibinfo{author}{\bibfnamefont{X.}~\bibnamefont{Li}} \bibnamefont{and}
  \bibinfo{author}{\bibfnamefont{E.}~\bibnamefont{Ma}},
  \bibinfo{journal}{Phys.Rev.Lett.} \textbf{\bibinfo{volume}{47}},
  \bibinfo{pages}{1788} (\bibinfo{year}{1981}).

\bibitem[{\citenamefont{Malkawi et~al.}(1996)\citenamefont{Malkawi, Tait, and
  Yuan}}]{Malkawi:1996fs}
\bibinfo{author}{\bibfnamefont{E.}~\bibnamefont{Malkawi}},
  \bibinfo{author}{\bibfnamefont{T.~M.} \bibnamefont{Tait}}, \bibnamefont{and}
  \bibinfo{author}{\bibfnamefont{C.}~\bibnamefont{Yuan}},
  \bibinfo{journal}{Phys.Lett.} \textbf{\bibinfo{volume}{B385}},
  \bibinfo{pages}{304} (\bibinfo{year}{1996}), \eprint{hep-ph/9603349}.

\bibitem[{\citenamefont{He et~al.}(2000)\citenamefont{He, Tait, and
  Yuan}}]{He:1999vp}
\bibinfo{author}{\bibfnamefont{H.-J.} \bibnamefont{He}},
  \bibinfo{author}{\bibfnamefont{T.~M.} \bibnamefont{Tait}}, \bibnamefont{and}
  \bibinfo{author}{\bibfnamefont{C.}~\bibnamefont{Yuan}},
  \bibinfo{journal}{Phys.Rev.} \textbf{\bibinfo{volume}{D62}},
  \bibinfo{pages}{011702} (\bibinfo{year}{2000}), \eprint{hep-ph/9911266}.

\bibitem[{\citenamefont{Chivukula et~al.}(2004)\citenamefont{Chivukula, He,
  Howard, and Simmons}}]{Chivukula:2003wj}
\bibinfo{author}{\bibfnamefont{R.~S.} \bibnamefont{Chivukula}},
  \bibinfo{author}{\bibfnamefont{H.-J.} \bibnamefont{He}},
  \bibinfo{author}{\bibfnamefont{J.}~\bibnamefont{Howard}}, \bibnamefont{and}
  \bibinfo{author}{\bibfnamefont{E.~H.} \bibnamefont{Simmons}},
  \bibinfo{journal}{Phys.Rev.} \textbf{\bibinfo{volume}{D69}},
  \bibinfo{pages}{015009} (\bibinfo{year}{2004}), \eprint{hep-ph/0307209}.

\bibitem[{\citenamefont{Du et~al.}(2012)\citenamefont{Du, He, Kuang, Zhang,
  Christensen et~al.}}]{Du:2012vh}
\bibinfo{author}{\bibfnamefont{C.}~\bibnamefont{Du}},
  \bibinfo{author}{\bibfnamefont{H.-J.} \bibnamefont{He}},
  \bibinfo{author}{\bibfnamefont{Y.-P.} \bibnamefont{Kuang}},
  \bibinfo{author}{\bibfnamefont{B.}~\bibnamefont{Zhang}},
  \bibinfo{author}{\bibfnamefont{N.~D.} \bibnamefont{Christensen}},
  \bibnamefont{et~al.}, \bibinfo{journal}{Phys.Rev.}
  \textbf{\bibinfo{volume}{D86}}, \bibinfo{pages}{095011}
  (\bibinfo{year}{2012}), \eprint{1206.6022}.

\bibitem[{\citenamefont{Abe et~al.}(2013)\citenamefont{Abe, Chen, and
  He}}]{Abe:2012fb}
\bibinfo{author}{\bibfnamefont{T.}~\bibnamefont{Abe}},
  \bibinfo{author}{\bibfnamefont{N.}~\bibnamefont{Chen}}, \bibnamefont{and}
  \bibinfo{author}{\bibfnamefont{H.-J.} \bibnamefont{He}},
  \bibinfo{journal}{JHEP} \textbf{\bibinfo{volume}{1301}}, \bibinfo{pages}{082}
  (\bibinfo{year}{2013}), \eprint{1207.4103}.

\bibitem[{\citenamefont{Wang et~al.}(2013{\natexlab{b}})\citenamefont{Wang, Du,
  and He}}]{Wang:2013jwa}
\bibinfo{author}{\bibfnamefont{X.-F.} \bibnamefont{Wang}},
  \bibinfo{author}{\bibfnamefont{C.}~\bibnamefont{Du}}, \bibnamefont{and}
  \bibinfo{author}{\bibfnamefont{H.-J.} \bibnamefont{He}},
  \bibinfo{journal}{Phys.Lett.} \textbf{\bibinfo{volume}{B723}},
  \bibinfo{pages}{314} (\bibinfo{year}{2013}{\natexlab{b}}),
  \eprint{1304.2257}.

\bibitem[{\citenamefont{Berger et~al.}(2011)\citenamefont{Berger, Cao, Yu, and
  Yuan}}]{Berger:2011xk}
\bibinfo{author}{\bibfnamefont{E.~L.} \bibnamefont{Berger}},
  \bibinfo{author}{\bibfnamefont{Q.-H.} \bibnamefont{Cao}},
  \bibinfo{author}{\bibfnamefont{J.-H.} \bibnamefont{Yu}}, \bibnamefont{and}
  \bibinfo{author}{\bibfnamefont{C.-P.} \bibnamefont{Yuan}},
  \bibinfo{journal}{Phys.Rev.} \textbf{\bibinfo{volume}{D84}},
  \bibinfo{pages}{095026} (\bibinfo{year}{2011}), \eprint{1108.3613}.

\bibitem[{\citenamefont{Abdallah et~al.}(2009)}]{Abdallah:2008ab}
\bibinfo{author}{\bibfnamefont{J.}~\bibnamefont{Abdallah}} \bibnamefont{et~al.}
  (\bibinfo{collaboration}{DELPHI Collaboration}),
  \bibinfo{journal}{Eur.Phys.J.} \textbf{\bibinfo{volume}{C60}},
  \bibinfo{pages}{1} (\bibinfo{year}{2009}), \eprint{0901.4461}.

\bibitem[{\citenamefont{Agashe et~al.}(2006)\citenamefont{Agashe, Contino,
  Da~Rold, and Pomarol}}]{Agashe:2006at}
\bibinfo{author}{\bibfnamefont{K.}~\bibnamefont{Agashe}},
  \bibinfo{author}{\bibfnamefont{R.}~\bibnamefont{Contino}},
  \bibinfo{author}{\bibfnamefont{L.}~\bibnamefont{Da~Rold}}, \bibnamefont{and}
  \bibinfo{author}{\bibfnamefont{A.}~\bibnamefont{Pomarol}},
  \bibinfo{journal}{Phys.Lett.} \textbf{\bibinfo{volume}{B641}},
  \bibinfo{pages}{62} (\bibinfo{year}{2006}), \eprint{hep-ph/0605341}.

\bibitem[{\citenamefont{Peskin and Takeuchi}(1992)}]{Peskin:1991sw}
\bibinfo{author}{\bibfnamefont{M.~E.} \bibnamefont{Peskin}} \bibnamefont{and}
  \bibinfo{author}{\bibfnamefont{T.}~\bibnamefont{Takeuchi}},
  \bibinfo{journal}{Phys.Rev.} \textbf{\bibinfo{volume}{D46}},
  \bibinfo{pages}{381} (\bibinfo{year}{1992}).

\bibitem[{\citenamefont{Lavoura and Silva}(1993)}]{Lavoura:1992np}
\bibinfo{author}{\bibfnamefont{L.}~\bibnamefont{Lavoura}} \bibnamefont{and}
  \bibinfo{author}{\bibfnamefont{J.~P.} \bibnamefont{Silva}},
  \bibinfo{journal}{Phys.Rev.} \textbf{\bibinfo{volume}{D47}},
  \bibinfo{pages}{2046} (\bibinfo{year}{1993}).

\bibitem[{\citenamefont{Anastasiou et~al.}(2009)\citenamefont{Anastasiou,
  Furlan, and Santiago}}]{Anastasiou:2009rv}
\bibinfo{author}{\bibfnamefont{C.}~\bibnamefont{Anastasiou}},
  \bibinfo{author}{\bibfnamefont{E.}~\bibnamefont{Furlan}}, \bibnamefont{and}
  \bibinfo{author}{\bibfnamefont{J.}~\bibnamefont{Santiago}},
  \bibinfo{journal}{Phys.Rev.} \textbf{\bibinfo{volume}{D79}},
  \bibinfo{pages}{075003} (\bibinfo{year}{2009}), \eprint{0901.2117}.

\bibitem[{\citenamefont{Cai}(2013)}]{Cai:2012ji}
\bibinfo{author}{\bibfnamefont{H.}~\bibnamefont{Cai}}, \bibinfo{journal}{JHEP}
  \textbf{\bibinfo{volume}{1302}}, \bibinfo{pages}{104} (\bibinfo{year}{2013}),
  \eprint{1210.5200}.

\bibitem[{\citenamefont{Baak et~al.}(2014)}]{Baak:2014ora}
\bibinfo{author}{\bibfnamefont{M.}~\bibnamefont{Baak}} \bibnamefont{et~al.}
  (\bibinfo{collaboration}{Gfitter Group}), \bibinfo{journal}{Eur.Phys.J.}
  \textbf{\bibinfo{volume}{C74}}, \bibinfo{pages}{3046} (\bibinfo{year}{2014}),
  \eprint{1407.3792}.

\bibitem[{\citenamefont{Khachatryan et~al.}(2015)}]{Khachatryan:2015axa}
\bibinfo{author}{\bibfnamefont{V.}~\bibnamefont{Khachatryan}}
  \bibnamefont{et~al.} (\bibinfo{collaboration}{CMS}) (\bibinfo{year}{2015}),
  \eprint{1503.01952}.

\bibitem[{The ATLAS collaboration({\natexlab{b}})}]{ATLAS:2015VLQ}
The ATLAS collaboration (\bibinfo{year}{2015}{\natexlab{b}}),
  \eprint{ATLAS-CONF-2015-012, ATLAS-COM-CONF-2015-012}.

\bibitem[{\citenamefont{Xiao and Yu}(2014)}]{Xiao:2014kba}
\bibinfo{author}{\bibfnamefont{M.-L.} \bibnamefont{Xiao}} \bibnamefont{and}
  \bibinfo{author}{\bibfnamefont{J.-H.} \bibnamefont{Yu}},
  \bibinfo{journal}{Phys.Rev.} \textbf{\bibinfo{volume}{D90}},
  \bibinfo{pages}{014007} (\bibinfo{year}{2014}), \eprint{1404.0681}.

\bibitem[{\citenamefont{Arkani-Hamed et~al.}(2002)\citenamefont{Arkani-Hamed,
  Cohen, Katz, and Nelson}}]{ArkaniHamed:2002qy}
\bibinfo{author}{\bibfnamefont{N.}~\bibnamefont{Arkani-Hamed}},
  \bibinfo{author}{\bibfnamefont{A.}~\bibnamefont{Cohen}},
  \bibinfo{author}{\bibfnamefont{E.}~\bibnamefont{Katz}}, \bibnamefont{and}
  \bibinfo{author}{\bibfnamefont{A.}~\bibnamefont{Nelson}},
  \bibinfo{journal}{JHEP} \textbf{\bibinfo{volume}{0207}}, \bibinfo{pages}{034}
  (\bibinfo{year}{2002}), \eprint{hep-ph/0206021}.

\bibitem[{\citenamefont{Reuter et~al.}(2013)\citenamefont{Reuter, Tonini, and
  de~Vries}}]{Reuter:2013zja}
\bibinfo{author}{\bibfnamefont{J.}~\bibnamefont{Reuter}},
  \bibinfo{author}{\bibfnamefont{M.}~\bibnamefont{Tonini}}, \bibnamefont{and}
  \bibinfo{author}{\bibfnamefont{M.}~\bibnamefont{de~Vries}}
  (\bibinfo{year}{2013}), \eprint{1307.5010}.

\bibitem[{\citenamefont{Cheng and Low}(2003)}]{Cheng:2003ju}
\bibinfo{author}{\bibfnamefont{H.-C.} \bibnamefont{Cheng}} \bibnamefont{and}
  \bibinfo{author}{\bibfnamefont{I.}~\bibnamefont{Low}},
  \bibinfo{journal}{JHEP} \textbf{\bibinfo{volume}{0309}}, \bibinfo{pages}{051}
  (\bibinfo{year}{2003}), \eprint{hep-ph/0308199}.

\bibitem[{\citenamefont{Cheng and Low}(2004)}]{Cheng:2004yc}
\bibinfo{author}{\bibfnamefont{H.-C.} \bibnamefont{Cheng}} \bibnamefont{and}
  \bibinfo{author}{\bibfnamefont{I.}~\bibnamefont{Low}},
  \bibinfo{journal}{JHEP} \textbf{\bibinfo{volume}{0408}}, \bibinfo{pages}{061}
  (\bibinfo{year}{2004}), \eprint{hep-ph/0405243}.

\bibitem[{\citenamefont{Low}(2004)}]{Low:2004xc}
\bibinfo{author}{\bibfnamefont{I.}~\bibnamefont{Low}}, \bibinfo{journal}{JHEP}
  \textbf{\bibinfo{volume}{0410}}, \bibinfo{pages}{067} (\bibinfo{year}{2004}),
  \eprint{hep-ph/0409025}.

\bibitem[{\citenamefont{Reuter et~al.}(2014)\citenamefont{Reuter, Tonini, and
  de~Vries}}]{Reuter:2013iya}
\bibinfo{author}{\bibfnamefont{J.}~\bibnamefont{Reuter}},
  \bibinfo{author}{\bibfnamefont{M.}~\bibnamefont{Tonini}}, \bibnamefont{and}
  \bibinfo{author}{\bibfnamefont{M.}~\bibnamefont{de~Vries}},
  \bibinfo{journal}{JHEP} \textbf{\bibinfo{volume}{1402}}, \bibinfo{pages}{053}
  (\bibinfo{year}{2014}), \eprint{1310.2918}.

\bibitem[{\citenamefont{Hubisz et~al.}(2006)\citenamefont{Hubisz, Meade, Noble,
  and Perelstein}}]{Hubisz:2005tx}
\bibinfo{author}{\bibfnamefont{J.}~\bibnamefont{Hubisz}},
  \bibinfo{author}{\bibfnamefont{P.}~\bibnamefont{Meade}},
  \bibinfo{author}{\bibfnamefont{A.}~\bibnamefont{Noble}}, \bibnamefont{and}
  \bibinfo{author}{\bibfnamefont{M.}~\bibnamefont{Perelstein}},
  \bibinfo{journal}{JHEP} \textbf{\bibinfo{volume}{0601}}, \bibinfo{pages}{135}
  (\bibinfo{year}{2006}), \eprint{hep-ph/0506042}.

\end{thebibliography}

\end{document}